\pdfoutput=1
\documentclass[12pt,a4paper]{article}
\usepackage{ifthen} % for conditional statements
\newboolean{pdflatex}
\setboolean{pdflatex}{true} % False for eps figures 

\newboolean{articletitles}
\setboolean{articletitles}{true} % False removes titles in references

\newboolean{uprightparticles}
\setboolean{uprightparticles}{true} %True for upright particle symbols

\def\paperauthors{LHCb collaboration} % Leave as is for PAPER, CONF and FIGURE
\def\paperasciititle{Study of 
the doubly charmed tetraquark Tcc+} 
\def\papertitle{Study of  
the~doubly charmed tetraquark~\Tcc} % Latex formatted title
\def\paperkeywords{{High Energy Physics}, {LHCb}} % Comma separated list
\def\papercopyright{\the\year\ CERN for the benefit of the LHCb collaboration} % new since 9/Apr/2018
\def\paperlicence{CC BY 4.0 licence}
\def\paperlicenceurl{https://creativecommons.org/licenses/by/4.0/}

\usepackage{lscape}
\usepackage{comment}
\usepackage{graphpap} 
\usepackage{multirow} 
\usepackage{rotating} 
\usepackage{mathrsfs}
\usepackage{mathtools} %% dcases 
\usepackage{dashrule}
\usepackage{tikz}
\usepackage{enumerate}
\usepackage{enumitem}
\usetikzlibrary{patterns}
\usepackage{nicefrac} %% 1/2, ...
\usepackage{afterpage}
\usepackage{xcolor}
\usepackage[toc,page]{appendix}
\usepackage[normalem]{ulem} 
\usepackage{wasysym}

\usepackage[top=1in, bottom=1.25in, left=1in, right=1in]{geometry}

\usepackage{microtype}
\usepackage{lineno}  % for line numbering during review
\usepackage{xspace} % To avoid problems with missing or double spaces after
\usepackage{caption} %these three command get the figure and table captions automatically small

\usepackage{graphicx}  % to include figures (can also use other packages)
\usepackage{color}
\usepackage{colortbl}
\graphicspath{{./figs/}{./figs/supplementary/}} % Make Latex search fig subdir for figures
\usepackage{amsmath} % Adds a large collection of math symbols
\usepackage{amssymb}
\usepackage{amsfonts}
\usepackage{upgreek} % Adds in support for greek letters in roman typeset

\newcommand*\patchAmsMathEnvironmentForLineno[1]{%
\expandafter\let\csname old#1\expandafter\endcsname\csname #1\endcsname
\expandafter\let\csname oldend#1\expandafter\endcsname\csname
end#1\endcsname
 \renewenvironment{#1}%
   {\linenomath\csname old#1\endcsname}%
   {\csname oldend#1\endcsname\endlinenomath}%
}
\newcommand*\patchBothAmsMathEnvironmentsForLineno[1]{%
  \patchAmsMathEnvironmentForLineno{#1}%
  \patchAmsMathEnvironmentForLineno{#1*}%
}
\AtBeginDocument{%
\patchBothAmsMathEnvironmentsForLineno{equation}%
\patchBothAmsMathEnvironmentsForLineno{align}%
\patchBothAmsMathEnvironmentsForLineno{flalign}%
\patchBothAmsMathEnvironmentsForLineno{alignat}%
\patchBothAmsMathEnvironmentsForLineno{gather}%
\patchBothAmsMathEnvironmentsForLineno{multline}%
\patchBothAmsMathEnvironmentsForLineno{eqnarray}%
}

\usepackage{hyperxmp}

\usepackage[pdftex,
            pdfauthor={\paperauthors},
            pdftitle={\paperasciititle},
            pdfkeywords={\paperkeywords},
            pdfcopyright={Copyright (C) \papercopyright},
            pdflicenseurl={\paperlicenceurl}]{hyperref}
\usepackage[colorinlistoftodos,textsize=scriptsize]{todonotes}

\usepackage[bottom,flushmargin,hang,multiple]{footmisc}

\usepackage[all]{hypcap} % Internal hyperlinks to floats.

\usepackage{xspace} 
\usepackage{upgreek}

\def\lhcb   {\mbox{LHCb}\xspace}

\def\MagUp {\mbox{\em Mag\kern -0.05em Up}\xspace}

\ifthenelse{\boolean{uprightparticles}}%
{
 
 \def\Pgamma      {\ensuremath{\upgamma}\xspace}

 \def\Pmu         {\ensuremath{\upmu}\xspace}

 \def\Ppi         {\ensuremath{\uppi}\xspace}                 
                  
 \def\Prho        {\ensuremath{\uprho}\xspace}

 \def\Pchi        {\ensuremath{\upchi}\xspace}                 
 \def\Ppsi        {\ensuremath{\uppsi}\xspace}

 \def\PDelta      {\ensuremath{\Delta}\xspace}                 
 \def\PXi         {\ensuremath{\Xi}\xspace}                 
 \def\PLambda     {\ensuremath{\Lambda}\xspace}                 
 \def\PSigma      {\ensuremath{\Sigma}\xspace}                 
 \def\POmega      {\ensuremath{\Omega}\xspace}                 
 \def\PUpsilon    {\ensuremath{\Upsilon}\xspace}

 \def\PB      {\ensuremath{\mathrm{B}}\xspace}                 
                  
 \def\PD      {\ensuremath{\mathrm{D}}\xspace}

 \def\PJ      {\ensuremath{\mathrm{J}}\xspace}                 
 \def\PK      {\ensuremath{\mathrm{K}}\xspace}

 \def\PP      {\ensuremath{\mathrm{P}}\xspace}                 
 \def\PQ      {\ensuremath{\mathrm{Q}}\xspace}                 
 \def\PR      {\ensuremath{\mathrm{R}}\xspace}                 
 \def\PS      {\ensuremath{\mathrm{S}}\xspace}                 
 \def\PT      {\ensuremath{\mathrm{T}}\xspace}

 \def\PX      {\ensuremath{\mathrm{X}}\xspace}                 
                  
 \def\PZ      {\ensuremath{\mathrm{Z}}\xspace}                 
                  
 \def\Pb      {\ensuremath{\mathrm{b}}\xspace}                 
 \def\Pc      {\ensuremath{\mathrm{c}}\xspace}                 
 \def\Pd      {\ensuremath{\mathrm{d}}\xspace}

 \def\Pi      {\ensuremath{\mathrm{i}}\xspace}

 \def\Pp      {\ensuremath{\mathrm{p}}\xspace}                 
 \def\Pq      {\ensuremath{\mathrm{q}}\xspace}                 
                  
 \def\Ps      {\ensuremath{\mathrm{s}}\xspace}                 
                  
 \def\Pu      {\ensuremath{\mathrm{u}}\xspace}

 \def\thebaroffset{0.0em}
}
{
 
 \def\Pgamma      {\ensuremath{\gamma}\xspace}

 \def\Pmu         {\ensuremath{\mu}\xspace}

 \def\Ppi         {\ensuremath{\pi}\xspace}                 
                  
 \def\Prho        {\ensuremath{\rho}\xspace}

 \def\Pchi        {\ensuremath{\chi}\xspace}                 
 \def\Ppsi        {\ensuremath{\psi}\xspace}                 
                  
 \mathchardef\PDelta="7101
 \mathchardef\PXi="7104
 \mathchardef\PLambda="7103
 \mathchardef\PSigma="7106
 \mathchardef\POmega="710A
 \mathchardef\PUpsilon="7107
                  
 \def\PB      {\ensuremath{B}\xspace}                 
                  
 \def\PD      {\ensuremath{D}\xspace}

 \def\PJ      {\ensuremath{J}\xspace}                 
 \def\PK      {\ensuremath{K}\xspace}

 \def\PP      {\ensuremath{P}\xspace}                 
 \def\PQ      {\ensuremath{Q}\xspace}                 
 \def\PR      {\ensuremath{R}\xspace}                 
 \def\PS      {\ensuremath{S}\xspace}                 
 \def\PT      {\ensuremath{T}\xspace}

 \def\PX      {\ensuremath{X}\xspace}                 
                  
 \def\PZ      {\ensuremath{Z}\xspace}                 
                  
 \def\Pb      {\ensuremath{b}\xspace}                 
 \def\Pc      {\ensuremath{c}\xspace}                 
 \def\Pd      {\ensuremath{d}\xspace}

 \def\Pi      {\ensuremath{i}\xspace}

 \def\Pp      {\ensuremath{p}\xspace}                 
 \def\Pq      {\ensuremath{q}\xspace}                 
                  
 \def\Ps      {\ensuremath{s}\xspace}                 
                  
 \def\Pu      {\ensuremath{u}\xspace}

 \def\thebaroffset{0.18em}
}
\newcommand{\offsetoverline}[2][\thebaroffset]{\kern #1\overline{\kern -#1 #2}}%

\makeatletter
\ifcase \@ptsize \relax% 10pt
  \newcommand{\miniscule}{\@setfontsize\miniscule{4}{5}}% \tiny: 5/6
\or% 11pt
  \newcommand{\miniscule}{\@setfontsize\miniscule{5}{6}}% \tiny: 6/7
\or% 12pt
  \newcommand{\miniscule}{\@setfontsize\miniscule{5}{6}}% \tiny: 6/7
\fi
\makeatother

\DeclareRobustCommand{\optbar}[1]{\shortstack{{\miniscule (\rule[.5ex]{1.25em}{.18mm})}
  \\ [-.7ex] $#1$}}

\def\mumu       {{\ensuremath{\Pmu^+\Pmu^-}}\xspace}

\def\g      {{\ensuremath{\Pgamma}}\xspace}

\def\quark     {{\ensuremath{\Pq}}\xspace}
\def\quarkbar  {{\ensuremath{\overline \quark}}\xspace}

\def\uquark    {{\ensuremath{\Pu}}\xspace}
\def\uquarkbar {{\ensuremath{\overline \uquark}}\xspace}

\def\dquark    {{\ensuremath{\Pd}}\xspace}
\def\dquarkbar {{\ensuremath{\overline \dquark}}\xspace}

\def\squark    {{\ensuremath{\Ps}}\xspace}
\def\squarkbar {{\ensuremath{\overline \squark}}\xspace}

\def\cquark    {{\ensuremath{\Pc}}\xspace}
\def\cquarkbar {{\ensuremath{\overline \cquark}}\xspace}

\def\bquark    {{\ensuremath{\Pb}}\xspace}
\def\bquarkbar {{\ensuremath{\overline \bquark}}\xspace}

\def\pion   {{\ensuremath{\Ppi}}\xspace}
\def\piz    {{\ensuremath{\pion^0}}\xspace}
\def\pip    {{\ensuremath{\pion^+}}\xspace}
\def\pim    {{\ensuremath{\pion^-}}\xspace}

\def\kaon    {{\ensuremath{\PK}}\xspace}
\def\KorKbar {\kern \thebaroffset\optbar{\kern -\thebaroffset \PK}{}\xspace}

\def\Kp      {{\ensuremath{\kaon^+}}\xspace}
\def\Km      {{\ensuremath{\kaon^-}}\xspace}

\def\KS      {{\ensuremath{\kaon^0_{\mathrm{S}}}}\xspace}

\def\Dbar    {{\ensuremath{\offsetoverline{\PD}}}\xspace}
\def\D       {{\ensuremath{\PD}}\xspace}

\def\DorDbar {\kern \thebaroffset\optbar{\kern -\thebaroffset \PD}\xspace}
\def\Dz      {{\ensuremath{\D^0}}\xspace}
\def\Dzb     {{\ensuremath{\Dbar{}^0}}\xspace}
\def\Dp      {{\ensuremath{\D^+}}\xspace}
\def\Dm      {{\ensuremath{\D^-}}\xspace}

\def\DpDm    {\ensuremath{\Dp {\kern -0.16em \Dm}}\xspace}
\def\Dstar   {{\ensuremath{\D^*}}\xspace}

\def\Dstarz  {{\ensuremath{\D^{*0}}}\xspace}

\def\Dstarp  {{\ensuremath{\D^{*+}}}\xspace}

\def\B       {{\ensuremath{\PB}}\xspace}
\def\Bbar    {{\ensuremath{\offsetoverline{\PB}}}\xspace}

\def\BorBbar {\kern \thebaroffset\optbar{\kern -\thebaroffset \PB}\xspace}

\def\Bd      {{\ensuremath{\B^0}}\xspace}

\def\BdorBdbar {\kern \thebaroffset\optbar{\kern -\thebaroffset \Bd}\xspace}
\def\Bu      {{\ensuremath{\B^+}}\xspace}

\def\Bs      {{\ensuremath{\B^0_\squark}}\xspace}

\def\BsorBsbar {\kern \thebaroffset\optbar{\kern -\thebaroffset \Bs}\xspace}

\def\jpsi     {{\ensuremath{{\PJ\mskip -3mu/\mskip -2mu\Ppsi}}}\xspace}
\def\psitwos  {{\ensuremath{\Ppsi{(2\PS)}}}\xspace}

\def\chiczero {{\ensuremath{\Pchi_{\cquark 0}}}\xspace}
\def\chicone  {{\ensuremath{\Pchi_{\cquark 1}}}\xspace}

\def\Y#1S{\ensuremath{\PUpsilon{(#1S)}}\xspace}

\def\proton      {{\ensuremath{\Pp}}\xspace}

\def\Lbar        {{\ensuremath{\offsetoverline{\PLambda}}}\xspace}
\def\LorLbar     {\kern \thebaroffset\optbar{\kern -\thebaroffset \PLambda}\xspace}

\def\Sigmares    {{\ensuremath{\PSigma}}\xspace}

\def\Sigmaresbar {{\ensuremath{\offsetoverline{\Sigmares}}}\xspace}

\def\Xires       {{\ensuremath{\PXi}}\xspace}

\def\Lcbar       {{\ensuremath{\Lbar{}^-_\cquark}}\xspace}

\def\Xiccpp      {{\ensuremath{\Xires^{++}_{\cquark\cquark}}}\xspace}

\def\Lbbar        {{\ensuremath{\Lbar{}^0_\bquark}}\xspace}

\def\Sigmabp      {{\ensuremath{\Sigmares_\bquark^+}}\xspace}

\def\Sigmabm      {{\ensuremath{\Sigmares_\bquark^-}}\xspace}

\newcommand{\decay}[2]{\ensuremath{#1\!\to #2}\xspace} 

\def\to                 {\ensuremath{\rightarrow}\xspace}

\def\AT#1     {\ensuremath{A_{\mathrm{T}}^{#1}}\xspace}           % 2

\def\C#1      {\ensuremath{\mathcal{C}_{#1}}\xspace}                       % 9
\def\Cp#1     {\ensuremath{\mathcal{C}_{#1}^{'}}\xspace}                    % 7
\def\Ceff#1   {\ensuremath{\mathcal{C}_{#1}^{\mathrm{(eff)}}}\xspace}        % 9  
\def\Cpeff#1  {\ensuremath{\mathcal{C}_{#1}^{'\mathrm{(eff)}}}\xspace}       % 7
\def\Ope#1    {\ensuremath{\mathcal{O}_{#1}}\xspace}                       % 2
\def\Opep#1   {\ensuremath{\mathcal{O}_{#1}^{'}}\xspace}                    % 7

\newcommand{\nospaceunit}[1]{\ensuremath{\text{#1}}}       
\newcommand{\aunit}[1]{\ensuremath{\text{\,#1}}}       
\newcommand{\tev}{\aunit{Te\kern -0.1em V}\xspace}
\newcommand{\gev}{\aunit{Ge\kern -0.1em V}\xspace}
\newcommand{\mev}{\aunit{Me\kern -0.1em V}\xspace}
\newcommand{\kev}{\aunit{ke\kern -0.1em V}\xspace}
\newcommand{\ev}{\aunit{e\kern -0.1em V}\xspace}
 
\newcommand{\mevc}{\ensuremath{\aunit{Me\kern -0.1em V\!/}c}\xspace}
\newcommand{\gevc}{\ensuremath{\aunit{Ge\kern -0.1em V\!/}c}\xspace}
\newcommand{\mevcc}{\ensuremath{\aunit{Me\kern -0.1em V\!/}c^2}\xspace}
\newcommand{\gevcc}{\ensuremath{\aunit{Ge\kern -0.1em V\!/}c^2}\xspace}
\def\mum  {\ensuremath{\,\upmu\nospaceunit{m}}\xspace}

\def\fm   {\aunit{fm}\xspace}

\def\fb   {\ensuremath{\aunit{fb}}\xspace}
\def\invfb   {\ensuremath{\fb^{-1}}\xspace}

\def\gsim{{~\raise.15em\hbox{$>$}\kern-.85em
          \lower.35em\hbox{$\sim$}~}\xspace}
\def\lsim{{~\raise.15em\hbox{$<$}\kern-.85em
          \lower.35em\hbox{$\sim$}~}\xspace}

\def\sPlot{\mbox{\em sPlot}\xspace}

\def\pt         {\ensuremath{p_{\mathrm{T}}}\xspace}

\def\ptot       {\ensuremath{p}\xspace}

\def\evtgen     {\mbox{\textsc{EvtGen}}\xspace}

\def\geant      {\mbox{\textsc{Geant4}}\xspace}

\def\photos     {\mbox{\textsc{Photos}}\xspace}

\def\pythia     {\mbox{\textsc{Pythia}}\xspace}

\def\tell1  {TELL1\xspace}
\def\ukl1   {UKL1\xspace}

\newcommand{\eg}{\mbox{\itshape e.g.}\xspace}
\newcommand{\ie}{\mbox{\itshape i.e.}\xspace}

\usepackage{cite} % Allows for ranges in citations
\usepackage{mciteplus}
\usepackage{longtable} % only for template; not usually to be used in PAPERs

\newboolean{wordcount}
\setboolean{wordcount}{false} % False for normal usage; true for 

\newboolean{preliminary}
\setboolean{preliminary}{false}

\usepackage{xpatch}

\newcounter{mybibstartvalue}
\xpatchcmd{\thebibliography}{%
  \usecounter{enumiv}%
}{%
  \usecounter{enumiv}%
  \setcounter{enumiv}{\value{mybibstartvalue}}%
}{}{}

\makeatletter
\g@addto@macro\bfseries{\boldmath}
\makeatother

\def\Tcc {{{\ensuremath{\PT_{\cquark\cquark}^+}}}\xspace}

\newcommand{\kevc}{\ensuremath{\aunit{ke\kern -0.1em V\!/}c}\xspace}
\newcommand{\kevcc}{\ensuremath{\aunit{ke\kern -0.1em V\!/}c^2}\xspace}

\usepackage{scalerel}

\def\XXint#1#2#3{{\setbox0=\hbox{$#1{#2#3}{\int}$}
     \vcenter{\hbox{$#2#3$}}\kern-.5\wd0}}

\def\Sigmac       {{\ensuremath{\Sigmares_\cquark}}\xspace}
\def\Sigmacz      {{\ensuremath{\Sigmares^{0}_\cquark}}\xspace}
\def\Sigmacp      {{\ensuremath{\Sigmares^{+}_\cquark}}\xspace}
\def\Sigmacpp     {{\ensuremath{\Sigmares^{++}_\cquark}}\xspace}
\def\Sigmacbar    {{\ensuremath{\Sigmaresbar_\cquark}}\xspace}

\def\mathclap#1{\text{\hbox to 0pt{\hss$\mathsurround=0pt#1$\hss}}}

\begin{document}

%% \numberwithin{equation}{section}
%% \numberwithin{table}{section}
%% \numberwithin{figure}{section}

%%%%%%%%%%%%%%%%%%%%%%%%%
%%%%% Title     %%%%%%%%%
%%%%%%%%%%%%%%%%%%%%%%%%%
\renewcommand{\thefootnote}{\fnsymbol{footnote}}
\setcounter{footnote}{1}

% %%%%%%% CHOOSE TITLE PAGE--------
\ifthenelse{\boolean{wordcount}}{}{
% ===============================================================================
% Purpose: LHCb-PAPER journal paper title page template
% Author: 
% Created on: 2010-09-25
% ===============================================================================

%%%%%%%%%%%%%%%%%%%%%%%%%
%%%%%  TITLE PAGE  %%%%%%
%%%%%%%%%%%%%%%%%%%%%%%%%
\begin{titlepage}
\pagenumbering{roman}

% Header ---------------------------------------------------
\vspace*{-1.5cm}
\centerline{\large EUROPEAN ORGANIZATION FOR NUCLEAR RESEARCH (CERN)}
\vspace*{1.5cm}
\noindent
\begin{tabular*}{\linewidth}{lc@{\extracolsep{\fill}}r@{\extracolsep{0pt}}}
\ifthenelse{\boolean{pdflatex}}% Logo format choice
{\vspace*{-1.5cm}\mbox{\!\!\!\includegraphics[width=.14\textwidth]{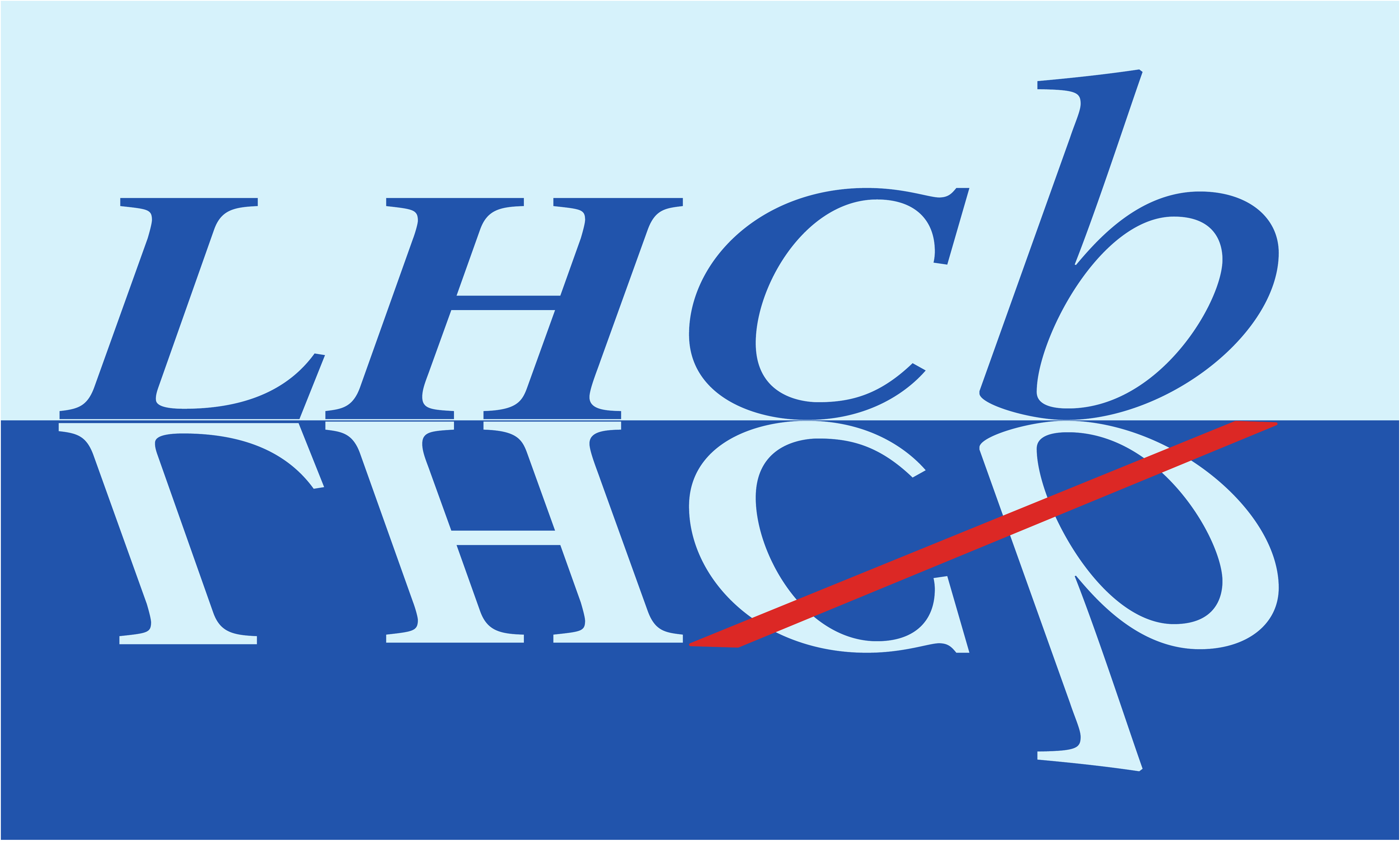}} & &}%
{\vspace*{-1.2cm}\mbox{\!\!\!\includegraphics[width=.12\textwidth]{lhcb-logo.eps}} & &}%
\\
 & & CERN-EP-2021-169 \\  % ID 
 & & LHCb-PAPER-2021-032 \\  % ID 
%% & & \today \\ % Date - Can also hardwire e.g.: 23 March 2010
  & & September 2, 2021 
%% & & v2.0   \\
% not in paper \hline
\end{tabular*}

\vspace*{2.0cm}

% Title --------------------------------------------------
{\normalfont\bfseries\boldmath\huge
\begin{center}
% DO NOT EDIT HERE. Instead edit macro in main.tex to keep metadata correct
  \papertitle 
\end{center}
}

\vspace*{1.0cm}

% Authors -------------------------------------------------
\begin{center}
\paperauthors\footnote{Authors are listed at the end of this paper.}
\end{center}

\vspace{\fill}

% Abstract -----------------------------------------------
\begin{abstract}
  \noindent
Quantum chromodynamics, 
the~theory of the~strong force, 
describes 
interactions of coloured quarks and gluons
and the~formation of hadronic matter.
%% , 
Conventional hadronic matter 
consists of 
baryons and mesons made of 
three quarks and quark-antiquark pairs, respectively.
Particles with  an~alternative quark content are known as exotic states.
Here a~study is reported of an~exotic narrow
state 
in the~\Dz\Dz\pip~mass spectrum 
just below 
the~\Dstarp\Dz~mass threshold
produced in %% 
proton\nobreakdash–proton collisions collected with 
the~LHCb detector at the~Large Hadron Collider.
  The~state is consistent with 
  the~ground isoscalar \Tcc~tetraquark with 
  a~quark content of  $\cquark\cquark\uquarkbar\dquarkbar$ 
  and spin\nobreakdash-parity quantum  
  numbers $\mathrm{J}^{\mathrm{P}}=1^+$.
 Study of the~\D\D~mass spectra disfavours 
 interpretation of 
 the~resonance as the~isovector state.~
 ~
 The decay structure via intermediate off\nobreakdash-shell 
 \Dstarp~mesons
 is consistent with the observed \Dz\pip mass distribution.
 To~analyse the~mass of the~resonance 
 and its coupling to the~\Dstar\D~system, 
 a~dedicated model is developed under 
 the~assumption of 
 an~isoscalar axial\nobreakdash-vector 
 \Tcc state decaying to the~\Dstar\D channel.
 Using this model, 
 resonance parameters including
 the~pole position, 
 scattering length, effective range 
 and compositeness are 
 %% {\color{red} 
 determined
 %% }
 to reveal important 
 information about the~nature of the~\Tcc~state. 
 In~addition, 
 an~unexpected dependence 
 of the~production rate on track multiplicity is observed.
\end{abstract}

\vspace*{2.0cm}

\begin{center}
  Published in \href{https://doi.org/10.1038/s41467-022-30206-w}{Nature Communications 13, 3351 (2022)} 
\end{center}

\vspace{\fill}

{\footnotesize 
% Edit macro in main.tex to keep metadata correct
\centerline{\copyright~\papercopyright. \href{\paperlicenceurl}{\paperlicence}.}}
\vspace*{2mm}

\end{titlepage}

%%%%%%%%%%%%%%%%%%%%%%%%%%%%%%%%
%%%%%  EOD OF TITLE PAGE  %%%%%%
%%%%%%%%%%%%%%%%%%%%%%%%%%%%%%%%

%  empty page follows the title page ----
\newpage
\setcounter{page}{2}
\mbox{~}

}
% %%%%%%%%%%%%% ---------

\renewcommand{\thefootnote}{\arabic{footnote}}
\setcounter{footnote}{0}

%%%%%%%%%%%%%%%%%%%%%%%%%%%%%%%%
%%%%%  Table of Content   %%%%%%
%%%%%%%%%%%%%%%%%%%%%%%%%%%%%%%%
%%%% Uncomment if desired
%\tableofcontents
\cleardoublepage

%%%%%%%%%%%%%%%%%%%%%%%%%
%%%%% Main text %%%%%%%%%
%%%%%%%%%%%%%%%%%%%%%%%%%

\pagestyle{plain} % restore page numbers for the main text
\setcounter{page}{1}
\pagenumbering{arabic}

%% \ifthenelse{\boolean{wordcount}}{}{
%% \linenumbers
%% }

\renewcommand{\figurename}{\bf{Figure}}

\section*{Introduction}

Hadrons with quark content other 
than that seen in mesons \,(\mbox{$\Pq_1\bar{\Pq}_2$}) 
and baryons \,(\mbox{$\Pq_1\Pq_2\Pq_3$})
have been actively discussed since the~birth 
of the~quark~model~\cite{GellMann:1964nj,
Zweig:352337,*Zweig:570209,
Jaffe:1976ig,
Jaffe:1976ih,
Rossi:1977cy,
Jaffe:1977cv,
Lipkin:1987sk}.
Since the~discovery of 
the~$\chicone(3872)$ state~\cite{Choi:2003ue} 
many 
tetraquark and pentaquark candidates,
listed
in Table~\ref{tab:TetraPenta},
have been 
observed~\cite{PDG2021,
Chen:2016qju,
Esposito:2016noz,
Ali:2017jda,
Hosaka:2016pey,
Lebed:2016hpi,
Guo:2017jvc,
Olsen:2017bmm,
Brambilla:2019esw,
Ali:2019roi}.
For~all but the~$\PX_0(2900)$
and $\PX_1(2900)$~states 
the minimal quark content 
implies  
the presence of 
either 
a~$\cquark\cquarkbar$ or 
$\bquark\bquarkbar$~quark\nobreakdash-antiquark pair.
The~masses of many tetra- and pentaquark states 
are close 
to
mass
thresholds, 
\eg $\D{}^{(*)}\Dbar{}^{(*)}$ or 
$\B{}^{(*)}\Bbar{}^{(*)}$, 
where $\D^{(*)}$ or $\B^{(*)}$ represents 
a~hadron containing a~charm or beauty quark, 
respectively. 
Therefore, 
these 
states are likely to be 
hadronic molecules~\cite{Richard_2016,
Guo:2017jvc, 
Oset:2019upy,
MartinezTorres:2020hus}
where colour\nobreakdash-singlet 
hadrons are bound by 
%% {\color{red}
residual nuclear forces,
such as  the~exchange of  a~pion 
or $\Prho$~meson~\cite{Tornqvist:1991ks}, 
similar to electromagnetic 
van der Waals 
forces attracting 
electrically neutral atoms and molecules.
These states are expected to have  
a~spatial extension 
significantly larger than a~typical compact hadron. 
Conversely, the~only hadron currently 
observed that contains a~pair of $\cquark\cquark$ 
quarks is
the~\Xiccpp (\cquark\cquark\uquark) baryon, 
a~long\nobreakdash-lived, 
weakly\nobreakdash-decaying 
compact object~\cite{LHCb-PAPER-2017-018,
LHCb-PAPER-2018-019}. 
The~recently observed
$\PX\mathrm{(6900)}$~structure 
in the~\jpsi\jpsi~mass 
spectrum~\cite{LHCb-PAPER-2020-011} belongs 
to both categories simultaneously. 
Its~proximity to 
the~$\chiczero\chicone$  threshold 
could indicate a~molecular
structure~\cite{Albuquerque:2020hio,
Dong:2020nwy}. 
Alternatively, it could be a~compact object, 
where all four quarks are within one
confinement volume and each quark interacts 
directly with the~other three
quarks via the~strong
force~\cite{Bedolla:2019zwg,
Karliner:2020dta,
Lu:2020cns,
Giron:2020wpx}.
\begin{table}[b]
	\centering
	\caption{\small 
	Tetra- and pentaquark
	candidates and their plausible 
	valence quark content.
	%Relevant references
	%are included inline.
	}
	\label{tab:TetraPenta}
	\vspace{2mm}
	\begin{tabular*}{0.85\textwidth}{@{\hspace{2mm}}m{10.5cm}
	@{\extracolsep{\fill}}c@{\hspace{2mm}}}
	\ \ \ \ \ \ \ \ \ \multirow{2}{*}{States}
	& Quark \\
	& content 
  \\[1mm]
  \hline 
  \\[-2mm]
  $\PX_0(2900)$,
  $\PX_1(2900)$~\cite{LHCb-PAPER-2020-024,LHCb-PAPER-2020-025} 
  & $\cquarkbar\dquark\uquark\squarkbar$ 
  \\[2mm]
  $\chicone\mathrm{(3872)}$~\cite{Choi:2003ue} 
  & $\cquark\cquarkbar\quark\quarkbar$ 
  \\[2mm]
  %% \hline 
  %% \\[-2mm]
  %%
  $\PZ_{\cquark}\mathrm(3900)$~\cite{Ablikim:2013mio,
  Belle:2013yex,
  BESIII:2013qmu,
  BESIII:2015cld,
  BESIII:2015ntl}
  $\PZ_{\cquark}(4020)$~\cite{BESIII:2013ouc,BESIII:2013mhi},
  $\PZ_{\cquark}(4050)$~\cite{Belle:2008qeq},
  $\PX(4100)$~\cite{LHcb-PAPER-2018-034},
  $\PZ_{\cquark}\mathrm(4200)$~\cite{Chilikin:2014bkk}, 
  $\PZ_{\cquark}\mathrm(4430)$~\cite{Choi:2007wga,
  Chilikin:2013tch,LHCb-PAPER-2014-014,LHCb-PAPER-2015-038},
  $\PR_{\cquark0}\mathrm{(4240)}$~\cite{LHCb-PAPER-2014-014} 
  & $\cquark\cquarkbar\uquark\dquarkbar$
  \\[4mm]
  %% \hline 
  %% \\[-2mm] 
  $\PZ_{\cquark\squark}(3985)$~\cite{Ablikim_2021},
  $\PZ_{\cquark\squark}(4000)$,
  $\PZ_{\cquark\squark}(4220)$~\cite{LHCb-PAPER-2020-044} 
  & $\cquark\cquarkbar\uquark\squarkbar$ 
  \\[2mm]
  %% \hline 
  %% \\[-2mm]
  %%
  $\chicone\mathrm{(4140)}$~\cite{Aaltonen:2009tz,
  Abazov:2013xda,Chatrchyan:2013dma,LHCb-PAPER-2016-018},  
  $\chicone\mathrm{(4274)}$,
  $\chiczero\mathrm{(4500)}$,
  $\chiczero\mathrm{(4700)}$~\cite{LHCb-PAPER-2016-018},
  $\PX(4630)$, 
  $\PX(4685)$~\cite{LHCb-PAPER-2020-044},
  $\PX(4740)$~\cite{LHCb-PAPER-2020-035}
  &  $\cquark\cquarkbar\squark\squarkbar$
  \\[4mm]
  %% \hline 
  %% \\[-2mm] 
  $\PX(6900)$~\cite{LHCb-PAPER-2020-011} 
  & $\cquark\cquarkbar\cquark\cquarkbar$
  \\[2mm]
  %% \hline 
  %% \\[-2mm]
  %%
  $\PZ_{\bquark}\mathrm{(10610)}$,
  $\PZ_{\bquark}\mathrm{(10650)}$~\cite{Belle:2011aa} 
  & $\bquark\bquarkbar\uquark\dquarkbar$ 
   \\[1mm]
  \hline 
  \\[-2mm]
  %%\hline 
  %% \\[-2mm]
  %%
  $\PP_{\cquark}\mathrm{(4312)}$~\cite{LHCb-PAPER-2019-014},
  $\PP_{\cquark}\mathrm{(4380)}$~\cite{LHCb-PAPER-2015-029},  
    % LHCb-PAPER-2019-009
  $\PP_{\cquark}\mathrm{(4440)}$,  
  $\PP_{\cquark}\mathrm{(4457)}$~\cite{LHCb-PAPER-2019-014},
  $\PP_{\cquark}\mathrm{(4357)}$~\cite{LHCb-PAPER-2021-018}
  & $\cquark\cquarkbar\uquark\uquark\dquark$
  \\[4mm]
  %%\hline 
  %% \\[-2mm]
  $\PP_{\cquark\squark}(4459)$~\cite{LHCb-PAPER-2020-039} 
  & $\cquark\cquarkbar\uquark\dquark\squark$
	\end{tabular*}
	\vspace{3mm}
\end{table}
The~existence and properties of 
\mbox{$\PQ_1\PQ_2\bar{\Pq}_1\bar{\Pq}_2$}~states 
with two heavy quarks and two light antiquarks
%, also referred to as {\em{dimesons}}, 
have been widely discussed 
for a long time~\cite{Ader:1981db,
Ballot:1983iv,
Zouzou:1986qh,
Lipkin:1986dw,
Heller:1986bt,
Manohar:1992nd}.
In~the~limit 
of  large masses of the heavy quarks
the~corresponding~ground state 
should be deeply bound. 
In~this~limit, the~two heavy 
quarks $\PQ_1\PQ_2$ form a~point\nobreakdash-like 
color-antitriplet object, 
analogous to an~antiquark, 
and as a~result
the~\mbox{$\PQ_1\PQ_2\bar{\Pq}_1\bar{\Pq}_2$} 
system 
has similar degrees of freedom for
its light quarks as 
an antibaryon with a~single heavy quark,
\eg the~\Lcbar or $\Lbbar$~antibaryons. 
The~beauty quark
is considered heavy enough to sustain 
the~existence of 
a~\mbox{$\bquark\bquark\uquarkbar\dquarkbar$}~state 
that is stable with 
respect to the~strong and 
electromagnetic interactions
with a~mass of 
about~$200\mev$
below the~$\B\PB^*$~mass threshold.
In~the~case of 
the~\mbox{$\bquark\cquark\uquarkbar\dquarkbar$}
and~\mbox{$\cquark\cquark\uquarkbar\dquarkbar$}~systems, there is currently 
no consensus in the~literature
whether such states exist 
%or have 
and if 
their~natural widths 
are 
narrow 
enough to allow for experimental observation.
The~theoretical predictions for the~mass of the~$\mbox{\cquark\cquark\uquarkbar\dquarkbar}$~ground~state
with  spin\nobreakdash-parity quantum 
numbers $\mathrm{J^P=1^+}$ and 
isospin $\mathrm{I}=0$,
denoted hereafter as \Tcc,
 %% \footnote{Theory predictions
 %% for the mass of the ground state isoscalar $\mathrm{J^P}=1^+$ \cquark\cquark\uquarkbar\dquarkbar
 %% tetraquark \Tcc are available
 %% at \url{https://cern.ch/lhcbproject/Publications/p/LHCb-PAPER-2021-031.html}.}
relative to the~$\Dstarp\Dz$~mass threshold
\ifthenelse{\boolean{wordcount}}{}{
\begin{equation}\label{eq:delta_prime}
 \updelta m  \equiv 
 m_{\Tcc} - \left( m_{\Dstarp} + m_{\Dz}\right) \,
\end{equation} 
}
lie in the~range 
\mbox{$-300< \updelta m < 300\mevcc$}~\cite{
Carlson:1987hh,
SilvestreBrac:1993ss,
Semay:1994ht,
Moinester:1995fk,
Pepin:1996id,
GELMAN2003296,
Vijande:2003ki,
Janc:2004qn,
Navarra:2007yw,
Vijande,
Ebert:2007rn,
Lee:2009rt,
Yang2009,
Li:2012ss,
Feng:2013kea,
Luo:2017eub,
Karliner:2017qjm,
Eichten:2017ffp,
Wang:2017uld,
Park:2018wjk,
Junnarkar:2018twb,
Deng:2018kly,
Liu:2019stu,
Maiani:2019lpu,
Yang:2019itm,
Tan:2020ldi,
Lu:2020rog,
Braaten:2020nwp,
Gao:2020ogo,
Cheng:2020wxa,
Noh:2021lqs,
Faustov:2021hjs},
where $m_{\Dstarp}$ and $m_{\Dz}$ denote 
the~known masses of the~$\Dstarp$ 
and $\Dz$~mesons~\cite{PDG2021}, 
with $\cquark\dquarkbar$ 
and 
$\cquark\uquarkbar$ quark content, respectively.
The~observation of 
a~narrow state in the~\Dz\Dz\pip~mass 
spectrum near the $\Dstarp\Dz$~mass threshold, 
compatible with being a~$\Tcc$~tetraquark 
state with 
$\cquark\cquark\uquarkbar\dquarkbar$~quark 
content is reported in Ref.~\cite{LHCb-PAPER-2021-031}.

In~the work presented here, 
the~properties of the~\Tcc~state are studied
by constructing a~dedicated amplitude 
model that accounts for
the~$\Dstarp\Dz$ and $\Dstarz\Dp$~decay 
channels.
In addition, the mass spectra of 
other $\D\D{}^{(*)}$ and 
opposite\nobreakdash-sign 
$\D\Dbar{}^{(*)}$ combinations are explored.
Furthermore, 
production-related observables,  
such as the~event multiplicity and 
transverse momentum\,(\pt) 
spectra
that are sensitive to the internal structure
of the~state, are discussed.
This~analysis is based on 
proton\nobreakdash-proton\,($\proton\proton$)~collision 
data, corresponding 
to  integrated luminosity of~9\invfb, collected
with the~LHCb detector at centre\nobreakdash-of\nobreakdash-mass 
energies of~7,~8 and~13\tev.
The~LHCb detector~\mbox{\cite{LHCb-DP-2008-001,LHCb-DP-2014-002}} 
is a~single-arm forward spectrometer covering the pseudorapidity
range $2 < \eta < 5$, designed for the study of 
particles containing \bquark\ or \cquark\ quarks and is further described in Methods.%

%% \section*{\Tcc~signal in the~$\Dz\Dz\pip$ mass spectrum} 

\section*{Results}
\subsection*{\Tcc~signal in the~$\Dz\Dz\pip$ mass spectrum}

\begin{figure}[tb]
  \centering
  \includegraphics[width=\textwidth]{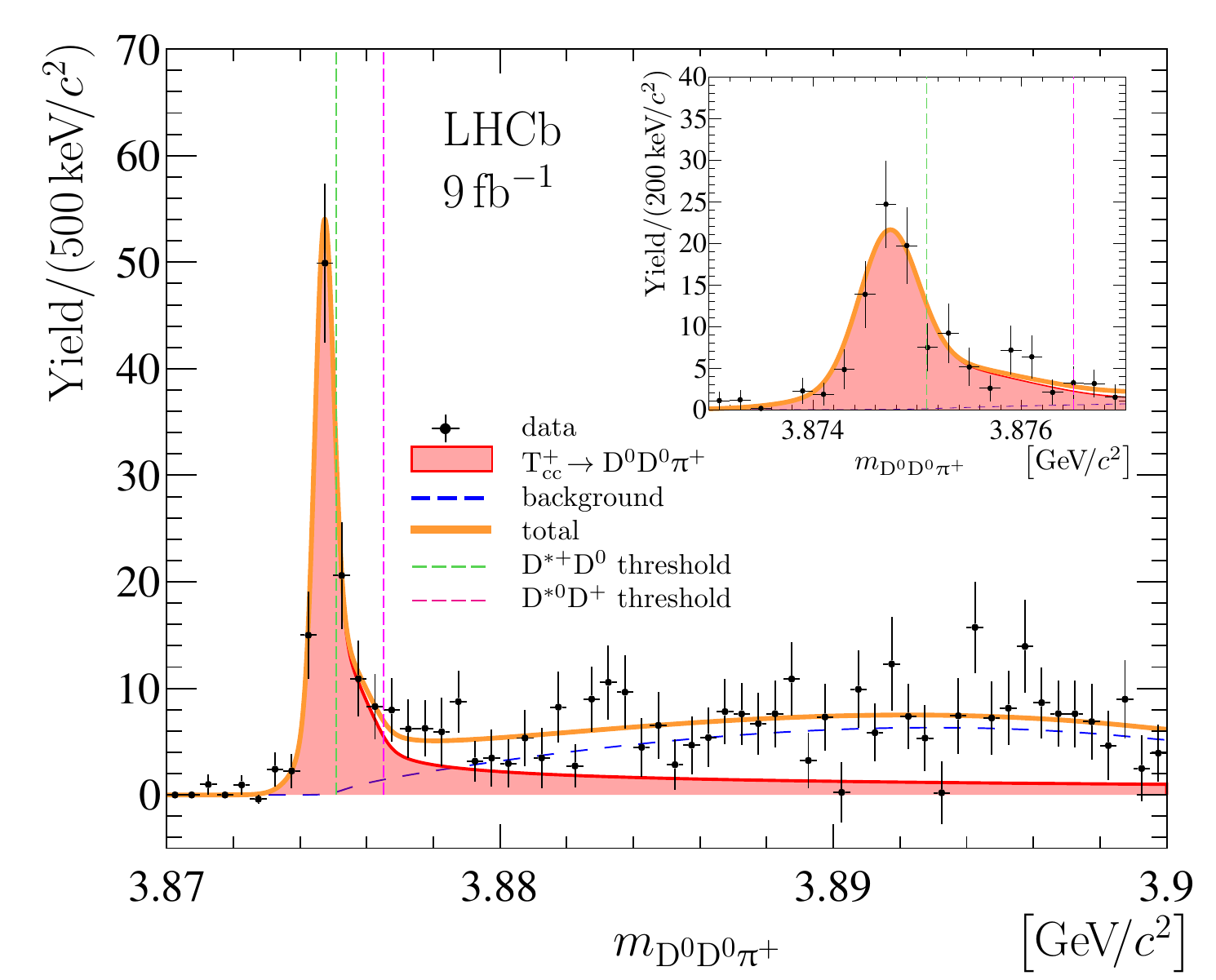}
  \caption { \small
  {\bf Distribution of \Dz\Dz\pip~mass}.
    Distribution of \Dz\Dz\pip~mass where 
   the~contribution of the~non\protect\nobreakdash-$\Dz$ 
   background has been statistically subtracted.
   The~result of the~fit described in the~text is overlaid. 
   %% 
   %% {\color{red}{
   Uncertainties on the~data points are statistical
 only and represent one standard deviation, 
 calculated as a~sum in quadrature of the~assigned weights from 
 the~background\protect\nobreakdash-subtraction procedure.
 %% }}
  }
  \label{fig:DATA_BWU_fit}
\end{figure}

The~$\Dz\Dz\pip$ final state 
is reconstructed using 
the~$\Dz\to\Km\pip$~decay channel
with two \Dz mesons and a~pion 
all produced 
promptly in the~same $\proton\proton$ collision.
The~inclusion of charge\nobreakdash-conjugated 
processes
is implied throughout the~paper.
The~selection criteria are similar to those used in 
Refs.~\cite{LHCb-PAPER-2012-003,
LHCb-PAPER-2013-062,
LHCb-PAPER-2015-046,
LHCb-PAPER-2019-005}
and described in detail in Methods.
The~background not originating from true \Dz~mesons
is subtracted  using 
an~extended unbinned 
maximum\nobreakdash-likelihood fit to
the~two\nobreakdash-dimensional distribution of 
the masses of the~two \Dz~candidates
%the~mass of one \Dz~candidate versus
%the~mass of the other \Dz~candidate 
from selected  $\Dz\Dz\pip$~combinations,
see Methods 
and 
%% Extended Data 
Supplementary~Fig.~\ref{fig:DATA_2D}(a).
%% Fig.~1(a).
%% 
The~obtained
$\Dz\Dz\pip$~mass distribution 
for selected $\Dz\Dz\pip$~combinations 
is shown in~Fig.~\ref{fig:DATA_BWU_fit}.

An~extended unbinned 
maximum\nobreakdash-likelihood fit 
to the~$\Dz\Dz\pip$~mass distribution 
is performed using a~model 
consisting of signal and background components.
The~signal component 
corresponds to 
the~\mbox{$\decay{\Tcc}{\Dz\Dz\pip}$}~decay
and is described as the~convolution of the~natural 
resonance profile 
with the~detector 
mass
resolution function.
A~relativistic P\nobreakdash-wave 
two\nobreakdash-body Breit\nobreakdash--Wigner 
function  $\mathfrak{F}^{\mathrm{BW}}$  
with a~Blatt\nobreakdash--Weisskopf 
form factor~\cite{Blatt:1952ije,VonHippel:1972fg} 
is used in Ref.~\cite{LHCb-PAPER-2021-031} 
as the~natural resonance profile.
That function, 
while sufficient to reveal 
the~existence of the state, 
%% however, 
does not account for the resonance being in close vicinity of
the~$\Dstar\D$ threshold. 
To~assess the~fundamental properties 
of resonances 
that are close to thresholds,
advanced parametrisations ought 
to be used~\cite{Hanhart:2007yq,
Braaten:2007dw,
Stapleton:2009ey,
Kalashnikova:2009gt,
Artoisenet:2010va,
Hanhart:2010wh,
Hanhart:2011jz,
Hanhart:2015cua,
Guo:2016bjq,
Hanhart:2016eyl,Guo:2017jpg}.
A~unitarised Breit\nobreakdash--Wigner 
profile $\mathfrak{F}^{\mathrm{U}}$, 
described in Methods Eq.~\eqref{eq:bwU}, 
is used in this analysis. 
The~function~$\mathfrak{F}^{\mathrm{U}}$,  
is built under two main assumptions. 
%%
%% {\color{red}
%% \begin{itemize}
\begin{description}
    \item[{\it{Assumption~1}}.] The~newly observed state has quantum 
numbers $\mathrm{J^P}=1^+$ and  
isospin $\mathrm{I}=0$ in accordance 
with the~theoretical expectation 
for the~$\Tcc$~ground state. 
    \item[{\it{Assumption~2}}.] The~\Tcc~state  
    is strongly coupled 
to the~$\Dstar\D$ channel, 
which is motivated by the~proximity 
of the~\Tcc~mass 
to the~$\Dstar\D$~mass threshold.
\end{description}
%% \end{itemize}
%%
The~derivation of 
the~$\mathfrak{F}^{\mathrm{U}}$~profile relies on the assumed~isospin 
symmetry for 
the~\mbox{$\decay{\Tcc}{\Dstar\D}$}~decays 
and the~coupled\nobreakdash-channel 
interaction of the~$\Dstarp\Dz$ and $\Dstarz\Dz$~system 
as required by unitarity and causality following
Ref.~\cite{Mikhasenko:2019vhk}.
The~resulting energy-dependent width of 
the~$\Tcc$~state accounts explicitly
for the~\mbox{$\decay{\Tcc}{\Dz\Dz\pip}$},
\mbox{$\decay{\Tcc}{\Dz\Dp\piz}$} and 
\mbox{$\decay{\Tcc}{\Dz\Dp\g}$}~decays.
The~modification of the~$\Dstar$~meson lineshape~\cite{Pasquier:1968zz} due to 
contributions from triangle diagrams~\cite{Aitchison:1979fj} to
the~final\nobreakdash-state interactions is neglected.
%% }
% 
%% The~model incorporates the one-particles exchange process, however, 
%% neglects the~modification of the~\Dstar propagator due to 
%% the~final-state interaction.
% 
%%
Similarly to the~$\mathfrak{F}^{\mathrm{BW}}$~profile, 
the~$\mathfrak{F}^{\mathrm{U}}$~function  
has two parameters: 
the~peak location $m_{\mathrm{U}}$, 
defined as the~mass value where 
the~real part of the~complex amplitude vanishes, 
and the~absolute value of the~coupling constant 
$g$ for the~\mbox{$\decay{\Tcc}{\Dstar\D}$}~decay.  

The~detector mass resolution, $\mathfrak{R}$,  
is modelled with the~sum of two Gaussian functions 
with a~common mean, and parameters taken from simulation,
see Methods. %%  Eq.~\eqref{eq:resolution}. 
The~widths 
of the~Gaussian functions are corrected by
a~factor of~1.05, that accounts for
a~small residual difference between 
simulation and data~\cite{LHCb-PAPER-2020-008,
LHCb-PAPER-2020-009,
LHCb-PAPER-2020-035}.
The~root mean square of the~resolution 
function is around 400\kevcc. 

A~study of the $\Dz\pip$~mass distribution
for selected $\Dz\Dz\pip$~combinations 
in the~region above 
the~$\Dstarz\Dp$~mass threshold and 
below $3.9\gevcc$
shows that 
approximately $90\%$ of all $\Dz\Dz\pip$~combinations 
contain a~true $\Dstarp$~meson.
Therefore, the~background component
is parameterised with a~product of  
the~two\nobreakdash-body phase space 
function $\Phi_{\Dstarp\Dz}$~\cite{Kallen} and 
a~positive polynomial function $P_n$, 
convolved with  
%% the~same 
the detector resolution function $\mathfrak{R}$ 
%% is further applied to obtain 
%% the~final background component
\begin{equation}
    B_n =   \left( \Phi_{\Dstarp\Dz} 
    \times  P_n \right) * \mathfrak{R}\,, \label{eq:background}
\end{equation}
where $n$~denotes the~order 
of the~polynomial function,
$n=2$ is used in the~default fit.

\begin{comment} 
\begin{table}[tb]
	\centering
	\caption{\small 
	Signal yield, $N$, and 
	the~resonance mass parameter with 
	respect to the~$\Dstarp\Dz$~mass threshold,
	$\updelta m_{\mathrm{U}}$,
	obtained
    from the~fit with a model based on 
	the~$\mathfrak{F}^{\mathrm{U}}$~signal profile. 
	Uncertainties are statistical only. }
	\label{tab:DATA_BWU_fits}
	\vspace{2mm}
	\begin{tabular*}{0.45\textwidth}
	{@{\hspace{3mm}}l@{\extracolsep{\fill}}l@{\hspace{3mm}}}
	Parameter  & ~~~~~Value 
   \\[1mm]
  \hline 
  \\[-2mm]
   %% 
   $N$                                      
   & $\phantom{-}186\pm24$   \\
   %% $B_2$     &           
   %% & $\phantom{-}238\pm26$   \\     
   %%
   $\updelta m_{\mathrm{U}}$  
 %%&  $-363\pm40$
 %%   &  $-360\pm40\kevcc$ 
   &  $-359\pm40\kevcc$ 
   \\
   $\left| g \right| $  
   & $\phantom{-00}3\times10^4 \gev 
   \,\mathrm{(fixed)}$  
   %%
	\end{tabular*}
\end{table}
\end{comment}

The~$\Dz\Dz\pip$~mass spectrum 
with  non\nobreakdash-\Dz~background subtracted 
is shown in Fig.~\ref{fig:DATA_BWU_fit}
with the~result of the~fit  
using a~model based on 
the~$\mathfrak{F}^{\mathrm{U}}$~signal profile 
overlaid.
%% 
%% The~signal yield, $N$, and mass parameter relative 
%% to the~$\Dstarp\Dz$~mass threshold, 
%% $\updelta m_{\mathrm{U}}$,
%% are listed in Table~\ref{tab:DATA_BWU_fits}. 
%% {\color{red}{
The~fit gives a~signal yield 
of $186\pm24$  
and a~mass parameter relative 
to the~$\Dstarp\Dz$~mass threshold, 
$\updelta m_{\mathrm{U}}$ of $-359\pm40\kevcc$. 
%% }}
%%
The~statistical significances of the~observed 
\mbox{$\decay{\Tcc}{\Dz\Dz\pip}$}~signal 
and for 
the~$\updelta m_{\mathrm{U}}<0$ hypothesis
are determined using Wilks' theorem 
to be $22$ and $9$~standard 
deviations, respectively.

\begin{figure}[t]
  \centering
  \includegraphics[width=\textwidth]{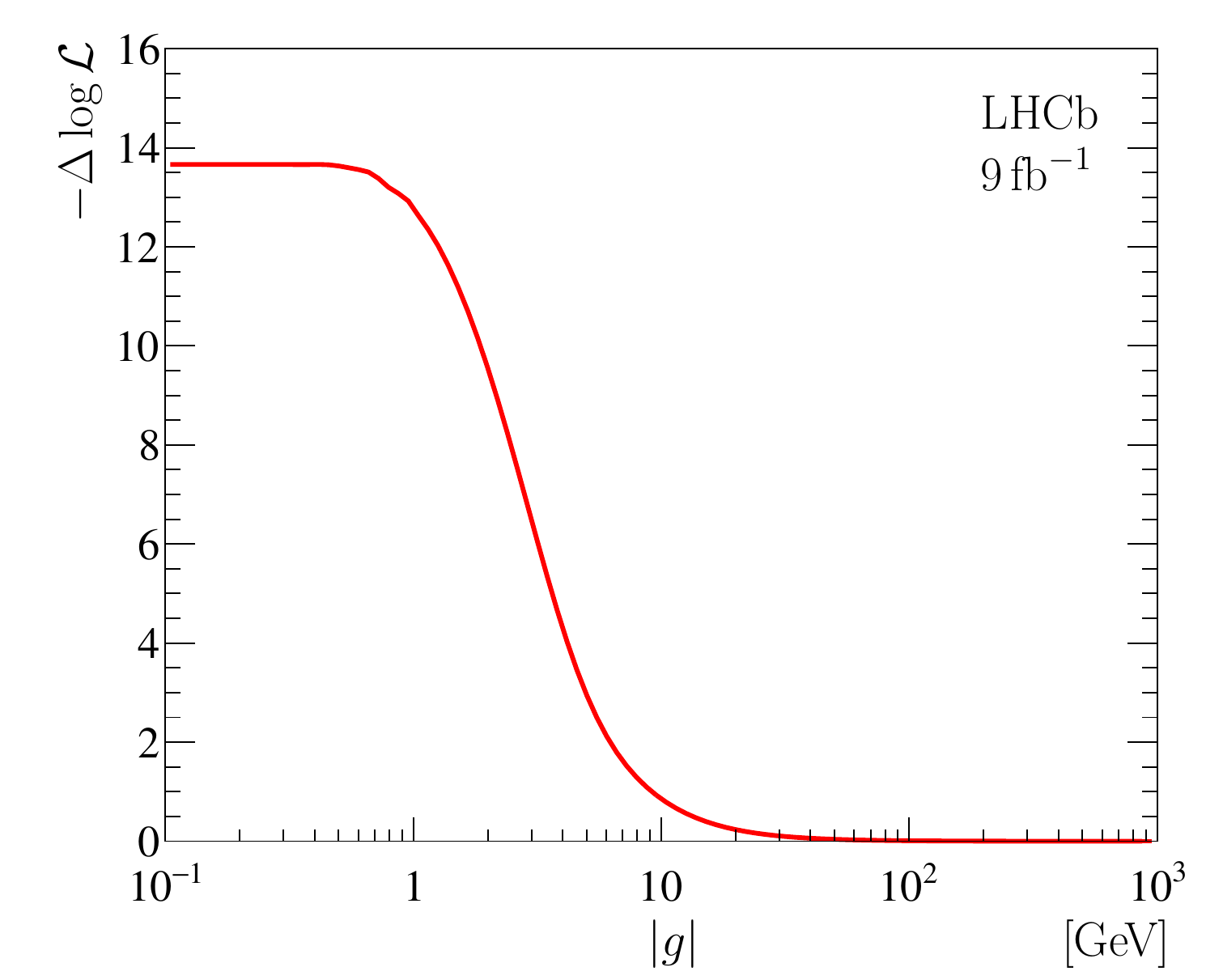}
  \caption { \small
  {\bf Likelihood profile for the~$\left|g\right|$~parameter.} 
  Likelihood profile for the~absolute value of 
  the~coupling constant $g$
  from the~fit to the~background\protect\nobreakdash-subtracted 
  $\Dz\Dz\pip$~mass spectrum with a~model based 
  on the~$\mathfrak{F}^{\mathrm{U}}$~signal profile.
  }
  \label{fig:DATA_NLL_g}
\end{figure}

The~width of the~resonance is determined by 
the coupling constant $g$ 
%the~value 
%of $\left|g\right|^2$ 
for small values 
of~$\left|g\right|$.
%% the~coupling parameter~$\left|g\right|$ is small.
%With the~increase of $\left|g\right|$, 
%the~width is motonically growing 
%and achieving 
With increasing $\left|g\right|$, 
the~width 
increases to an~asymptotic value 
determined by the~width of 
the~$\Dstarp$~meson, 
see Methods and 
%% Extended Data 
Supplementary~Fig.~\ref{fig:SCALING}.
%%  Fig.~7.
%%
In~this regime of large $\left|g\right|$,
the~$\mathfrak{F}^{\mathrm{U}}$~signal profile
exhibits a~scaling 
property similar 
to 
the~Flatt\'e~function~\mbox{\cite{Flatte:1976xu, 
Baru:2004xg,
LHCb-PAPER-2020-008}}.
The~parameter $\left|g\right|$ effectively 
decouples from the~fit model, 
and the~model 
resembles the~scattering\nobreakdash-length
approximation~\cite{Braaten:2007dw}.
The~likelihood profile 
for the~parameter $\left| g \right|$
is shown in Fig.~\ref{fig:DATA_NLL_g}, where 
one can see a~plateau at large values.
% \footnote{
% For small values of 
At~small values 
of the~$\left| g\right|$~parameter,
$\left|g\right|<1\gev$, 
the~likelihood function is 
independent of $\left|g\right|$
because the~resonance 
is too narrow 
for the~details of 
the~$\mathfrak{F}^{\mathrm{U}}$~signal 
profile to be resolved 
by the~detector.
The~lower limits 
on the~$\left|g\right|$~parameter
of \mbox{$\left| g \right| > 7.7\,(6.2)\gev$} 
at 90\,(95)\,\%~confidence level\,(CL) 
are obtained as  
the~values where 
the~difference in the~negative 
log\nobreakdash-likelihood 
$-\Delta\log \mathcal{L}$ 
is equal to 1.35 and 1.92, 
respectively.
Smaller values for 
$\left|g\right|$ are further used 
for systematic uncertainty evaluation. 
%%%% 

%% {\color{red} 
%% Variant to replace the text in green:

%% The~modes 
%% relative to the~\Dstarp\Dz~mass threshold, 
%% $\updelta \mathfrak{m}$, 
%% and the~full widths at half 
%% maximum\,(FWHM), $\mathfrak{w}$, 
%% for the~$\mathfrak{F}^{\mathrm{BW}}$ 
%% and $\mathfrak{F}^{\mathrm{U}}$~signal profiles 
%% are compared in Table~\ref{tab:DATA_visible} 
%% and appear to be rather different. 

%% {\color{red}{
The~mode 
relative to the~\Dstarp\Dz~mass threshold, 
$\updelta \mathfrak{m}$, 
and the~full width at half 
maximum\,(FWHM), $\mathfrak{w}$, 
for the~$\mathfrak{F}^{\mathrm{BW}}$ 
profile are found 
to be 
\mbox{$\updelta \mathfrak{m} = -361 \pm 40\kevcc$}
and 
\mbox{$\mathfrak{w}=47.8\pm1.9\kevcc$}, 
to be compared with those quantities
determined for 
the~$\mathfrak{F}^{\mathrm{U}}$~signal profile
of 
\mbox{$\updelta \mathfrak{m} = -279 \pm 59\kevcc$}
and 
\mbox{$\mathfrak{w}=409\pm163\kevcc$}.
They appear to be rather different. 
%% }}
%%
\begin{comment}
 \begin{table}[bt]
	\centering
	\caption{\small 
	Mode 
	relative to the~\Dstarp\Dz~mass threshold,
	$\updelta\mathfrak{m}$, 
	and FWHM, 
	$\mathfrak{w}$, 
	for
	the~$\mathfrak{F}^{\mathrm{U}}$ and 
	$\mathfrak{F}^{\mathrm{BW}}$ signal profiles.
	Uncertainties are statistical only. 
	}
	\label{tab:DATA_visible}
	\vspace{2mm}
	\begin{tabular*}{0.45\textwidth}{@{\hspace{3mm}}l@{\extracolsep{\fill}}cc@{\hspace{3mm}}}
	   & $\updelta \mathfrak{m}~\left[\!\kevcc\right]$ 
	   & $\mathfrak{w}~\left[\!\kevcc\right]$ 
   \\[1mm]
  \hline 
  \\[-2mm]
   %% $\mathfrak{F}^{\mathrm{BW}}$   &  $-283 \pm 59$               &   $409\pm163$   
      $\mathfrak{F}^{\mathrm{BW}}$   &  $-279 \pm 59$               &   $409\pm163$   
    \\
  %% $\mathfrak{F}^{\mathrm{U}}$   &   $-365 \pm 40$     &   $47.8\pm1.9\phantom{0}$ 
    $\mathfrak{F}^{\mathrm{U}}$   &   $-361 \pm 40$     &   $47.8\pm1.9\phantom{0}$ 
 	\end{tabular*}
\end{table}
\end{comment}
%%
Nonetheless, both functions properly
describe the data given 
the~limited sample size,
and accounting for the~detector resolution, 
and residual background. 
To~quantify 
the~impact of these experimental effects,
two ensembles of pseudoexperiments are performed. 
Firstly, pseudodata samples are generated with 
a~model based on the~$\mathfrak{F}^{\mathrm{U}}$ profile.
The parameters used here are obtained from the~default fit,
and the size of the sample corresponds to 
the~size of data sample.
Each pseudodata sample is then analysed with 
a~model based on the~$\mathfrak{F}^{\mathrm{BW}}$~function.
The~obtained mean and root mean square\,(RMS) values 
for the~parameters~$\updelta m_{\mathrm{BW}}$
and~$\Gamma_{\mathrm{BW}}$ over the~ensemble
are shown in Table~\ref{tab:DATA_consistency}.
% They agree the the parameters determined
% from the data Ref.~\cite{LHCb-PAPER-2021-031}.
The~mass parameter $\updelta m_{\mathrm{BW}}$ agrees well 
with the~value determined
from data~\cite{LHCb-PAPER-2021-031}.
The~difference for the~parameter~$\Gamma_{\mathrm{BW}}$ does 
not exceed one standard deviation.
Secondly, an~ensemble of pseudodata samples 
generated with a~model based on 
the~$\mathfrak{F}^{\mathrm{BW}}$~profile 
is analysed with a~model based on the~$\mathfrak{F}^{\mathrm{U}}$~function.
The~obtained mean and RMS values for 
the~$\updelta m_{\mathrm{U}}$ 
parameter over an~ensemble 
are also reported in Table~\ref{tab:DATA_consistency}.
These values agree well with the result of the default fit to data.
The~results of these pseudoexperiments
explains the~seeming inconsistency between 
the~models
and illustrate the~importance of 
an~accurate description of 
the~detector resolution and 
residual background
given the~limited sample size.
%% }

\begin{table}[tb]
	\centering
	\caption{\small 
	Mean and root mean square\,(RMS) values for 
	the~$\updelta m_{\mathrm{BW}}$,
    $\Gamma_{\mathrm{BW}}$ and 
    $\updelta m_{\mathrm{U}}$~parameters 
    obtained from pseudoexperiments produced as 
    a~consistency check. 
	}
	\label{tab:DATA_consistency}
	\vspace{2mm}
	\begin{tabular*}{0.75\textwidth}{@{\hspace{3mm}}l
	@{\extracolsep{\fill}}lccc@{\hspace{-3mm}}l@{\hspace{3mm}}}
    \multicolumn{2}{l}{\multirow{2}{*}{Parameter}}
	     &  \multicolumn{2}{l}{Pseudoexperiments} &  \multirow{2}{*}{Data} &  
	    \\[1mm]
	    &  &  mean  & RMS &   & \\[1mm]
  \hline 
  \\[-2mm]
   $\updelta m_{\mathrm{BW}}$  &  $\left[\!\kevcc\right] $ 
    &  $-301$   & $\phantom{0}50$   & $-273\pm61\phantom{0}$ & \multirow{2}{*}{\cite{LHCb-PAPER-2021-031}}           \\ 
  $\Gamma_{\mathrm{BW}}$              &  $\left[\!\kev\right] $     
   &  $\phantom{-}222$  & $121$  & $\phantom{-}410\pm165$ & \\[0.5mm] \hline \\[-4.5mm]
   $\updelta m_{\mathrm{U}}$              &  $\left[\!\kevcc\right] $ 
   &  $-378$   & $\phantom{0}46$    
%   & $-360\pm40\phantom{0}$ 
   & $-359\pm40\phantom{0}$ 
   & 
 	\end{tabular*}
\end{table}

%% \section*{Systematic uncertainties}\label{sec:systematics}
\subsection*{Systematic uncertainties}\label{sec:systematics}

Systematic uncertainties
for the~$\updelta m_{\mathrm{U}}$~parameter
are summarised in Table~\ref{tab:systematic}
and described in greater detail below. 
The~systematic uncertainty related to the~fit model 
is studied using pseudoexperiments
with a~set of alternative
parameterisations.
For~each alternative
model an~ensemble of pseudoexperiments 
is performed
with parameters obtained from a~fit to data.
A~fit with the~baseline model is performed
to each pseudoexperiment, 
and the~mean values of
the parameters of interest are evaluated 
over the ensemble.
The~absolute values of the~differences between 
these~mean values  and 
the~corresponding parameter values obtained from 
the~fit to data are used to
assess the~systematic uncertainty
due to the~choice of the~fit model.
The~maximal value of such 
differences over the~considered set of 
alternative models is taken as 
the~corresponding systematic uncertainty.
The~following sources of systematic uncertainty
related to the~fit model are considered: 
\begin{itemize} 
\item Imperfect knowledge of the detector resolution model. 
To~estimate the~associated systematic uncertainty 
a~set of alternative resolution functions 
is tested:
a~symmetric variant 
of an~Apollonios function\cite{Santos:2013gra},  
a~modified Gaussian function with 
symmetric power\nobreakdash-law tails on 
both sides of the~distribution~\cite{Skwarnicki:1986xj,LHCb-PAPER-2011-013},  
a~generalised symmetric Student's 
$t$\nobreakdash-distribution~\cite{Student,Jackman},  
a~symmetric Johnson's 
$\mathrm{S_U}$~distribution~\mbox{\cite{JohnsonSU1,
JohnsonSU2}}, 
and a~modified Novosibirsk function~\cite{Bukin}.
\item A~small difference in the~detector 
resolution between data and simulation.
A~correction~factor of 1.05 is applied 
to account for known discrepancies in 
modelling the~detector resolution in simulation.  
This factor was studied for 
different decays~\mbox{\cite{LHCb-PAPER-2017-036,
LHCb-PAPER-2019-045,
LHCb-PAPER-2020-008,
LHCb-PAPER-2020-009,
LHCb-PAPER-2020-035,
LHCb-PAPER-2021-023}} and found 
to lie between~1.0 and 1.1. 
For~decays with 
relatively 
low momentum 
tracks, 
this factor is close to~1.05,
which is the nominal value used in this analysis. 
This~factor is also 
cross\nobreakdash-checked using large samples of 
\mbox{$\decay{\Dstarp}{\Dz\pip}$}~decays,
where a~value of~1.06 is obtained.
To~assess the~systematic uncertainty 
related to this factor,
detector resolution models with correction 
factors of 1.0 and 1.1 are studied 
as alternatives. % models (repeating models two times)
\item Parameterisation of the~background component. 
To assess the associated systematic uncertainty,  
the~order of the~positive polynomial function 
of Eq.~\eqref{eq:background} is 
varied. In~addition, to estimate a~possible 
effect from a~small contribution 
from three\nobreakdash-body $\Dz\Dz\pip$~combinations 
without an~intermediate \Dstarp~meson, 
a~more general family of  background models is tested
\ifthenelse{\boolean{wordcount}}{}{
\begin{equation}\label{eq:background_ext}
    B^\prime_{nm} = B_n  + 
    \Phi_{\Dz\Dz\pip}\times P_m \,,
\end{equation}
}
where $\Phi_{\Dz\Dz\pip}$~denotes 
the three\nobreakdash-body phase-space 
function~\cite{Byckling}. %% ,Davydychev:2002cu,Davydychev:2003cw}. 
The~functions $B_0$, $B_1$, $B_3$ and 
$B^\prime_{nm}$ with $n\le2, m\le1$
are used as alternative models for the~estimation 
of the~systematic uncertainty.

\item Values of the~coupling  constants 
for the~\mbox{$\decay{\Dstar}{\D\Ppi}$}
and \mbox{$\decay{\Dstar}{\D\g}$}~decays 
affecting the~shape of 
the~$\mathfrak{F}^{\mathrm{U}}$~signal profile. 
These coupling constants 
are calculated from the~known branching fractions 
of the~\mbox{$\decay{\Dstar}{\D\Ppi}$}
and \mbox{$\decay{\Dstar}{\D\g}$}~decays~\cite{PDG2021},
the~measured natural width of 
the~\Dstarp~meson~\cite{Lees:2013zna,PDG2021}
and the~derived value for the~natural width of 
the~\Dstarz~meson~\cite{Braaten:2007dw,
Guo:2019qcn,
Braaten:2020nwp}.
To~assess the~associated systematic uncertainty,  
a~set of~alternative models 
built around the~$\mathfrak{F}^{\mathrm{U}}$~profiles, 
obtained with coupling constants varying within 
their calculated uncertainties,
is studied.

\item Unknown value of the $\left|g\right|$
parameter. In the~baseline fit 
the~value of the~$\left|g\right|$~parameter
is fixed to a~large value. To~assess the~effect of this 
constraint the~fit is repeated using the~value of 
$\left|g\right|=8.08\gev$, that corresponds 
to $-2\Delta\log\mathcal{L}=1$
for the~most conservative likelihood 
profile for~$\left|g\right|$ 
that accounts for 
the~systematic 
uncertainty.
The~change of $7\kevcc$ of 
the~$\updelta m_{\mathrm{U}}$~parameter
is assigned as the~systematic uncertainty.

\end{itemize}

\ifthenelse{\boolean{wordcount}}{}{
\begin{table}[tb]
	\centering
	\caption{\small 
	Systematic uncertainties for 
	the~$\updelta m_{\mathrm{U}}$~parameter. 
	The~total uncertainty is calculated 
	as the sum in quadrature of all components. 
	%% except model parameters.
	}
	\label{tab:systematic}
	\vspace{2mm}
	\begin{tabular*}{0.50\textwidth}{@{\hspace{3mm}}l@{\extracolsep{\fill}}c@{\hspace{3mm}}}
	Source
    & $\upsigma_{\updelta m_{\mathrm{U}}}~\left[\!\kevcc\right]$
   \\[1.5mm]
  \hline 
  \\[-2mm]
  Fit model                           &                 \\
  ~~Resolution model                  &  $2$            \\
  ~~Resolution correction factor      &  $2$            \\
  ~~Background model                  &  $2$            \\ 
  ~~Coupling constants                &  $1$            \\ 
  ~~Unknown value of $\left|g\right|$  &  ${}^{+\,7}_{-\,0}$ \\ 
  %% ~~{\em{Model parameters}}     &  ---  & ${}^{+11}_{-14}$ & ${}^{+18}_{-38}$ \\ 
  Momentum scaling                    &  3              \\
  Energy loss                         &  1              \\
  $\Dstarp-\Dz$ mass difference       &  2        
  \\[1.5mm]
  \hline 
  \\[-2mm]
  Total    &   ${}^{+\,9}_{-\,6}$ 
   \end{tabular*}
	\vspace{3mm}
\end{table}
}

%% \end{itemize} 

The~calibration of the momentum scale of the tracking system 
is based upon large 
%% calibration 
samples of 
\mbox{$\decay{\Bu}{\jpsi\Kp}$} and 
\mbox{$\decay{\jpsi}{\mumu}$}~decays~\cite{LHCb-PAPER-2012-048}. 
The~accuracy of the~procedure has been checked using 
fully reconstructed $\B$~decays together 
with two\nobreakdash-body
$\PUpsilon(\mathrm{nS})$ and 
$\KS$~decays and the~largest deviation of 
the~bias in the~momentum  scale
of $\updelta\upalpha=3\times10^{-4}$~is taken as 
the~uncertainty~\cite{LHCb-PAPER-2013-011}.
This~uncertainty is propagated 
%% to uncertainties 
for the~parameters of  interest 
using simulated samples,
with momentum scale 
corrections of $\left(1\pm\updelta\upalpha\right)$
applied. Half of the~difference 
between the~obtained peak locations
is taken as an~estimate of the~systematic uncertainty.

In the~reconstruction
%% step, 
the~momenta of the~charged tracks 
are corrected for energy loss in the~detector material 
using 
the~Bethe\nobreakdash--Bloch 
formula~\cite{Bethe,BLOCH}.
The~amount of the~material traversed 
in the~tracking
system by a~charged particle 
is known to~10\% accuracy~\cite{LHCb-PAPER-2010-001}.
To~assess 
the corresponding uncertainty
the~magnitude of the~calculated corrections 
is varied by $\pm10\%$.
Half of the~difference 
between the~obtained peak locations 
is taken as an~estimate of the~systematic 
uncertainty 
due to energy loss corrections.

The~mass of $\Dz\Dz\pip$~combinations is calculated 
with the mass of each \Dz meson constrained 
to the~known value of the \Dz mass~\cite{PDG2021}.
This procedure produces negligible~uncertainties for 
the~$\updelta m_{\mathrm{U}}$~parameter
due to imprecise knowledge of the~\Dz~mass.
However, the~small uncertainty of 2\kevcc for 
the~known 
$\Dstarp-\Dz$~mass difference~\cite{Lees:2013zna,
Anastassov:2001cw,PDG2021}
directly affects the~values of these parameters
and is assigned as corresponding systematic uncertainty. 

For the lower limit on the~parameter $\left|g\right|$, 
only systematic  uncertainties related to the~fit 
model are considered. 
For~each alternative model 
the~likelihood profile curves
are built and corresponding 90 and 95\,\% CL lower limits 
are calculated using the~procedure described above. 
%Conservatively, 
The~smallest
of the resulting values
%value of lower limit 
is taken
as the~lower limit that accounts 
for the~systematic uncertainty:
\mbox{$\left| g \right| > 
    %% 5.07\,(4.26)\gev   
    5.1\,(4.3)\gev
$} at~90\,(95)\,\%~CL.

%% \section*{Discussion}\label{sec:discussion}
\subsection*{Results}\label{sec:discussion}

\begin{figure}[t]  
  \centering
  \includegraphics[width=\textwidth]{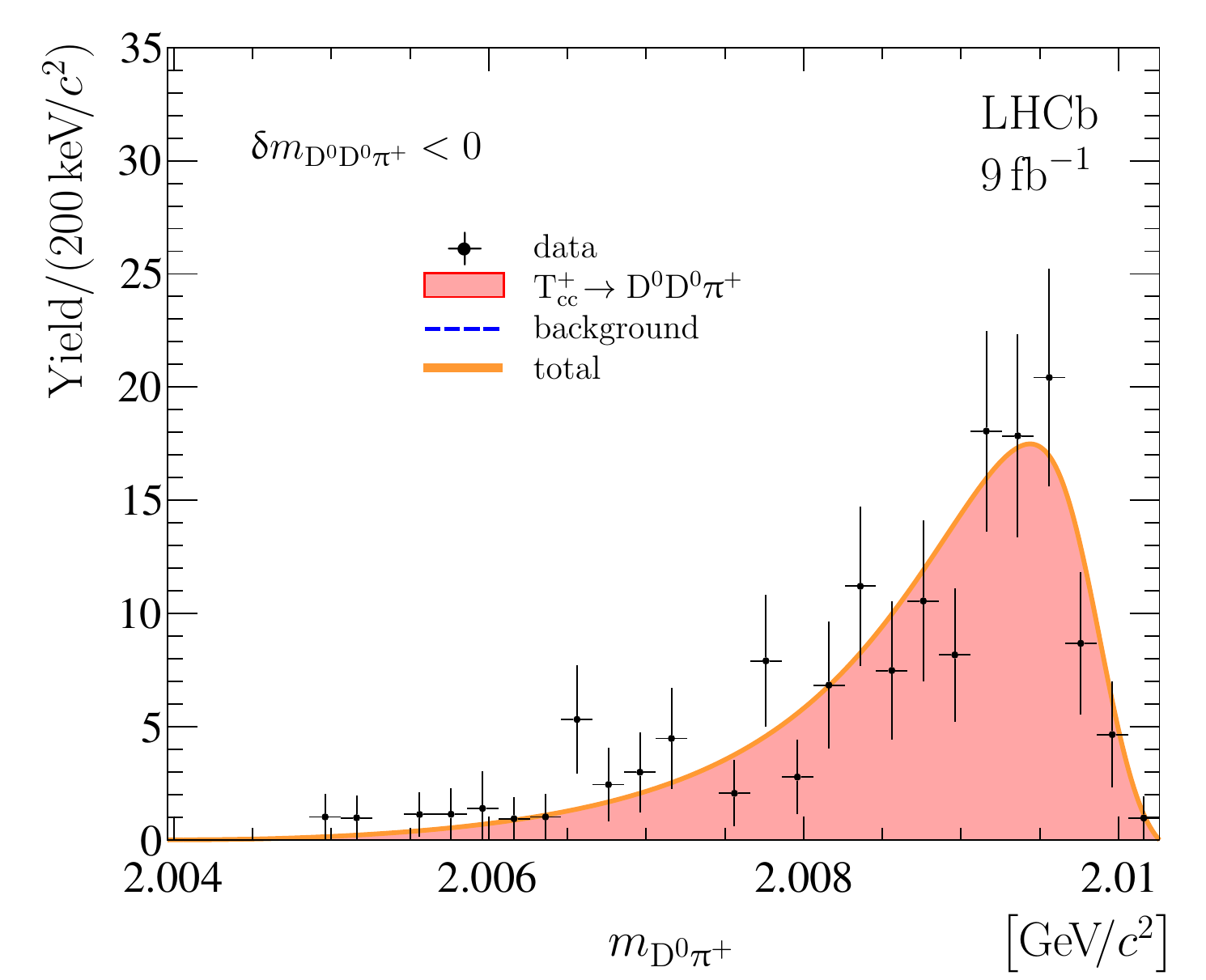}
  \caption { \small
  {\bf Mass distribution for \Dz\pip~pairs. }
  Mass distribution for \Dz\pip~pairs
  from  selected $\Dz\Dz\pip$~candidates 
  with a~mass below the $\Dstarp\Dz$~mass threshold
  with  non\protect\nobreakdash-\Dz~background subtracted. 
  The~overlaid fit result
  is described in the~text. 
  %%
  %% {\color{red}{
  The~background component vanishes in the~fit.
  %% }} 
  %% 
   %% {\color{red}{
   Uncertainties on the~data points are statistical
 only and represent one standard deviation, 
 calculated as a~sum in quadrature of the~assigned weights from 
 the~background\protect\nobreakdash-subtraction procedure.
 %% }}
  }
  \label{fig:DATA_BWU_Dpi}
\end{figure}

%%%%%%%%%%%%%%%%%%%%%%%%%%%%%%%%%%%%%%%%%%%%%%%%%%%%%%%%%%%%%%%%%%%%%%%%%%
%%%%%%%%%%% spin %%%%%%%%%%%%%%%%%%%%%%%%%%%%%%%%%%%%%%%%%%%%%%%%%%%%%%%%%
%%%%%%%%%%%%%%%%%%%%%%%%%%%%%%%%%%%%%%%%%%%%%%%%%%%%%%%%%%%%%%%%%%%%%%%%%%
Studying the~$\Dz\pip$~mass 
distribution 
for  
\mbox{$\decay{\Tcc}{\Dz\Dz\pip}$}~decays 
allows 
testing 
the~hypothesis 
that the~\mbox{$\decay{\Tcc}{\Dz\Dz\pip}$}~decay
proceeds through an~intermediate 
off\nobreakdash-shell $\Dstarp$~meson.
The~background\nobreakdash-subtracted 
$\Dz\pip$ mass 
distribution for selected 
$\Dz\Dz\pip$~candidates 
%% \footnote{Both $\Dz\pip$ 
%% combinations are included.
%% The~two\nobreakdash-dimensional
%% %% {\em{Dalitz\nobreakdash-like}} 
%% distribution  
%% of the~mass of 
%% one $\Dz\pip$~combination
%% versus the~mass of another $\Dz\pip$
%% combination
%% is presented in 
%% Extended Data Fig.~\ref{fig:DALITZ}.} 
with 
the~\Dz\Dz\pip mass
with respect 
to the~$\Dstarp\Dz$~mass threshold, 
$\updelta m_{\Dz\Dz\pip}$,
below zero
is shown in Fig.~\ref{fig:DATA_BWU_Dpi}. 
Both $\Dz\pip$ 
combinations are included
in this plot.
The~two\nobreakdash-dimensional
%% {\em{Dalitz\nobreakdash-like}} 
distribution  
of the~mass of 
one $\Dz\pip$~combination
versus the~mass of another $\Dz\pip$
combination
is presented in 
%% Extended Data 
Supplementary~Fig.~\ref{fig:DALITZ}.
%%Information
%%Fig.~10.

%%
%%
A~fit is performed 
to this~distribution with a~model containing 
signal and background components.
The~signal component is derived 
from the~$\mathcal{A}_{\mathrm{U}}$~amplitude,
see Methods Eq.~\eqref{eq:bwA}, 
and is convolved with 
a~detector resolution for the~$\Dz\pip$~mass.
This~detector resolution function 
is modelled with 
a~modified Gaussian function with 
power\nobreakdash-law tails on both 
sides of the~distribution~\cite{Skwarnicki:1986xj,
LHCb-PAPER-2011-013}
and parameters taken from simulation.
Similarly to the~correction 
used for the~$\Dz\Dz\pip$~mass resolution 
function $\mathfrak{R}$,
the~width of the~Gaussian function 
is corrected  by a~factor of~1.06 
which is determined by studying 
large samples of
\mbox{$\decay{\Dstarp}{\Dz\pip}$}~decays.
%%
%% {\color{red}{
The~root mean square of the~resolution 
function is around 220\kevcc.
%% }} 
%%
The~shape of the~background component is derived
from data for 
\mbox{$\updelta m_{\Dz\Dz\pip}>0.6\mevcc$}\,.
% \end{enumerate}
%%
The~fit results are overlaid 
in Fig.~\ref{fig:DATA_BWU_Dpi}. 
The~background component vanishes in the~fit, 
and the~\Dz\pip~spectrum is consistent with 
the~hypothesis 
that the~\mbox{$\decay{\Tcc}{\Dz\Dz\pip}$}~decay
proceeds through an intermediate 
off\nobreakdash-shell $\Dstarp$~meson.
This in turn favours 
the~$1^+$ assignment for the~spin\nobreakdash-parity
of the~state.
%% 

%% STANDARD 

%%%%%%%%%%%%%%%%%%%%%%%%%%%%%%%%%%%%%%%%%%%%%%%%%%%%%%%%%%%%%%%%%%%%%%%%%%
%%%%%%%%%%% Isospin %%%%%%%%%%%%%%%%%%%%%%%%%%%%%%%%%%%%%%%%%%%%%%%%%%%%%%
%%%%%%%%%%%%%%%%%%%%%%%%%%%%%%%%%%%%%%%%%%%%%%%%%%%%%%%%%%%%%%%%%%%%%%%%%%

\begin{figure}[t]
  \centering
  \includegraphics[width=\textwidth]{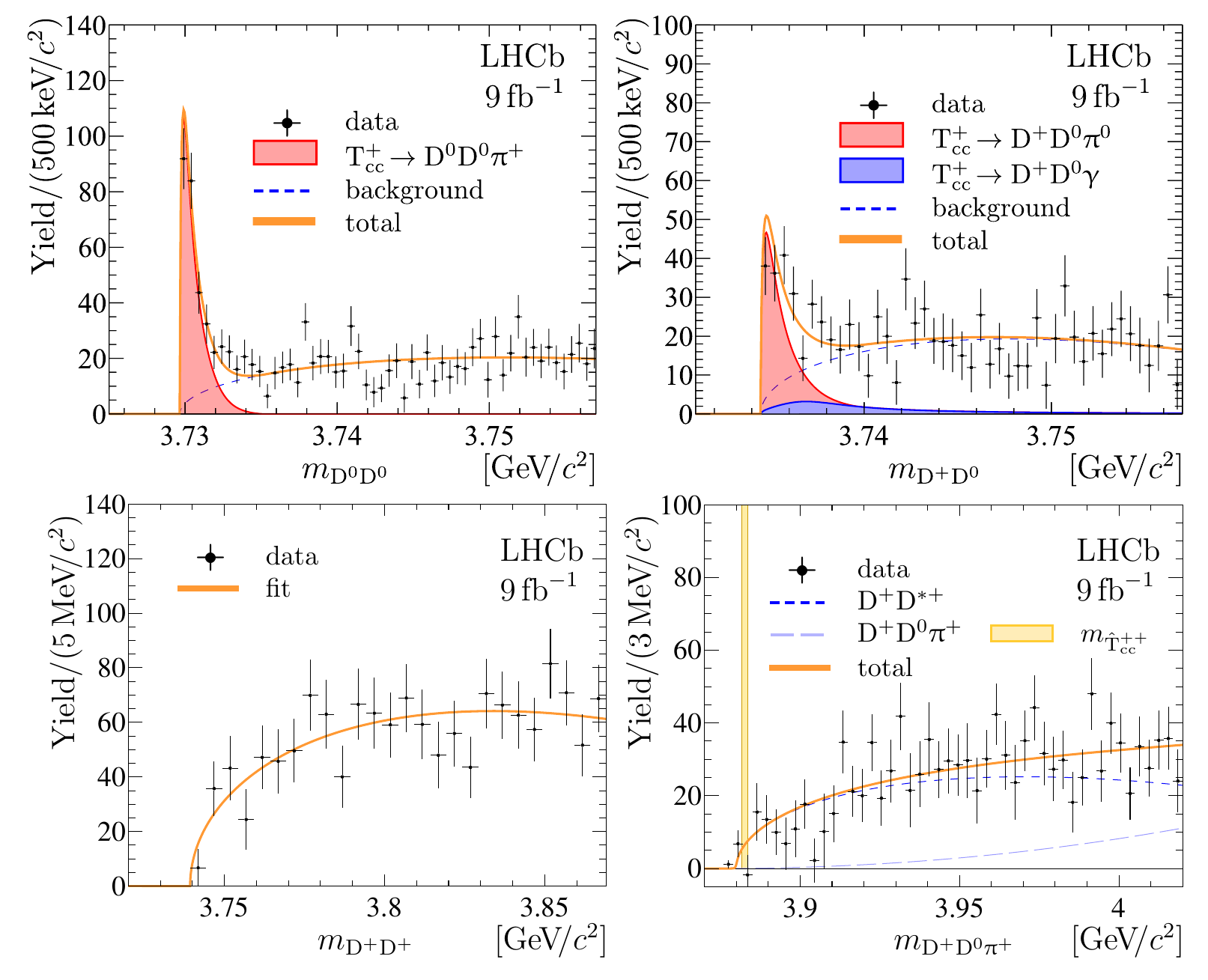}
  \caption { \small
  {\bf
  Mass distributions for selected 
  $\Dz\Dz$, 
  $\Dp\Dz$
  $\Dp\Dp$ and $\Dp\Dz\pip$~combinations.
  }
  (Top) \D\D~and $\D\D\pip$~mass distributions
  for
  selected 
  (left)~\Dz\Dz
  and (right)~\Dp\Dz~candidates 
  with the~non\protect\nobreakdash-\D~background subtracted. 
  The~overlaid fit results
  are described in the~text.
  For~visibility
  the~\mbox{$\decay{\Tcc}{\Dp\Dz\piz}$}
  is stacked
  on top of the~\mbox{$\decay{\Tcc}{\Dp\Dz\g}$}~component.
  (Bottom) 
  Mass distributions 
  for
  selected 
  (left)~\Dp\Dp
  and (right)~\Dp\Dz\pip~candidates 
  with the
  non\protect\nobreakdash-\D~background subtracted. 
  The~vertical coloured band indicates 
  the~expected mass
  for  the~hypothetical
  $\hat{\PT}^{++}_{\cquark\cquark}$~state. 
  The~overlaid fit results
  with background\protect\nobreakdash-only 
  functions
  are described in the~text. 
  %%
  %% {\color{red}{
  Uncertainties on the~data points are statistical
 only and represent one standard deviation, 
 calculated as a~sum in quadrature of the~assigned weights from 
 the~background\protect\nobreakdash-subtraction procedure.
 %%}}
  %%
  }
  \label{fig:DATA_DD}
\end{figure}

Due to the proximity of the~observed  
\Tcc signal
%% in the~$\Dz\Dz\pip$~mass distribution 
to the~$\Dstarp\Dz$~mass threshold,
and the~small energy release in 
the~\mbox{$\decay{\Dstarp}{\Dz}{\pip}$}~decay, 
the~$\Dz\Dz$~mass distribution from 
the~\mbox{$\decay{\Tcc}{\Dz\Dz\pip}$}~decay %also 
forms a~narrow peak 
just above the~$\Dz\Dz$~mass threshold.
In~a~similar way, %% for an~isoscalar \Tcc~state
a~peaking structure
in the~$\Dp\Dz$~mass spectrum 
just above the~$\Dp\Dz$~mass threshold 
is expected 
from \mbox{$\decay{\Tcc}{\Dp\Dz\piz}$}
and 
\mbox{$\decay{\Tcc}{\Dp\Dz\g}$}~decays,
both
proceeding via off\nobreakdash-shell 
intermediate $\Dstarp\Dz$ 
and $\Dstarz\Dp$~states.
The~$\Dz\Dz$ and $\Dp\Dz$~final 
states are reconstructed and selected 
similarly to the~$\Dz\Dz\pip$ final state,
where 
the~\mbox{$\decay{\Dp}{\Km\pip\pim}$}~decay
channel is used.
The~background-subtracted 
$\Dz\Dz$ and $\Dp\Dz$~mass distributions 
are shown in Fig.~\ref{fig:DATA_DD}(top), where  
narrow structures are clearly visible just above 
the~$\D\D$~thresholds. 
Fits to these distributions are performed using 
models consisting of two components:
a~signal component $F_{\D\D}$ 
described in Methods 
%% Eq.~\eqref{eq:dd_spectra}
Eqs.~\eqref{eq:fDzDz} and~\eqref{eq:fDpDz}
and 
obtained via integration of the~matrix elements 
for the~\mbox{$\decay{\Tcc}{\D\D\Ppi/\g}$}~decays 
with the~$\mathfrak{F}^{\mathrm{U}}$~profile,
and a~background component, 
parameterised as a~product of 
the~two\nobreakdash-body phase 
space function $\Phi_{\D\D}$  
and a~positive linear function $P_1$.
The~fit results are 
overlaid in  Fig.~\ref{fig:DATA_DD}(top).
%% and  summarised in Table~\ref{tab:DATA_DD}. 
%% {\color{red}{
The~signal yields
in the~$\Dz\Dz$ and 
$\Dp\Dz$~spectra are found to be 
$263\pm23$ and $171\pm26$, respectively.
%% }}
The~statistical significance of the~observed 
\mbox{$\decay{\Tcc}{\Dz\Dz\PX}$}
and \mbox{$\decay{\Tcc}{\Dp\Dz\PX}$}~signals, 
where \PX stands for non-reconstructed 
pions or photons, 
is estimated 
using Wilks' theorem~\cite{Wilks:1938dza} 
and is found to be in excess
of 20 and 10 standard deviations,
respectively.
The~relative yields for 
the~signals observed in the~$\Dz\Dz\pip$, 
$\Dz\Dz$ and $\Dz\Dp$~mass spectra 
agree with the expectations 
of the~model 
described in Methods
where the~decay of 
an~isoscalar \Tcc~state 
via 
the~$\Dstar\D$~channel with 
an~intermediate off-shell~$\Dstar$~meson is assumed.

%%

\begin{comment}
\begin{table}[tb]
	\centering
	\caption{\small 
	Signal and background yields, $N_S$ and $N_B$, 
	from the~fits to $\D\D$~mass spectra.
	The~uncertainties are statistical only. 
	}
	\label{tab:DATA_DD}
	\vspace{2mm}
	\begin{tabular*}{0.45\textwidth}{@{\hspace{3mm}}l@{\extracolsep{\fill}}cc@{\hspace{3mm}}}
	%% 
	    & $N_S$  
	    & $N_B$    
	    %% &  $\mathcal{S}~\left[\sigma \right]$
    %%   
   \\[1mm]
  \hline 
  \\[-2mm]
  %%
  $\Dz\Dz$   &   $263\pm23$  & $962\pm45$  \\ %% & 20  
  $\Dp\Dz$   &   $171\pm26$  & $763\pm47$ \\ %%  & 10 
  %%
	\end{tabular*}
	\vspace{3mm}
\end{table}
\end{comment}

The~observation of the~near\nobreakdash-threshold
signals in the~$\Dz\Dz$ and $\Dp\Dz$~mass spectra, along with 
the~signal shapes and yields, 
all agree with the~isoscalar $\Tcc$~hypothesis
for the~narrow signal observed in the~\Dz\Dz\pip~mass 
spectrum.
However,
%% if the~observed state is 
%% not an~isoscalar $\Tcc$~state, 
an~alternative interpretation
could be that this state
is the~$\mathrm{I}_3=0$ component of 
a~$\hat{\PT}_{\cquark\cquark}$ 
isotriplet ($\hat{\PT}^{0}_{\cquark\cquark}$,
$\hat{\PT}^{+}_{\cquark\cquark}$, 
$\hat{\PT}^{++}_{\cquark\cquark}$)
with 
$\cquark\cquark\uquarkbar\uquarkbar$,
$\cquark\cquark\uquarkbar\dquarkbar$
and $\cquark\cquark\dquarkbar\dquarkbar$
quark content, respectively.
Assuming that the~observed peak corresponds 
to the~$\hat{\PT}^{+}_{\cquark\cquark}$~component 
and using the~estimates for 
the~$\hat{\PT}_{\cquark\cquark}$~mass splitting 
from Methods 
%% Eq.~\eqref{eq:tcc},
Eqs.~\eqref{eq:mTccone}
and~\eqref{eq:mTcctwo},
the~masses of the~$\hat{\PT}^{0}_{\cquark\cquark}$
and $\hat{\PT}^{++}_{\cquark\cquark}$~states 
are estimated 
to be  
slightly below 
the~$\Dz\Dstarz$ and slightly above 
the~$\Dp\Dstarp$~mass 
thresholds,  
respectively: 
%% \begin{subequations}
\begingroup
\allowdisplaybreaks
\begin{eqnarray}
    m_{\hat{\PT}^0_{\cquark\cquark}} 
    - \left( m_{\Dz} + m_{\Dstarz}\right) 
    & = & 
    -2.8 \pm 1.5\mevcc \,, 
    \\
    m_{\hat{\PT}^{++}_{\cquark\cquark}} 
    - \left( m_{\Dp} + m_{\Dstarp}\right) 
    & = & 
    \phantom{-}2.7 \pm 1.3 \mevcc \,.
\end{eqnarray}
\endgroup
%% \end{subequations}
With these mass assignments, 
assuming equal production 
of all 
three~$\hat{\PT}_{\cquark\cquark}$~components, 
the~$\hat{\PT}^{0}_{\cquark\cquark}$~state would be 
an~extra narrow state that decays 
into the~$\Dz\Dz\piz$ and $\Dz\Dz\g$~final 
states via
an~off\nobreakdash-shell \Dstarz~meson. 
These decays would contribute
to the~narrow near\nobreakdash-threshold 
enhancement in the~$\Dz\Dz$~spectrum,
and %% Such contribution would 
increase 
the~signal in the~$\Dz\Dz$~mass spectrum by 
almost a~factor of~three. 
The~$\hat{\PT}^{++}_{\cquark\cquark}$~state 
would decay via an~on-shell \Dstarp meson 
\mbox{$\decay{\hat{\PT}^{++}_{\cquark\cquark}}{\Dp\Dstarp}$}, 
therefore it could be a~relatively wide state, 
with width up to a~few \mev~\cite{DelFabbro:2004ta}. 
Therefore, it~would manifest itself 
as a~peak with a~moderate width 
in the~\Dp\Dz\pip~mass 
spectrum with %% signal yield comparable with 
a~yield comparable to 
that of  
the~\mbox{$\decay{\hat{\PT}^{+}_{\cquark\cquark}}
{\Dz\Dz\pip}$}~decays.
In addition, it would contribute to the~\Dp\Dz~mass spectrum,
tripling the~contribution 
from the~$\hat{\PT}^{+}_{\cquark\cquark}$~decays.
However, due to the larger~mass 
of the~$\hat{\PT}^{++}_{\cquark\cquark}$~state 
and its larger width, this contribution should 
be %% significantly 
wider, 
making it more difficult to disentangle  from the~background. 
Finally, 
the~$\hat{\PT}^{++}_{\cquark\cquark}$~state 
would make a~contribution to the~\Dp\Dp~spectrum
with a~yield similar to 
the~contribution from 
\mbox{$\decay{\hat{\PT}^{+}_{\cquark\cquark}}
{\Dz\Dp\piz/\g}$}~decays
to the~\Dz\Dp~spectrum, but wider. 
The~mass spectra for $\Dp\Dp$ and $\Dp\Dz\pip$~combinations 
are shown in Fig.~\ref{fig:DATA_DD}(bottom).
%%
\begin{comment}
\begin{figure}[t]
  \centering
  \includegraphics[width=\textwidth]{Fig_5.pdf}
  \caption { \small
  {\bf
  Mass distributions for
  selected 
  \Dp\Dp
  and \Dp\Dz\pip~candidates.
  %% 
  }
  Mass distributions 
  for
  selected 
  (left)~\Dp\Dp
  and (right)~\Dp\Dz\pip~candidates 
  with the
  non\protect\nobreakdash-\D~background subtracted. 
  %% 
  The~vertical coloured band indicates 
  the~expected mass
  for  the~hypothetical
  $\hat{\PT}^{++}_{\cquark\cquark}$~state. 
  The~overlaid fit results
  with background\protect\nobreakdash-only 
  functions
  are described in the~text. 
  %%
  %% {\color{red}{
  Uncertainties on the~data points are statistical
 only and represent one standard deviation, 
 calculated as a~sum in quadrature of the~assigned weights from 
 the~background\protect\nobreakdash-subtraction procedure.
 %% }}
  %%
  }
  \label{fig:DATA_DpDp}
\end{figure}
\end{comment}
%%
Neither distribution exhibits any narrow 
signal\nobreakdash-like structure.
Fits to these spectra are performed 
using the~following background\nobreakdash-only functions: 
%%
%% \begin{subequations}
\begingroup
\allowdisplaybreaks
\begin{eqnarray}
 B_{\Dp\Dp}     & = &  
 \Phi_{\Dp\Dp} \times P_{1}   \,, \\ 
 B_{\Dp\Dz\pip} & = &  
 \left (\Phi_{\Dp\Dstarp} \times P_{1} \right) * \mathfrak{R} + 
 \Phi_{\Dp\Dz\pip} \times P_{0}\,.
\end{eqnarray}
\endgroup
%% \end{subequations}
%%
The~results of these fits 
are overlaid in Fig.~\ref{fig:DATA_DD}(bottom). 
The~absence of any signals in the~$\Dp\Dp$ and $\Dp\Dz\pip$~mass 
spectra is therefore a~strong argument in favour of 
the~isoscalar nature of the~observed peak
in the~$\Dz\Dz\pip$~mass spectrum.

The~interference 
between two virtual channels 
for the~\mbox{$\decay{\Tcc}{\Dz\Dz\pip}$} decay,
corresponding to two amplitude terms,
see Methods Eq.~\eqref{eq:pipdzdz},
is studied by 
setting the~term proportional to $C$ 
in Methods~Eq.~\eqref{eq:ampsqpip} 
to be equal to zero.
This~causes a~43\% reduction 
in the~decay rate, 
pointing to a~large 
interference. 
The~same procedure applied to 
the~\mbox{$\decay{\Tcc}{\Dp\Dz\piz}$}~decays
gives 
the~contribution of~$45\%$
for the~interference
between 
the~\mbox{$\left(\decay{\Dstarp}{\Dp\piz}\right)\Dz$}
and 
\mbox{$\left(\decay{\Dstarz}{\Dz\piz}\right)\Dp$}
channels. 
For \mbox{$\decay{\Tcc}{\Dp\Dz\g}$}~decays
the~role of the~interference 
between 
the~\mbox{$\left(\decay{\Dstarp}{\Dp\g}\right)\Dz$}
and 
\mbox{$\left(\decay{\Dstarz}{\Dz\g}\right)\Dp$}
channels 
is estimated
by equating to zero 
the~\mbox{$\mathfrak{F}_+\mathfrak{F}^{*}_0$} and
\mbox{$\mathfrak{F}^{*}_+\mathfrak{F}_0$} 
terms in 
Methods~Eqs.~\eqref{eq:msq_gamma_one}
and~\eqref{eq:msq_gamma_two}.
The~interference contribution 
is found to be~33\%.

%%%%%%%%%%%%%%%%%%%%%%%%%%%%%%%%%%%%%%%%%%%%%%%%%%%%%%%%%%%%%%%%%%%%%%%%%%
%%%%%%%%%%% Pole %%%%%%%%%%%%%%%%%%%%%%%%%%%%%%%%%%%%%%%%%%%%%%%%%%%%%%%%% 
%%%%%%%%%%%%%%%%%%%%%%%%%%%%%%%%%%%%%%%%%%%%%%%%%%%%%%%%%%%%%%%%%%%%%%%%%%

\begin{figure}[tb]
  \centering
  \includegraphics[width=\textwidth]{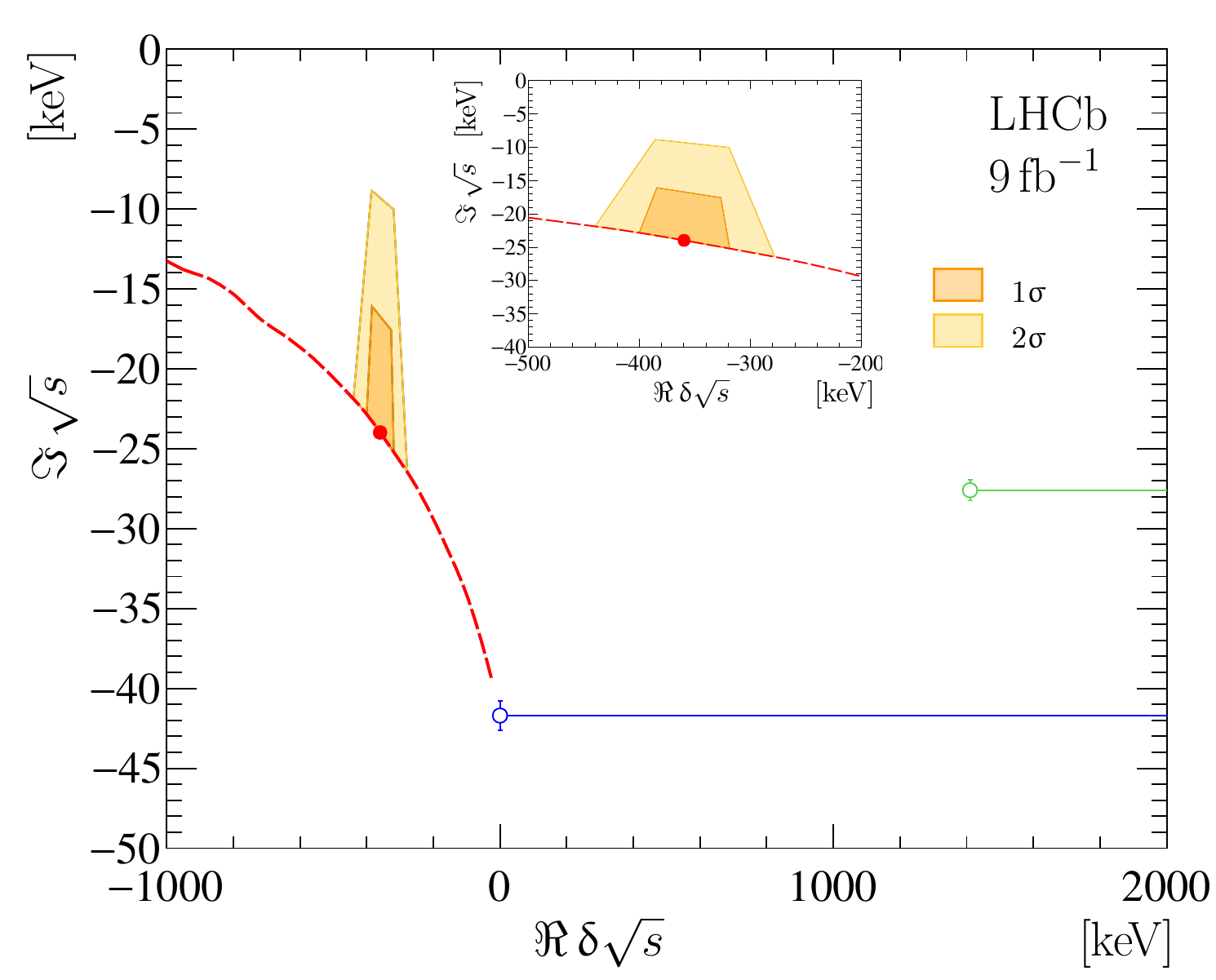}
  \caption { \small
  {\bf
  Complex plane of 
  the~$\updelta \sqrt{s}$ 
  variable.
  }
  Complex plane of 
  the~$\updelta \sqrt{s}$~variable.
  The~dashed red line shows the~allowed region
  for large $\left| g\right|$~values.
  The~filled red circle indicates 
  the~best estimate for 
  the~pole location and the filled regions show 
  $1\upsigma$ and $2\upsigma$~confidence regions. 
  Open blue and green circles show 
  the~branch points corresponding 
  to the~$\Dstarp\Dz$ and 
  $\Dstarz\Dp$~channels, respectively, and 
  the corresponding blue and green lines 
  indicate branch cuts.  
  Three other branch points at 
  $\sqrt{s}$ of \mbox{$m_{\Dz}+m_{\Dp}$},
  \mbox{$m_{\Dz}+m_{\Dp}+m_{\piz}$} 
  and \mbox{$2m_{\Dz}+m_{\pip}$},
  corresponding to the~openings of 
  the~$\Dz\Dp\gamma$, $\Dz\Dp\piz$
  and $\Dz\Dz\pip$~decay channels, 
  are outside of the~displayed region. 
  }
  \label{fig:POLE}
\end{figure}

%% {\color{red}{
Using the~model described earlier
and results of the fit to 
the~\Dz\Dz\pip~mass spectrum, 
the~position of the~amplitude pole $\hat{s}$
in the~complex plane, 
responsible for the~appearance of 
the~narrow structure in the~\Dz\Dz\pip~mass spectrum
is determined.
%% }}
%% 
%% A model-independent characteristic 
%% of the~state is the~position of 
%% the~amplitude pole $\hat{s}$ 
%% in the~complex plane, 
%% responsible for the~appearance of 
%% the~narrow structure in the~\Dz\Dz\pip~mass spectrum.
The~pole parameters, 
mass $m_{\mathrm{pole}}$ and width 
$\Gamma_{\mathrm{pole}}$, are defined 
through the pole location $\hat{s}$  as
\begin{equation}
    \sqrt{ \hat{s} } \equiv 
    m_{\mathrm{pole}} - \frac{i}{2}\Gamma_{\mathrm{pole}} \,.
\end{equation}
The~pole location $\hat{s}$~is a~solution of the~equation
\begin{align}
  \dfrac{1}{ \mathcal{A}^{I\! I}_{\mathrm{U}} 
  (\hat{s})}  = 0\,,
\end{align}
where $\mathcal{A}^{I\! I}_{\mathrm{U}}(s)$ 
denotes the~amplitude 
on the~second Riemann sheet
defined in Methods~Eq.~\eqref{eq:ABWII}. 
For~large coupling $\left|g\right|$ the~position of 
the~resonance pole is uniquely determined 
by the~parameter $\updelta m_{\mathrm{U}}$,
\ie the binding energy
and the~width of the~$\Dstarp$ meson. 
Figure~\ref{fig:POLE} shows 
the~complex plane of the~$\updelta \sqrt{s}$~variable, 
defined as 
  \begin{equation}
    \updelta \sqrt {s} \equiv \sqrt{s} - \left( m_{\Dstarp} + m_{\Dz}\right)\,.
\end{equation}
All~possible positions of the~pole 
for 
%% $\left|g\right|\to+\infty$ 
$\left| g \right| \gg m_{\Dz}+m_{\Dstarp}$ 
%% $\tfrac{ \left|g\right|}{m_{\Dstarp}+m_{\Dz}}\to+\infty$ 
are located on a~red dashed curve
in Fig.~\ref{fig:POLE}. 
The~behaviour of the curve
can be understood as follows:
with an~increase 
of the~binding energy\,(distance to 
the~\Dstarp\Dz~mass threshold), 
the~width gets narrower;
and 
when the~parameter $\updelta m_{\mathrm{U}}$ 
approaches zero, 
the~pole touches the~$\Dz\Dstarp$ cut and 
moves to the~other complex 
sheet, \ie the~state becomes virtual.
For~smaller values of $\left|g\right|$, 
the~pole is located between
the limiting curve and the~$\Re\,s=0$~line. 
The~pole parameters are found to be 
%%\begin{subequations}
%\label{eq:bwupole}
 \begingroup
\allowdisplaybreaks
\begin{eqnarray}
     \updelta  m_{\mathrm{pole}} 
     & = & 
%% -364 \pm 40  ^{\,+\,5\phantom{0}}_{\,-\,0} \kevcc \,, \\ 
     -360 \pm 40  ^{\,+\,4\phantom{0}}_{\,-\,0} \kevcc \,, \\ 
    \Gamma_{\mathrm{pole}}              & = &  
     \phantom{-0}48 \pm \phantom{0}2 ^{\,+\,0}_{\,-\,14}\kev \,,
  \end{eqnarray}
  \endgroup
%%\end{subequations}
where the~first uncertainty 
is due to the~$\updelta m_{\mathrm{U}}$~parameter
and the~second is due to the~unknown
value of the~$\left|g\right|$~parameter.
The~peak is well separated 
from the~$\Dstarp\Dz$ threshold 
in the~$\Dz\Dz\pip$~mass spectrum.
Hence, as for an~isolated narrow resonance,
the~parameters of the~pole are similar
to the~visible peak parameters, namely 
the~mode $\updelta {\mathfrak{m}}$
and FWHM ${\mathfrak{w}}$.
%% from Table~\ref{tab:DATA_visible}.
%%

%% {\color{red}{
The~systematic uncertainties 
quoted here do not account for 
the~possibility that any of 
the~underlying assumptions on which 
the~model is built are not~valie. 
For~example, 
as~shown earlier 
the~data are consistent with  a~wide 
range of~FHWM $\mathfrak{w}$~values for the~signal profile.
Therefore the~pole width $\Gamma_{\mathrm{pole}}$
is based  mainly on the~\Tcc~amplitude model 
and the~value of the~$m_{\mathrm{U}}$~parameter
determined from the fit to
the~$\Dz\Dz\pip$~mass spectrum.
%% }}

%%%%%%%%%%%%%%%%%%%%%%%%%%%%%%%%%%%%%%%%%%%%%%%%%%%%%%%%%%%%%%%%%%%%%%%%%%
%%%%%%%%%%% Scattering length and effective range  %%%%%%%%%%%%%%%%%%%%%%
%%%%%%%%%%%%%%%%%%%%%%%%%%%%%%%%%%%%%%%%%%%%%%%%%%%%%%%%%%%%%%%%%%%%%%%%%%

A~study of the~behaviour of 
the~$\mathcal{A}_{\mathrm{U}}(s)$~amplitude 
in the vicinity of the~$\Dstarp\Dz$~mass threshold 
leads to the~determination of 
the~low\nobreakdash-energy scattering parameters, 
namely the~scattering length, $a$, and 
the~effective range, $r$.
These parameters are defined via 
the~coefficients of 
the~first two terms of 
the~Taylor expansion of 
the~inverse non\nobreakdash-relativistic 
amplitude~\cite{Bethe:1949yr}, \ie 
\ifthenelse{\boolean{wordcount}}{}{
\begin{equation}\label{eq:A_NR}
  \mathcal{A}_\text{NR}^{-1}  = 
  \dfrac{1}{a}+ r\dfrac{k^2}{2} - i k + \mathcal{O}(k^4) \,,
\end{equation}
}
where $k$ is the wave number.
For $\updelta \sqrt {s} \lesssim -\Gamma_\Dstarp$
the~inverse amplitude from Eq.~\eqref{eq:bwU}
matches Eq.~\eqref{eq:A_NR}
%% starts matching 
%% the~inverse amplitude from Eq.~\eqref{eq:bwU} at 
%% $\left|\updelta m\right| \gtrsim \Gamma_\Dstarp$
up to a~scale parameter obtained numerically, 
see Methods~Eq.~\eqref{eq:w_scattering}.  
The~value of the~scattering length is found to be 
\ifthenelse{\boolean{wordcount}}{}{
\begin{equation}
    a = \Bigl[ -\left(7.16 \pm 0.51\right) 
    + i\left(1.85 \pm 0.28 \right) \Bigr] \fm\,.
\end{equation}
}
Typically, a~non\nobreakdash-vanishing 
imaginary part of 
the~scattering length indicates the~presence 
of inelastic channels~\cite{Balakrishnan:1997xyz};
however, 
in~this case the 
non\nobreakdash-zero imaginary part is related to the~lower threshold, 
\mbox{$\decay{\Tcc}{\Dz\Dz\pip}$}, 
and is determined by the width of the~$\Dstarp$~meson.
The~real part of the~scattering length $a$ is 
negative indicating attraction.
This~can be interpreted as the~characteristic~size 
of the~state~\cite{Guo:2017jvc},
\begin{equation} \label{eq:Ra}
R_a \equiv - \Re\,a = 7.16\pm0.51\fm\,.
\end{equation}
For the $\mathcal{A}_\mathrm{U}$ amplitude
the~effective range $r$ is non\nobreakdash-positive and proportional to $\left|g\right|^{-2}$,
see Methods Eq.~\eqref{eq:eff_range}.
Its~value is consistent with zero for the~baseline fit. 
An~upper limit on the~$-r$~value is set as 
\begin{equation}
0 \,\,\leq \,\, -r 
 <  11.9\,(16.9)\fm~\text{at}~90\,(95)\%~\text{CL}\,.
\end{equation}
The~Weinberg compositeness
criterion~\cite{Weinberg:1965zz,Matuschek:2020gqe}
makes use of the~relation between 
the~scattering length and the~effective range
to construct the~compositeness variable $Z$,
\ifthenelse{\boolean{wordcount}}{}{
\begin{eqnarray}
Z = 1-\sqrt{\dfrac{1}{ 1 + 2 \left|r/\Re\,a\right|}}\,,
\end{eqnarray}
}
for which $Z=1$ corresponds 
to a~compact state 
that does not interact 
with the~continuum,
while $Z=0$ indicates a~composite state 
formed by compound interaction.
% {\color{red}
Using the~relation between $r$ and $|g|$ 
from Methods~Eq.~\eqref{eq:eff_range},
one finds $Z\propto \left|g\right|^{-2}$ 
for large values of~$\left|g\right|$.
The~default fit corresponds 
to large values of $\left|g\right|$,
%% to $\left|g\right| \to +\infty$,
and thus, $Z$ approaching to zero.
A~non\nobreakdash-zero value of $Z$ would require 
a smaller value of $\left|g\right|$, 
\ie smaller resonance width, 
see 
%% Extended Data 
Supplementary~Fig.~\ref{fig:SCALING}.
%%Information
%%Fig.~7.
%% }
The~following upper limit of 
the~compositeness parameter $Z$ is set:
\ifthenelse{\boolean{wordcount}}{}{
\begin{equation}
Z  < 0.52\,(0.58)~\text{at}~90\,(95)\%~\text{CL}\,.
\end{equation}
}
%%

%%%%%%%%%%%%%%%%%%%%%%%%%%%%%%%%%%%%%%%%%%%%%%%%%%%%%%%%%%%%%%%%%%%%%%%%%%
%%%%%%%%%%% binding energy %%%%%%%%%%%%%%%%%%%%%%%%%%%%%%%%%%%%%%%%%%%%%%%
%%%%%%%%%%%%%%%%%%%%%%%%%%%%%%%%%%%%%%%%%%%%%%%%%%%%%%%%%%%%%%%%%%%%%%%%%%

Another estimate of 
the~characteristic size is obtained 
from the~value of the~binding energy $\Delta E$.
Within the~interpretation of the~\Tcc~state 
as a~bound 
\Dstarp\Dz~molecular\nobreakdash-like state,
the~binding energy 
is $\Delta E=-\updelta m_{\mathrm{U}}$.
The~characteristic momentum 
scale $\gamma$~\cite{Guo:2017jvc} is estimated to be 
\ifthenelse{\boolean{wordcount}}{}{
\begin{equation}
    \gamma = \sqrt{2 \upmu \Delta E} = 
%    26.36 \pm 1.47\mevcc\,, 
    26.4 \pm 1.5\mevc\,, 
\end{equation}
}
where $\upmu$ is the~reduced mass of the~\Dstarp\Dz~system.
This value of the~momentum scale  
in turn corresponds to 
a~characteristic~size $R_{\Delta E}$ of the~molecular-like  
state,  
\ifthenelse{\boolean{wordcount}}{}{
\begin{equation}
    R_{\Delta E} \equiv \dfrac{1}{\gamma}=
    %% 7.4853 +- 0.418228 
    % 7.49 \pm 0.42\fm\,,
    7.5\pm0.4\fm \,,
\end{equation}
}
which is consistent with the $R_{a}$~estimate from 
the~scattering length.

\begin{figure}[tb]
  \centering
   \includegraphics[width=\textwidth]{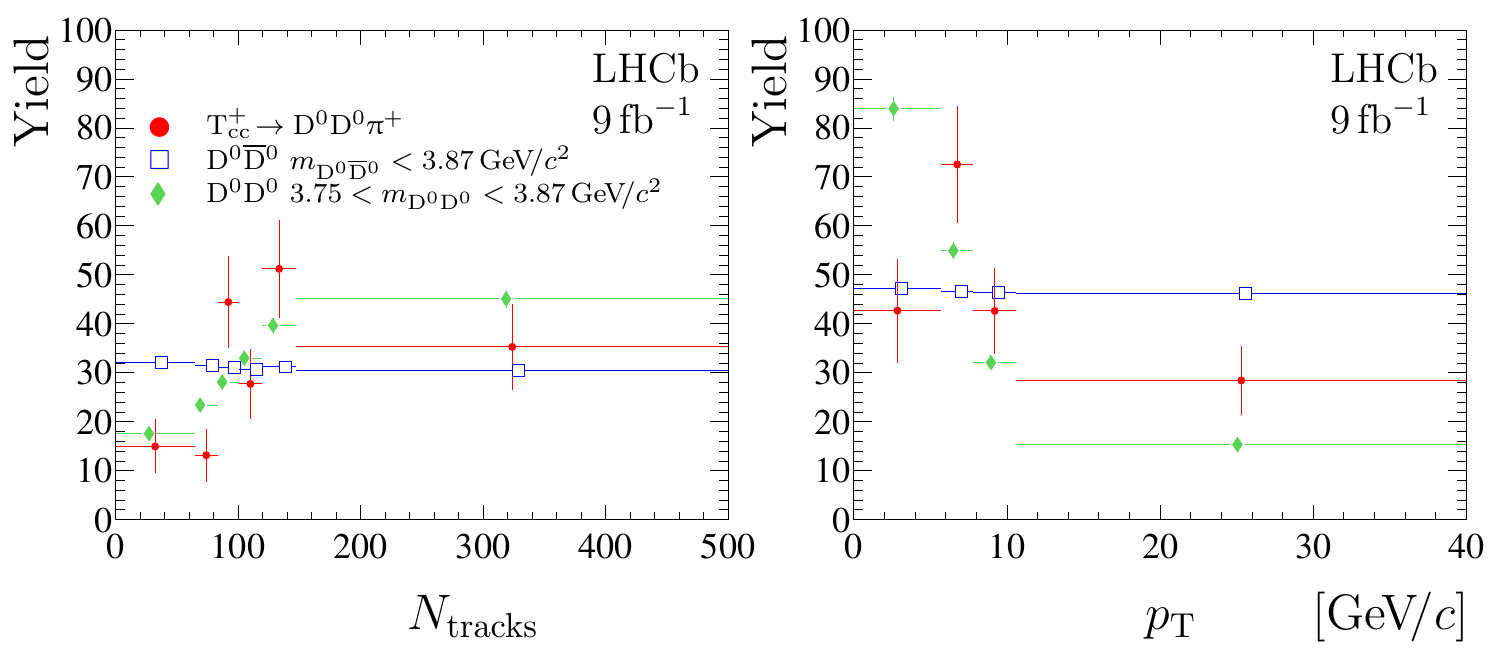}
  \caption { \small
  {\bf Track multiplicity and transverse momentum distributions.}
  (left)~Background\protect\nobreakdash-subtracted distributions
  for the~multiplicity of tracks 
  reconstructed 
  in the~vertex detector 
  for (red circles)~\mbox{$\decay{\Tcc}{\Dz\Dz\pip}$}~signal,
  low\protect\nobreakdash-mass~(blue open squares) $\Dz\Dzb$~and (green filled diamonds)
  $\Dz\Dz$~pairs.
  The~binning scheme is chosen to have 
  an~approximately uniform 
  distribution for \Dz\Dzb~pairs.
  The~distributions for 
  the~$\Dz\Dzb$~and 
  $\Dz\Dz$~pairs are 
  normalised to the~same yields 
  as the~$\decay{\Tcc}{\Dz\Dz\pip}$~signal.
  (right)~Background-subtracted
  transverse momentum spectra 
  for (red circles)~\mbox{$\decay{\Tcc}
  {\Dz\Dz\pip}$}~signal,
  (blue open squares)~low-mass $\Dz\Dzb$
  and (green filled diamonds)
  $\Dz\Dz$~pairs.
  The binning scheme is chosen to have 
  an approximately uniform distribution for \Dz\Dzb~pairs.
  The distributions for 
  the~$\Dz\Dzb$~and 
  $\Dz\Dz$~pairs are 
  normalised to the~same yields 
  as $\decay{\Tcc}{\Dz\Dz\pip}$~signal.
  For~better visualisation, the
  points are slightly displaced from 
  the~bin centres. 
  For~better visualisation, 
  the~points are slightly 
  displaced from the~bin centres.
  %%
  %% {\color{red}{
  Uncertainties on the~data points are statistical
 only and represent one standard deviation, 
 %% }}
  %%
  }
  \label{fig:DATA_nVELO_6}
\end{figure}

\begin{comment}
\begin{figure}[htb]
  \centering
  \includegraphics[width=\textwidth]{Fig_7.pdf}
  \caption { \small
  {\bf Transverse momentum spectra.}  
  Background-subtracted
  transverse momentum spectra 
  for (red circles)~\mbox{$\decay{\Tcc}
  {\Dz\Dz\pip}$}~signal,
  (blue open squares)~low-mass $\Dz\Dzb$
  and (green filled diamonds)
  $\Dz\Dz$~pairs.
  The binning scheme is chosen to have 
  an approximately uniform distribution for \Dz\Dzb~pairs.
  The distributions for 
  the~$\Dz\Dzb$~and 
  $\Dz\Dz$~pairs are 
  normalised to the~same yields 
  as $\decay{\Tcc}{\Dz\Dz\pip}$~signal.
  For~better visualisation, the
  points are slightly displaced from 
  the~bin centres. 
  %%
   %%{\color{red}{
   Uncertainties on the~data points are statistical
 only and represent one standard deviation.
 %%}}
  %%
  }
  \label{fig:DATA_PT_4}
\end{figure}
\end{comment}

\begin{figure}[t]
  \centering
  \includegraphics[width=\textwidth]{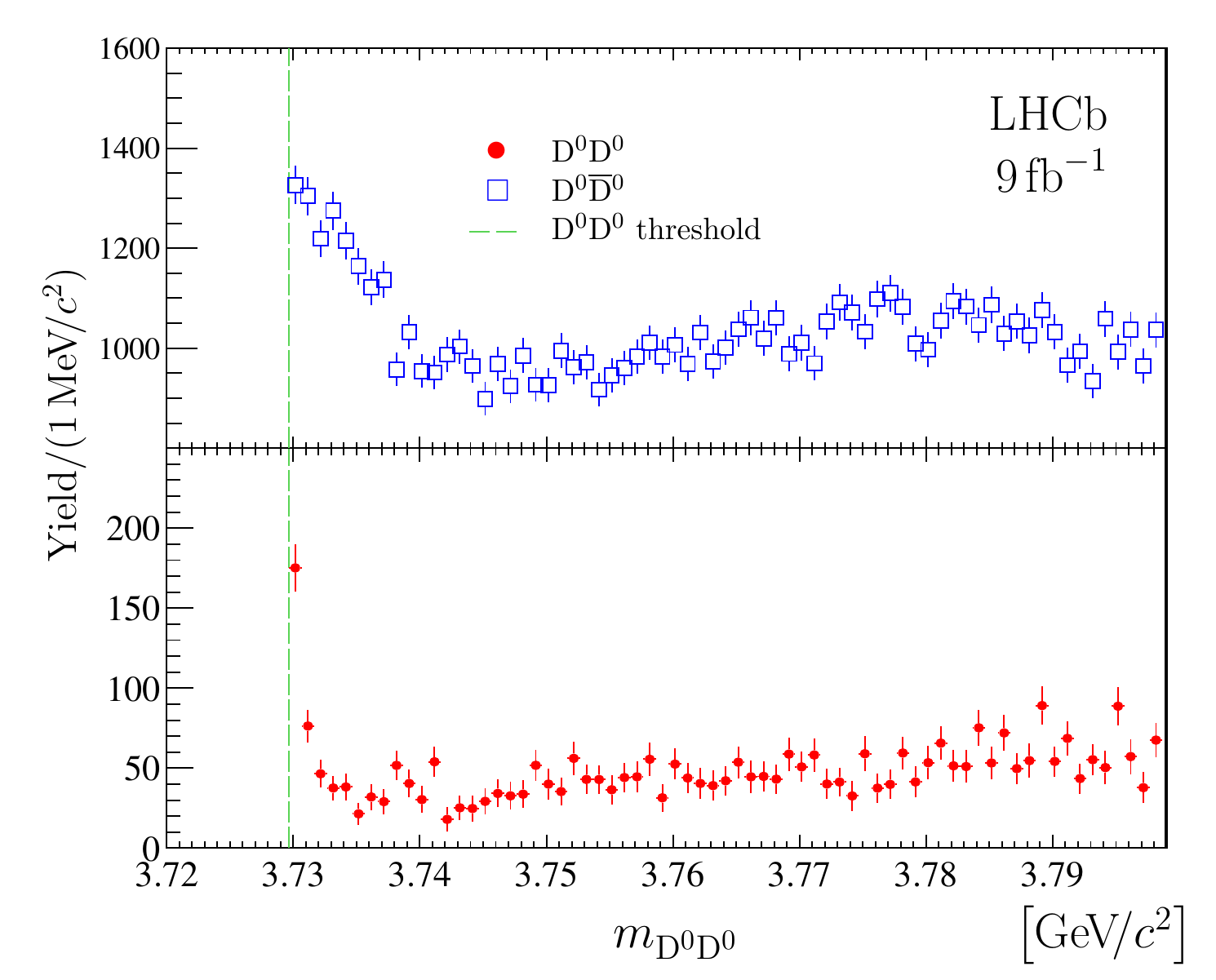}
  \caption {\small
  {\bf Mass distributions for \Dz\Dz and \Dz\Dzb~candidates.}
  Background\protect\nobreakdash-subtracted 
  \Dz\Dz and \Dz\Dzb~mass distributions.
  The~near\protect\nobreakdash-threshold enhancement in 
   the $\Dz\Dz$ channel corresponds to partially reconstructed 
   ${\Tcc\to\Dz\Dz\pip}$ decays, while in the~$\Dz\Dzb$ channel 
   the~threshold enhancement
   corresponds to partially reconstructed 
   $\chicone(3872)\to\Dz\Dzb\piz$ decays.
   %%
   %% {\color{red}{
   The~$\Dz\Dzb$~mass distribution
   is zero\protect\nobreakdash-suppressed for better visualisation. 
   %% }}
   %% 
   %% {\color{red}{
   Uncertainties on the~data points are statistical
 only and represent one standard deviation, 
 calculated as a~sum in quadrature of the~assigned weights from 
 the~background\protect\nobreakdash-subtraction procedure.
 %% }}
   %%
  }
  \label{fig:D0D0OS}
\end{figure}

For high\nobreakdash-energy hadroproduction of 
a~state with such a~large size, $R_a$ or $R_{\Delta E}$, 
one expects 
a~strong dependency of 
the~production rate 
on event multiplicity, 
similar to that observed for
the~$\chicone(3872)$~state~\cite{LHCb-PAPER-2020-023}. 
The~background\nobreakdash-subtracted 
distribution of the~number of tracks 
reconstructed in the~vertex detector, $N_{\mathrm{tracks}}$,  
is shown 
in Fig.~\ref{fig:DATA_nVELO_6}(left) 
together 
with the~distributions for
low\nobreakdash-mass 
$\Dz\Dzb$~pairs
%% \footnote{The chosen 
%% interval for $\Dz\Dzb$~pairs includes 
%% the~region populated by 
%% the~\mbox{$\decay{\chicone(3872)}
%% {\Dz\Dzb\piz/\g}$}~decays; 
%% however, this contribution is small, 
%% see Fig.~\ref{fig:D0D0OS}.}
with $m_{\Dz\Dzb}<3.87\gevcc$
and 
low\nobreakdash-mass $\Dz\Dz$~pairs
with mass $3.75<m_{\Dz\Dz}<3.87\gevcc$.
The~former is dominated by 
\mbox{$\decay{\proton\proton}
{\cquark\cquarkbar\PX}$}~production, 
while the~latter is presumably dominated by 
the~double parton scattering
process~\cite{LHCb-PAPER-2012-003,DPSBOOK}.
The~chosen 
interval for $\Dz\Dzb$~pairs includes 
the~region populated by 
the~\mbox{$\decay{\chicone(3872)}
{\Dz\Dzb\piz/\g}$}~decays; 
however, this contribution is small, 
see Fig.~\ref{fig:D0D0OS}.
The~$\chicone(3872)$~production cross\nobreakdash-section 
is suppressed with respect to the~conventional 
charmonium state $\psitwos$ at large track 
multiplicities~\cite{LHCb-PAPER-2020-023}.
It~is noteworthy that the~track multiplicity distribution 
for the~\Tcc state differs from that of 
the~low\nobreakdash-mass $\Dz\Dzb$~pairs,
in particular 
no suppression 
at large multiplicity 
is observed.
A~$p$\nobreakdash-value for 
the~consistency of 
the~track multiplicity distributions
for \Tcc~production and low\nobreakdash-mass $\Dz\Dzb$~pairs 
is found 
%%% to be~0.70\%.  %% BW0!
to be~0.1\%.      %% BWU! 
It~is~interesting to note that 
the~multiplicity 
distribution for \Tcc~production 
and the~one for $\Dz\Dz$-pairs with $3.75<m_{\Dz\Dz}<3.87$ 
are consistent with 
a~corresponding $p$\nobreakdash-value 
%% of 27.7\%.  %% BW0! 
of 12\%.     %% BWU!
The~similarity between \Tcc~production, 
which is inherently a~single parton 
scattering process,  
and the~distribution 
for process dominated by 
a~double parton scattering 
is surprising.

The~transverse momentum spectrum for the~\Tcc~state
is compared with those for the~low\nobreakdash-mass 
$\Dz\Dzb$ and $\Dz\Dz$~pairs
%% in Fig.~\ref{fig:DATA_PT_4}. 
in Fig.~\ref{fig:DATA_nVELO_6}(right). 
The~$p$\nobreakdash-values for the~consistency of the~$\pt$~spectra 
for the~\Tcc~state and low-mass $\Dz\Dzb$~pairs 
%% is 5\%,  %% BW0 
are 1.4\%, %% BWU 
and 0.02\% for low-mass $\Dz\Dz$~pairs.
%% is  0.6\%. %% BW0! 
%is  0.02\%. %% BWU! 
More data are needed for further conclusions.

The background\nobreakdash-subtracted $\Dz\Dz$~mass distribution 
in a wider mass range is shown in Fig.~\ref{fig:D0D0OS} together 
with a similar distribution for $\Dz\Dzb$~pairs. 
In~the~\Dz\Dzb~mass spectrum 
the~near\nobreakdash-threshold  enhancement 
is due to  \mbox{$\decay{\chicone(3872)}{\Dz\Dzb\piz}$}
and  \mbox{$\decay{\chicone(3872)}{\Dz\Dzb\g}$}~decays
via intermediate $\Dstarz$~mesons~\cite{LHCb-PAPER-2019-005}.
This~structure is significantly wider 
than the~structure in the~\Dz\Dz~mass spectrum 
from \mbox{$\decay{\Tcc}{\Dz\Dz\pip}$}~decays
primarily due to the~larger natural width 
and smaller binding energy for 
the~$\chicone(3872)$~state~\cite{LHCb-PAPER-2020-008,
LHCb-PAPER-2020-009}.
With more data, and
with a~better understanding of 
the~dynamics of
\mbox{$\decay{\chicone(3872)}
{\Dz\Dzb\piz/\g}$}~decays, 
and therefore of the~corresponding 
shape in  the~\Dz\Dzb~mass spectrum, 
it will be possible to estimate 
the~relative production rates 
for the~\Tcc and $\chicone(3872)$~states.
Background\nobreakdash-subtracted
$\Dz\Dz\pip$ and $\Dz\Dp$~mass 
distributions 
together with those for $\Dzb\Dz\pip$
and $\Dz\Dm$ are shown 
in 
%% Extended Data 
Supplementary~Figs.~\ref{fig:D0D0piOS} and~\ref{fig:D0DpOS}. 
%%Information
%% Figs.~3 and~4.
%%

%% \section*{Summary and conclusion}
\section*{Discussion}

The~exotic narrow tetraquark state 
\Tcc~observed in Ref.~\cite{LHCb-PAPER-2021-031}~is studied 
using a dataset
corresponding to an integrated luminosity of $9\invfb$,
collected by the~\lhcb experiment 
in $\proton\proton$~collisions
at centre-of-mass energies of 7, 8 and 13\tev.
%during the~2011--2018 period. 
%%
The~observed~$\Dz\pip$~mass distribution 
%from the~signal \mbox{$\decay{\Tcc}{\Dz\Dz\pip}$}~decays 
indicates that 
the~$\decay{\Tcc}{\Dz\Dz\pip}$~decay proceeds via an~intermediate 
off\nobreakdash-shell $\Dstarp$~meson.
Together with the~proximity of the~state 
to the~\Dstar\Dz~mass threshold 
this favours the~spin\nobreakdash-parity 
quantum numbers $\mathrm{J^P}$ to be~$1^+$. 
Narrow  near\nobreakdash-threshold structures are observed
in the~$\Dz\Dz$ and $\Dz\Dp$~mass spectra 
 with high significance.
These are found to be consistent 
with originating from off\nobreakdash-shell 
\mbox{$\decay{\Tcc}{\Dstar\D}$}~decays 
followed by the~\mbox{$\decay{\Dstar}{\D\Ppi}$} and 
\mbox{$\decay{\Dstar}{\D\g}$}~decays.
No signal is observed in the~$\Dp\Dz\pip$~mass spectrum,
and no structure is observed in the~$\Dp\Dp$~mass spectrum. 
These non-observations 
provide a~strong argument in favour 
of the~isoscalar  nature for the~observed state, 
supporting its interpretation as  
the~isoscalar $\mathrm{J^P}=1^+$ 
$\cquark\cquark
\uquarkbar\dquarkbar$\nobreakdash-tetraquark ground state.  
%%
% The model of the isoscalar resonance coupled to \Dstar\D
%% {\color{red}
A~dedicated unitarised three\nobreakdash-body
Breit\nobreakdash--Wigner amplitude
is built on the assumption of strong isocalar coupling of 
the~axial\nobreakdash-vector \Tcc state to 
the~\Dstar\D~channel.
This assumption is supported by 
the~data, however, alternative 
models are not excluded by 
the~distributions studied 
in this analysis. 
Probing alternative models 
and the~validity of 
the~underlying 
assumptions of this analysis will be a~subject 
for future studies.
%% }

%% {\color{red}
Using the developed amplitude model,
%% }
%% 
the~mass of the~\Tcc~state, 
relative to the~$\Dstarp\Dz$~mass
threshold, is determined to be
\begin{equation}
    \updelta m_{\mathrm{U}} = -359 \pm 40 ^{\,+\,9}_{\,-\,6} \kevcc\,,
\end{equation}
where the~first uncertainty is statistic and 
the~second systematic.
The~lower limit on the~absolute value of 
the~coupling constant of the~$\Tcc$~state
to the~$\Dstar\D$~system is 
\begin{equation} 
     \left| g \right| > 
   5.1\,(4.3)\gev~\mathrm{at}~90\,(95)\,\%\ \mathrm{CL}\,.
\end{equation}
%% With assumption of molecular nature of 
%% the~\Tcc~state
%% {\color{red} {
Using the~same model,
the estimates for the~scattering length $a$,
effective range $r$, and the~compositeness, 
$Z$ are obtained 
from the~low\nobreakdash-energy limit of the~amplitude
to be
%% }}
%% From the~low\nobreakdash-energy limit of the~amplitude,  
%% the~estimates for the~scattering length $a$,
%% effective range $r$, and the~compositness, 
%% $Z$, are obtained to be
%%
%%\begingroup
%%\allowdisplaybreaks
\begin{eqnarray}
    a & = & \Bigl[ -\left(7.16 \pm 0.51\right) 
    + i\left(1.85 \pm 0.28 \right) \Bigr] \fm\,,\\
    -r            &  < & 
    11.9\,(16.9)\fm~\text{at}~90\,(95)\%\,\mathrm{CL}     \,,  \\
    Z            &  < &  
    0.52\,(0.58)~\text{at}~90\,(95)\%\,\mathrm{CL}  \,.
\end{eqnarray}%%\endgroup
%%
%% \noindent
The~characteristic size calculated from 
the~binding energy is $R_{\Delta E} = 7.49 \pm 0.42\fm$.
This~value is consistent with the~estimation from 
the~scattering length, $R_a = 7.16 \pm 0.51\fm$.
Both $R_{\Delta E}$  and $R_{a}$ correspond to
a~spatial extension 
significantly exceeding 
the~typical scale for  heavy\nobreakdash-flavour hadrons.
%%
% 
%% {\color{red}
%% Given the model assumptions
% 
Within this model
the~resonance
%% }
pole is found to be located 
on the~second Riemann sheet with respect to 
the~$\Dz\Dz\pip$ threshold,
at
$\hat{s}=m_{\mathrm{pole}}-\tfrac{i}{2}
\Gamma_{\mathrm{pole}}$, 
where 
%%\ifthenelse{\boolean{wordcount}}{}{
%%\begingroup
%%\allowdisplaybreaks
\begin{eqnarray}
    \updelta m_{\mathrm{pole}} 
     & = & 
 %% -364 \pm 40  ^{\,+\,5\phantom{0}}_{\,-\,0} \kevcc \,, \\ 
    -360 \pm 40  ^{\,+\,4\phantom{0}}_{\,-\,0} \kevcc \,, \\ 
    \Gamma_{\mathrm{pole}}              & = &  
     \phantom{-0}48 \pm \phantom{0}2 ^{\,+\,0}_{\,-\,14}\kev \,,
\end{eqnarray} 
%%\endgroup
%%
where the~first uncertainty  
accounts for statistical
and systematic uncertainties for
the~$\updelta m_{\mathrm{U}}$~parameters, 
and the~second is due to 
the~unknown value of 
the~$\left|g\right|$~parameter.
The~pole position, scattering 
length, effective range and compositeness 
%% reveals 
form a~complete set 
of observables
related to the~\mbox{$\decay{\Tcc}{\Dz\Dz\pip}$}
reaction 
amplitude,
which 
%% is vital 
are crucial 
for inferring
the~nature of the~\Tcc~tetraquark.
% 
% Supported by
% 

Unlike in the~prompt production 
of the~$\chicone(3872)$~state,
no suppression of the~\Tcc~production
at high track multiplicities
is observed
relative to 
the~low\nobreakdash-mass $\Dz\Dzb$~pairs. 
The~observed similarity 
with the~multiplicity distribution for 
the~low\nobreakdash-mass $\Dz\Dz$~production process, 
that is presumably 
double-parton-scattering dominated, 
is unexpected.
In~the~future with a~larger dataset and
including other decay modes, 
\eg\mbox{$\decay{\Dz}
{\Km\pip\pip\pim}$}, detailed studies 
of  the~properties of this~new state and its 
production mechanisms
could  be possible.

%\begin{comment}
In conclusion, the \Tcc tetraquark  
observed in $\Dz\Dz\pip$ decays is studied in detail,
using a~unitarised model that
accounts for the~relevant thresholds by
taking into account the~$\Dz\Dz\pip$ and 
$\Dz\Dp\piz(\g)$~decay channels with 
intermediate $\Dstar$~resonances.
This model is found to give an~excellent 
description of the~$\Dz\pip$ mass distribution 
in the~\mbox{$\decay{\Tcc}{\Dz\Dz\pip}$} decay and 
of the~threshold enhancements observed in 
the~$\Dz\Dz$ and $\Dz\Dp$ spectra. 
Together with the~absence of a~signal 
in the~$\Dz\Dp$ and $\Dp\Dz\pip$ mass distributions 
this provides a~strong argument for interpreting the~observed state as
the~isoscalar \Tcc tetraquark with spin-parity
$\mathrm{J^P}=1^+$.
% 
% Currently an alternative 
%% interpretations are not ruled out by the data.
% 
The~precise \Tcc mass measurement will 
rule out or improve on a~considerable range 
of theoretical models on heavy quark systems.
%and 
%helps improving others.
The~determined pole position and physical quantities 
derived from low\nobreakdash-energy scattering parameters 
reveal important information about the~nature of 
the~\Tcc tetraquark. 
In~addition, the~counter\nobreakdash-intuitive dependence 
of the~production rate on track multiplicity
%% and transverse momentum 
will pose a~challenge for theoretical explanations.

%\end{comment}

%% \clearpage

%% \renewcommand{\thetable}{M\arabic{table}}
%% \renewcommand{\thefigure}{M\arabic{figure}}

%% \renewcommand{\theequation}{M\arabic{equation}}
%% \setcounter{figure}{0}
%% \setcounter{table}{0}
%% \setcounter{equation}{0}

\section*{Methods}

%% \cite{FAKE} 

\subsection*{Experimental setup}\label{sec:detector}

The \lhcb detector~\cite{LHCb-DP-2008-001,LHCb-DP-2014-002} is a single-arm forward
spectrometer covering the~pseudorapidity range $2<\eta <5$,
designed for the study of particles containing \bquark or \cquark
quarks. The detector includes a high-precision tracking system
consisting of a~silicon\nobreakdash-strip vertex detector surrounding 
the~$\proton\proton$~interaction region,
%%~\cite{LHCb-DP-2014-001}\verb!*!, 
a~large-area silicon-strip detector located
upstream of a~dipole magnet with a~bending power of about
$4{\mathrm{\,Tm}}$, and three stations of silicon\nobreakdash-strip detectors and straw
drift tubes
placed downstream of the magnet.
The~tracking system provides a measurement of the momentum, \ptot, 
of charged particles with
a~relative uncertainty that varies from 0.5\% at low momentum to 1.0\% at 200\gevc.
The minimum distance of a track to 
a~primary $\proton\proton$~collision vertex\,(PV), 
the~impact parameter\,(IP), 
is measured with a resolution of $(15+29/\pt)\mum$,
where \pt is the component of the momentum transverse to the beam, in\,\gevc.
Different types of charged hadrons are distinguished using information
from two ring-imaging Cherenkov detectors~\cite{LHCb-DP-2012-003}. 
Photons, electrons and hadrons are identified by 
a~calorimeter system consisting of
scintillating-pad and preshower detectors, an electromagnetic
% calorimeter
and a hadronic calorimeter. 
Muons are identified by a~system composed of 
alternating layers of iron and multiwire
proportional chambers.
%% ~\cite{LHCb-DP-2012-002}\verb!*!.
%% 
The~online event selection is performed by 
a~trigger, 
%% ~\cite{LHCb-DP-2012-004}, 
which consists of a~hardware stage, based on information from 
the~calorimeter and muon systems, 
followed by a~software stage, which applies a full event
reconstruction.

\subsection*{Simulation}
Simulation is required to model 
the~effects of the~detector acceptance, 
resolution, 
and the~efficiency of 
the~imposed selection requirements.
In~the~simulation, $\proton\proton$~collisions 
are generated using
  \pythia~\cite{Sjostrand:2007gs}   %% ,*Sjostrand:2006za} 
  with a~specific 
  \lhcb configuration~\cite{LHCb-PROC-2010-056}.
  Decays of unstable particles
  are described by \evtgen~\cite{Lange:2001uf}, 
  in which final-state
  radiation is generated 
  using \photos~\cite{davidson2015photos}.
  The~interaction of the~generated particles 
  with the~detector, and its response,
  are implemented using the~\geant
  toolkit~\cite{Allison:2006ve, 
  *Agostinelli:2002hh} as described in
  Ref.~\cite{LHCb-PROC-2011-006}.

\subsection*{Event selection}\label{sec:selection}

%The~\Dz, \Dp and \Dstarp~mesons are reconstructed in decay channels \mbox{$\decay{\Dz}{\Km\pip}$}, \mbox{$\decay{\Dp}{\Km\pip\pip}$} and  \mbox{$\decay{\Dstarp}{\Dz\pip}$} correspondingly.
%% The~selection  of~\Dp~and~\Dz~mesons largely follows 
%% the~one used in  
%% Refs.~\cite{LHCb-PAPER-2012-003,LHCb-PAPER-2013-062,LHCb-PAPER-2015-046,LHCb-PAPER-2019-005}
%% and is shortly summarized in this  section.
%%
The~$\Dz\Dz$, $\Dz\Dp$ and $\Dz\Dz\pip$ final states 
are reconstructed using the
\mbox{$\decay{\Dz}{\Km\pip}$} and 
\mbox{$\decay{\Dp}{\Km\pip\pip}$}~decay channels. 
The~selection criteria are similar to those used in 
Refs.\mbox{\cite{LHCb-PAPER-2012-003,
LHCb-PAPER-2013-062,
LHCb-PAPER-2015-046,
LHCb-PAPER-2019-005}}. 
Kaons and pions are selected 
from well\nobreakdash-reconstructed tracks
within the~acceptance of 
the~spectrometer that are identified using information from
the~ring-imaging Cherenkov detectors.
The~kaon and pion candidates that have transverse momenta 
larger than $250\mevc$ 
and are inconsistent with being produced
at a~$\proton\proton$~interaction 
vertex are combined together to form $\Dz$ and $\Dp$ candidates, 
referred to as \D~hereafter. 
The~resulting $\D$ candidates are required to have 
good vertex quality,
mass within $\pm 65\mevcc$ and $\pm 50\mevcc$ 
of the~known \Dz and \Dp masses~\cite{PDG2021}, respectively, 
transverse momentum larger than $1\gevc$, 
decay time larger than $100\mum/c$ and 
a~momentum  direction that is consistent 
with the~vector from the~primary to~secondary vertex. 
Selected \Dz and \Dp~candidates 
consistent with originating 
from a~common primary vertex 
are combined to 
form $\Dz\Dz$ and $\Dz\Dp$ candidates. 
The~resulting $\Dz\Dz$ candidates are
combined with a~pion to form $\Dz\Dz\pip$~candidates.
At~least one  of the two $\Dz\pip$ 
combinations is required to have good vertex quality and mass 
not exceeding the known $\Dstarp$ mass by more than $155\mevcc$.
For~each $\Dz\Dz$, $\Dz\Dp$ and $\Dz\Dz\pip$ candidate 
a~kinematic fit~\cite{Hulsbergen:2005pu} is performed.
This~fit 
constrains the mass of the $\D$ candidates to their known values and 
requires both $\D$ mesons, 
and a~pion in the case of $\Dz\Dz\pip$,
to originate from the~same primary vertex.
A~requirement is applied to the~quality
of this fit to further suppress combinatorial background 
and reduce 
background from $\D$~candidates produced 
in two independent $\proton\proton$~interactions or 
in the~decays of beauty hadrons~\cite{LHCb-PAPER-2012-003}.
To~suppress background from kaon and pion candidates reconstructed from 
a~common track, all track pairs of the~same charge are required to have 
an opening angle inconsistent with  zero 
and the~mass of the~combination must be inconsistent with the~sum 
of the~masses of the~two constituents.
For cross-checks additional final states 
$\Dp\Dp$, $\Dp\Dz\pip$,
$\Dz\Dzb$, $\Dz\Dm$ and $\Dzb\Dz\pip$ are 
reconstructed, selected and treated in the~same way.

\subsection*{Non\nobreakdash-\D background subtraction}
Two-dimensional distributions 
of the~mass of one \D~candidate versus
the~mass of the~other \D~candidate from selected $\Dz\Dz\pip$, 
$\Dz\Dz$ and $\Dz\Dp$~combinations
are shown in 
%% Extended Data 
Supplementary~Fig.~\ref{fig:DATA_2D}.
%% Information
%% Fig.~1.
These distributions illustrate the relatively
small combinatorial background levels due to fake \D~candidates. 
This background is subtracted 
using the~\sPlot technique~\cite{Pivk:2004ty}, which is based on 
an~extended unbinned  
maximum\nobreakdash-likelihood  fit to 
these~two\nobreakdash-dimensional distributions with the function described in Ref.~\cite{LHCb-PAPER-2012-003}.
%%
\begin{comment}
with  a~four\nobreakdash-component 
function from Ref.~\cite{LHCb-PAPER-2012-003}.
\begin{eqnarray*}
  f(x, y )  & \propto &  N_{\D\D} {S}_x(x) {S}_y(y)  \\
  &   +     &  N_{\D\&\mathrm{b}} {S}_x(x) {B}_y(y)            \\
  &   +     &  N_{\mathrm{b}\&\D} {B}_x(x) {S}_y(y)            \\
  &   +     &  N_{\mathrm{b\&b}}  {B}_{2D}(x,y),
\end{eqnarray*}
where $N_{{\D\&\D}}$ is  
the~number of genuine $\D_{1}\D_{2}$ pairs,
$N_{\mathrm{\D\&b(b\&\D)}}$ is 
the~number of combinations of 
$\D_{1(2)}$~mesons with combinatorial 
background, $N_{\mathrm{b\&b}}$ is 
the~number of pure background pairs
and $x(y)$ is the combined mass of $\D$ daughters: 
$m_{\Km\pip}$ for $\Dz$ and $m_{\Km\pip\pip}$ for $\Dp$.
%%
\end{comment} 
%%
This~function consists of four components:
\begin{itemize}
    \item a~component corresponding 
    to genuine $\D_1\D_2$~pairs and described 
    as a~product of two~signal functions, 
    each parameterised with 
    a~modified Novosibirsk function~\cite{Bukin};
    \item two components corresponding
    to combinations of 
    one of the~\D~mesons
    %$\D_{1,2}$~mesons
    with combinatorial background,  
    described as a~product
    of the~signal function and 
    a background function, which is
    parameterised with a~product 
    of an~exponential function 
    and a~positive first\nobreakdash-order 
    polynomial; 
    \item a~component corresponding to pure background 
    pairs and described by a~product of exponential functions
    and a~positive two\nobreakdash-dimensional 
    non\nobreakdash-factorisable second\nobreakdash-order 
    polynomial function.
\end{itemize}
Based on the~results of the~fit,  
each candidate is assigned 
a~positive weight for being signal\nobreakdash-like or
a~negative weight for being background\nobreakdash-like, 
with the~masses of the~two \Dz~candidates 
as discriminating variables. 
The~$\Dz\Dz\pip$ mass distributions for each of 
the~subtracted background components 
are presented in 
%% Extended Data 
Supplementary~Fig.~\ref{fig:SIDEBANDS}, 
%%Information
%% Fig.~2.
where fit results 
with background\nobreakdash-only 
functions $B^{\prime}_{10}$, 
defined in Eq.~\eqref{eq:background_ext}
are overlaid.

\subsection*{Resolution model for the~$\Dz\Dz\pip$~mass}

In the vicinity of the~$\Dstarp\Dz$~mass threshold 
the~resolution function $\mathfrak{R}$ for 
the~$\Dz\Dz\pip$~mass 
is parametrised with the~sum of two Gaussian 
functions with a~common mean.
The~widths of the~Gaussian functions 
are 
$\upsigma_1=1.05\times 263\kevcc$ and $\upsigma_2=2.413\times\upsigma_1$ 
for the~narrow and wide components, respectively, 
and the~fraction of the~narrow Gaussian is $\upalpha=0.778$.
The~parameters $\upalpha$ and  $\upsigma_{1,2}$  
are taken from simulation,  
and $\upsigma_{1,2}$ are 
corrected with a~factor of~1.05  
that accounts for a~small difference 
between simulation and data for the~mass 
resolution~\cite{LHCb-PAPER-2020-008,
LHCb-PAPER-2020-009,
LHCb-PAPER-2020-035}.
The~root mean square of the~resolution 
function is around 400\kevcc. 

\subsection*{Matrix elements 
for $\decay{\Tcc}{\D\D\Ppi/\g}$~decays }\label{sec:Xdecays}

Assuming isospin symmetry, the~isoscalar vector state \Tcc that decays into the
$\Dstar\D$~final state can be expressed as 
\ifthenelse{\boolean{wordcount}}{}{
\begin{equation}
    \left| \Tcc \right\rangle = 
    \dfrac{1}{\sqrt{2}} \Bigl(
    \left| \Dstarp \Dz \right\rangle - 
    \left| \Dstarz\Dp \right\rangle \Bigr)\,.
\end{equation}
}
Therefore, 
the~S\nobreakdash-wave 
amplitudes
%% \footnote{The S-wave\,(corresponding 
%% to orbital angular momentum equal to zero)
%% approximation
%% is valid for a~near\nobreakdash-threshold peak. 
%% For \Tcc~masses significantly above 
%% the~\Dstar\D~threshold, 
%% higher\nobreakdash-order waves 
%% also need to be considered. }
for the~$\decay{\Tcc}{\Dstarp\Dz}$
and $\decay{\Tcc}{\Dstarz\Dp}$~decays
have different signs 
%% \begin{subequations} \label{eq:xdd:swave}
\begin{eqnarray}
\mathcal{A}^{\mathrm{S-wave}}_{\decay{\Tcc}{\Dstarp\Dz}} & = &  
+ \dfrac{g}{\sqrt{2}}  \epsilon_{\Tcc\mu}  \epsilon^{\ast\mu}_{\Dstar} \,, \\
\mathcal{A}^{\mathrm{S-wave}}_{\decay{\Tcc}{\Dstarz\Dp}} & = & 
-  \dfrac{g}{\sqrt{2}}  \epsilon_{\Tcc\mu}  \epsilon^{\ast\mu}_{\Dstar} \,,
\end{eqnarray}
%% \end{subequations}
where $g$~is a~coupling constant,
$\epsilon_{\Tcc}$~is the~polarisation vector of the~\Tcc~particle
and $\epsilon_{\Dstar}$~is the~polarisation vector of \Dstar meson,
and the~upper and 
lower Greek indices imply the~summation in the Einstein notation.
The~S\nobreakdash-wave\,(corresponding 
to orbital angular momentum equal to zero)
approximation
is valid for a~near\nobreakdash-threshold peak. 
For \Tcc~masses significantly above 
the~\Dstar\D~threshold, 
higher\nobreakdash-order waves 
also need to be considered. 
The~amplitudes for  
the~$\decay{\Dstar}{\D\pion}$~decays are written as
%% \begin{subequations}\label{eq:A.Dstar.decays}
\begingroup
\allowdisplaybreaks
\begin{eqnarray}
\mathcal{A}_{\decay{\Dstarp}{\Dz\pip}} 
& = &  \phantom{+}  
f                 \epsilon^{\alpha}_{\Dstar}p_{\D\alpha} 
\label{eq:DA_one}
\\
\mathcal{A}_{\decay{\Dstarp}{\Dp\piz}} 
& = &   - \frac{f}{\sqrt{2}} \epsilon^{\alpha}_{\Dstar}p_{\D\alpha} \\
\mathcal{A}_{\decay{\Dstarz}{\Dz\piz}} 
& = &   + \frac{f}{\sqrt{2}}  \epsilon^{\alpha}_{\Dstar}p_{\D\alpha}\,, 
\end{eqnarray}
\endgroup
where $f$~denotes a coupling constant, and
$p_{\D}$~stands for the~momentum of the \D~meson.
The~amplitude for 
the~$\decay{\Dstar}{\D\gamma}$~decays is 
\begin{equation} \label{eq:radiative} 
  \mathcal{A}_{\decay{\Dstar}{\g\D}} = i \upmu h 
  \epsilon_{\alpha\beta\eta\xi}\epsilon_{\Dstar}^{\alpha} p_{\Dstar}^{\beta}
  \epsilon^{\ast\eta}_{\g}p_{\g}^{\xi}\,,
\end{equation}
%% \end{subequations}
where $h$ denotes a~coupling constant,
$\upmu$~stands for the~magnetic moment 
for \mbox{$\decay{\Dstar}{\D\g}$}~transitions,
$p_{\Dstar}$ and $p_{\g}$
are the the~\Dstar-meson and photon momenta, 
respectively, and 
$\epsilon_{\g}$~is the~polarisation 
vector of the~photon.  
The~three amplitudes
for $\decay{\Tcc}{\pion\D\D}$
and $\decay{\Tcc}{\g\D\D}$~decays
%% \footnote{{\color{red}{
%% The~decays of the~\Tcc~state into the $\Dp\Dp\pim$~final  state 
%% via off\nobreakdash-shell \mbox{$\decay{\Dstarz}{\Dp\pim}$}~decays
%% are highly suppressed and are not considered here.   
%% }}}
are
%%
%%
%% \begin{subequations}\label{eq:bw3_C}
\begingroup
\allowdisplaybreaks
\begin{eqnarray}
\mathcal{A}_{{\pip\Dz\Dz}} & =  &  
\phantom{+} \frac{fg}{\sqrt{2}}\epsilon_{\Tcc\nu}\left[ 
\mathfrak{F}_{+}(s_{12}) \times \left( -p_2^{\nu} + 
\dfrac{(p_2p_{12}) p_{12}^{\nu}  }{s_{12}} \right)  +  (p_2\leftrightarrow p_3) \right]\,,      \label{eq:pipdzdz}
\\
\mathcal{A}_{{\piz\Dp\Dz}} & =  &  
- \frac{fg}{2}\epsilon_{\Tcc\nu}\left[ 
\mathfrak{F}_{+}(s_{12}) \times \left( -p_2^{\nu} + 
\dfrac{ (p_2p_{12}) p_{12}^{\nu}  }{s_{12}} \right)  +  \left( 
\begin{array}{c} p_2\leftrightarrow p_3 \\  
\mathfrak{F}_{+} \leftrightarrow \mathfrak{F}_{0} \end{array}\right)  \right]\,,      \label{eq:pizdzdp}\\
\mathcal{A}_{{\g\Dp\Dz}} & =  &  
i \frac{hg}{\sqrt{2}}
\epsilon_{\alpha\beta\eta\xi}
\epsilon_{\Tcc}^{\beta}
\epsilon_{\g}^{\eta}p_{\gamma}^{\xi}
\left[ 
\upmu_{+}\mathfrak{F}_{+}(s_{12}) p_{12}^{\alpha} -  
\upmu_{0}\mathfrak{F}_{0}(s_{13}) p_{13}^{\alpha} 
\right]\,,  
\end{eqnarray}
\endgroup
%% \end{subequations}
%% 
where $s_{ij}=p^2_{ij}=\left(p_i+p_j\right)^2$
%, $\upmu_{+/0}$ 
and the~$\mathfrak{F}$ 
functions that
denote 
%magnetic moments and 
the~Breit\nobreakdash--Wigner 
amplitude for the~$\Dstar$~mesons are
\begin{equation} \label{eq:amp_bw}
    \mathfrak{F}(s) =
    \dfrac{1}{ m^2_{\Dstar} - s - im_{\Dstar}\Gamma_{\Dstar} }\,.
\end{equation}
A~small~possible distortion of   the~Breit\nobreakdash--Wigner shape 
of the~\Dstar~meson due to three\nobreakdash-body 
final\nobreakdash-state interactions 
is neglected in the~model.
The~impact  of 
the~energy\nobreakdash-dependence 
of  the~\Dstar~meson self\nobreakdash-energy
is found to be insignificant.
%% 
%% \footnote{{\color{red}{
The~decays of the~\Tcc~state into the $\Dp\Dp\pim$~final  state 
via off\nobreakdash-shell \mbox{$\decay{\Dstarz}{\Dp\pim}$}~decays
are highly suppressed and are not considered here.   
%% }}}
%%
The~last terms in Eqs.~\eqref{eq:pipdzdz}  
and~\eqref{eq:pizdzdp} imply the~same amplitudes 
with swapped momenta. 

The~\Tcc~state is assumed to be produced unpolarized,
therefore  the~squared absolute value of 
the~decay amplitudes with pions 
in the~final state, 
averaged over 
the~initial spin-state 
are 
%% \begin{subequations} \label{eq:msq_pions}
\begingroup
\allowdisplaybreaks
\begin{eqnarray}
\left|\mathfrak{M}_{{\pip\Dz\Dz}}\right|^2 & = &   
\frac{1}{3} \frac{f^2g^2}{2! \cdot 2 } \left[ 
\left| \mathfrak{F}_{+}(s_{12})\right|^2 A    
 + \left| \mathfrak{F}_{+}(s_{13}) \right|^2 B   
 +2\Re\left\{ \mathfrak{F}_{+}(s_{12})\mathfrak{F}^{\ast}_{+}(s_{13}) \right\}  C 
\right] \,, \label{eq:ampsqpip}\\
\left|\mathfrak{M}_{{\piz\Dp\Dz}}\right|^2 & = &   
\frac{1}{3} \frac{f^2g^2}{4} \left[ 
\left| \mathfrak{F}_{+}(s_{12})\right|^2 A    
 + \left| \mathfrak{F}_{0}(s_{13}) \right|^2 B   
 +2\Re\left\{ \mathfrak{F}_{+}(s_{12})\mathfrak{F}^{\ast}_{0}(s_{13}) \right\} C 
\right] \,, \label{eq:ampsqpiz}
\end{eqnarray}
\endgroup
%% \end{subequations}
where 
%%
%% \begin{subequations}\label{eq:bw3_c7}
\begingroup
\allowdisplaybreaks
\begin{eqnarray}
A & = &  \dfrac{\uplambda( s_{12}, m_1^2, m_2^2)}{4s_{12}} +  
\dfrac{1}{s} \left(  \dfrac{  (pp_{12})(p_2p_{12}) }  {s_{12}}  
-  pp_2 \right)^2  \,,  \label{eq:aaa_A}\\
B & = &  \dfrac{\uplambda( s_{13}, m_1^2, m_3^2)}{4s_{13}} +  
\dfrac{1}{s} \left(  \dfrac{  (pp_{13})(p_3p_{13}) }  {s_{13}}  
-  pp_3 \right)^2  \,,  \label{eq:aaa_B} \\
C & = & D + E \nonumber \\ 
D & = &  
\dfrac{ (p_3p_{12}) (p_2p_{12}) }{s_{12}} + 
\dfrac{ (p_2p_{13}) (p_3p_{13}) }{s_{13}} - 
\dfrac{ (p_{12}p_{13}) (p_2p_{12}) (p_3p_{13})  }{s_{12}s_{13}} - p_2p_3 \,,\\
E & = & 
\dfrac{ (pp_{12})(pp_{13})(p_2p_{12})(p_3p_{13})}{ss_{12}s_{13}}  +
\dfrac{ (pp_2)(pp_3)}{s}  \nonumber \\ 
& - &  
\dfrac{(pp_{12})(p_2p_{12})(pp_3)  }{ss_{12}} - 
\dfrac{(pp_{13})(p_3p_{13})(pp_2)  }{ss_{13}} \,,
\end{eqnarray} 
\endgroup
%% \end{subequations}
%% 
and $\uplambda\left(x,y,z\right)$ stands for the
K\"all\'en function~\cite{Kallen}. 
The~additional factor of $2!$ in 
the~denominator of Eq.~\eqref{eq:ampsqpip} is due to the
presence of two identical particles\,($\Dz$) in the~final state. 
The~squared absolute values 
of the~decay amplitude with a~photon in the~final state, 
averaged over the~initial spin state is
%%
%% \begin{subequations}\label{eq:msq_gamma}
\begingroup
\allowdisplaybreaks
\begin{eqnarray}
\left|\mathfrak{M}_{\decay{\Tcc}{\g\Dp\Dz}}\right|^2 & = &   
\frac{1}{3} \left| gh\right|^2 
 \left| \upmu_{+}\mathfrak{F}_{+}(s_{12}) (p_1p_2) 
 - \upmu_0\mathfrak{F}_0(s_{13})(p_1p_3)\right|^2  \\ %% \nonumber \\ 
 & + & 
 \frac{1}{3} \left| gh\right|^2 
 \left| \upmu_+\mathfrak{F}_{+}(s_{12}) + 
 \upmu_0\mathfrak{F}_0(s_{13}) \right|^2  G  \,,  
 \label{eq:msq_gamma_one} \\
 G & = &  \dfrac{1}{2s}\left[ 
2(p_1p_2) (p_1p_3)(p_2p_3) - m_2^2 (p_1p_3)^2 - m_3^2(p_1p_2)^2  \right] \,.
\label{eq:msq_gamma_two}
\end{eqnarray}
\endgroup
%% \end{subequations}
%%
%% 
The~coupling  constants $f$ and $h$ 
for the~\mbox{$\decay{\Dstar}{\D\Ppi}$}
and \mbox{$\decay{\Dstar}{\D\g}$}~decays 
are calculated using 
%% Eqs.~\eqref{eq:A.Dstar.decays} 
Eqs.~\eqref{eq:DA_one}--\eqref{eq:radiative},
from the~known branching fractions 
of the~\mbox{$\decay{\Dstar}{\D\Ppi}$}
and \mbox{$\decay{\Dstar}{\D\g}$}~decays~\cite{PDG2021},
the~measured natural width of 
the~\Dstarp~meson~\cite{Lees:2013zna,PDG2021}
and the~derived value for the~natural width for 
the~\Dstarz~meson~\cite{Braaten:2007dw,
Guo:2019qcn,
Braaten:2020nwp}.
The~magnetic moment $\upmu_+$ is taken to be 1
and the~ratio of magnetic moments $\upmu_0/\upmu_+$ 
is calculated according to 
Refs.~\cite{Rosner:1980bd,Gasiorowicz:1981jz,Rosner_2013}.

\subsection*{Unitarised Breit--Wigner shape}\label{sec:BWU}

A~unitarised three\nobreakdash-body Breit\nobreakdash--Wigner 
function is defined as 
%%
%% \begin{subequations}\label{eq:bw_u}
\begingroup
\allowdisplaybreaks
\begin{eqnarray}
\mathfrak{F}^{\mathrm{U}}_{f}\left(s\right) 
& =  &  \varrho_f\left(s\right) \left| \mathcal{A}_{\mathrm{U}} \left(s\right)\right|^2  \,, \label{eq:bwU}
\\ 
\mathcal{A}_{\mathrm{U}}\left(s \right) 
& = & 
\dfrac{1}{ m_{\mathrm{U}}^2 - s  -im_{\mathrm{U}} \hat{\Gamma}(s) }\,, \label{eq:bwA}
\end{eqnarray}
\endgroup
%% \end{subequations}
%% 
where $f \in \left\{ \Dz\Dz\pip, 
\Dz\Dp\piz,  
\Dz\Dp\g \right\}$ 
denotes the~final state.
The~decay matrix element for each channel 
integrated over the~three\nobreakdash-body phase space
is denoted by 
%% $\varrho_f(s)$:
\ifthenelse{\boolean{wordcount}}{}{
\begin{equation} 
    \varrho_f(s) = 
    \dfrac{1}{ ( 2\pi)^5}  \dfrac{\pi^2}{4s} \iint ds_{12} ds_{23} 
    \dfrac{ \left|\mathfrak{M}_f\left(s,s_{12},s_{23}\right)\right|^2 }
    {\left| g \right|^2}  \,, \label{eq:rho_i}
\end{equation}
}
where $\left|\mathfrak{M}_f\right|^2$ is defined by
%% Eqs.\eqref{eq:msq_pions} 
%% and~\eqref{eq:msq_gamma},
Eqs.~\eqref{eq:ampsqpip}--\eqref{eq:msq_gamma_two}
and 
the~unknown coupling constant $g$ is taken 
out of the~expression for $\left|\mathfrak{M}_f\right|^2$.
For~large values of  $s$,  in excess of 
$s^\ast$, such  as 
\mbox{$\sqrt{s^*}-
\left(m_{\Dstar}+m_{\D}\right)
\gg\Gamma_{\Dstar}$},  
the~functions $\varrho_f(s)$ are 
defined as 
%% 
%% \begin{subequations}\label{eq:rho_infty}
\begingroup
\allowdisplaybreaks
\begin{eqnarray}
 \left. \varrho_{\Dz\Dz\pip}(s) \right|_{s>s^*}
  & = & c_1 \Phi_{\Dstarp\Dz} (s)  \label{eq:c_one}\,, \\ 
 \left. \varrho_{\Dz\Dp\piz}(s) \right|_{s>s^*}
  & = & c_2 \Phi_{\Dstarz\Dp} (s)  \,, \\ 
 \left. \varrho_{\Dz\Dp\g}(s) \right|_{s>s^*} 
  & = & c_3 \Phi_{\Dstarz\Dp} (s)  \,, 
\end{eqnarray}
\endgroup
%%\end{subequations}
%%
where $\Phi_{\Dstar\D}(s)$~denotes
the~two\nobreakdash-body 
phase\protect\nobreakdash-space function, 
the~constants 
$c_1$, $c_2$ and $c_3$ are chosen
to ensure the continuity of 
the~functions $\varrho_f(s)$,
%% In~this analysis 
and a~value of 
\mbox{$\sqrt{s^*}=3.9\gevcc$} is used. 
The~functions $\varrho_f(s)$ are shown 
in 
%% Extended Data 
Supplementary~Fig.~\ref{fig:RHOS}.
%%Information
%%Fig.~5.
%% In~this analysis 
%% a~value of $\sqrt{s^*}=3.9\gevcc$ is used.
%%
The~complex\nobreakdash-valued width 
$\hat{\Gamma}(s)$ is defined via
the self-energy function $\Sigma(s)$~\cite{Peskin:1995ev}
\begin{equation} \label{eq:def.Sigma}
   i m_{\mathrm{U}} 
   \hat{\Gamma}(s) \equiv \left| g\right|^2 
   \Sigma(s) \,,
\end{equation}
where $\left| g\right|^2$ is again factored out for convenience.
The~imaginary part of $\Sigma(s)$ for real physical 
values of $s$ is computed through 
the~optical theorem as half of the sum of 
the~decay probability to all available channels~\cite{Gribov:2009zz}:
%% \begin{subequations}\label{eq:Sigma_real}
\begingroup
\allowdisplaybreaks
  \begin{eqnarray} \label{eq:imSigma}
\Im\, \left. \Sigma(s) \right|_{\Im s = 0^+ }  
& = & \dfrac{1}{2} \varrho_{\mathrm{tot}}(s)\,,
    \\ \label{eq:rho_tot}
 \varrho_{\mathrm{tot}}\left(s\right)  
 & \equiv&
 \sum\limits_f \varrho_f \left( s \right) \,.
  \end{eqnarray}
\endgroup
%%\end{subequations}
%%
The~real part of the~self\nobreakdash-energy 
function is computed using 
Kramers\nobreakdash--Kronig dispersion
relations with a~single 
subtraction~\cite{Martin:102663,Eden:1966dnq}, 
%%\begin{subequations}
\begingroup
\allowdisplaybreaks
\begin{eqnarray} \label{eq:reSigma}
    \left.  \Re\,\Sigma(s) \right|_{\Im s = 0^+ } 
    & = & \xi(s) - \xi(m_{\mathrm{U}}^2)\,,
    \\ \label{eq:xi}
    \xi(s) & = & \dfrac{s}{2\pi}\,\mathrm{p.v.}\!\!\!
    \int\limits^{+\infty}_{s^*_{\mathrm{th}}}
    \dfrac{\varrho_{\mathrm{tot}} (s^\prime)} 
    {s^{\prime}\left(s^\prime-s\right)} ds^{\prime}\,,
\end{eqnarray}
\endgroup
%% \end{subequations}
where the~Cauchy principal value (p.v.) integral 
over $\varrho_{\mathrm{tot}}(s)$ is understood as 
\ifthenelse{\boolean{wordcount}}{}{
\begin{equation} \label{eq:sth.inf.limits}
    \mathrm{p.v.}\!\!\!\int\limits^{+\infty}_{s^*_{\mathrm{th}}}
    ds\, \varrho_{\mathrm{tot}}(s) \, ...
    \equiv 
    \sum_f 
    \mathrm{p.v.}\!\!\!
    \int\limits^{+\infty}_{s_f} 
    ds\, \varrho_f(s) \, ... \,,
\end{equation}
}
and $s_f$ denotes the~threshold value for 
the~channel~$f$. 
The~subtraction is needed since the~integral 
$\int\varrho_{\mathrm{tot}}(s) / s\,ds$~diverges.
The~term $\xi(m_{\mathrm{U}}^2)$ in 
Eq.~\eqref{eq:reSigma} corresponds to 
the~choice of subtraction constant such that 
$\Re\,\mathcal{A}_{\mathrm{U}}(m_{\mathrm{U}}^2) =0$. 
The~function $\xi(s)$ is shown 
in 
%% Extended Data
Supplementary~Fig.~\ref{fig:xi}.
%%Information
%%Fig.~6.

% \todo[inline,size=large]{Add diagrams to the equation}
%% 
Alternatively, the~isoscalar amplitude $\mathcal{A}_{\mathrm{U}}$
is constructed using the~$K$-matrix approach~\cite{Aitchison:1972ay}
with two coupled channels, $\Dstarp\Dz$ and $\Dstarz\Dp$.
The relation reads:
\begin{equation}
    \mathcal{A}_{\mathrm{U}} \begin{pmatrix} \phantom{-} g\\ - g  \end{pmatrix}
= [1- K G]^{-1}P \,, 
\end{equation}
where a~production vector $P$ and an~isoscalar potential $K$ 
are defined as
\begin{align}
    P &= \frac{1}{m_{\mathrm{U}}^2-s} \begin{pmatrix} \phantom{-} g\\- g  \end{pmatrix}\,,&
    K(s) &= \dfrac{1}{m^2_{\mathrm{U}}-s}
    \begin{pmatrix}
    \phantom{-} \left| g \right|^2   &           -   \left| g \right|^2 \\   
    - \left| g \right|^2             &  \phantom{-}  \left| g \right|^2  
    \end{pmatrix}\,.
\end{align}
The propagation matrix $G$ describes 
the~$\Dstar\D \to \Dstar\D$ rescattering via 
the~virtual loops 
including the~one\nobreakdash-particle exchange process~\cite{Mikhasenko:2019vhk}
and expressed in a~symbolical way in 
Supplementary~Eq.~\eqref{eq:G}.
%%Eq.~(1).
%%Information
%% 
\begin{comment}
\begin{equation}
      G  = \left[
 \begin{array}{cc}
   \raisebox{-0.35\height}{\includegraphics[width=0.20\textwidth]{Graph1.pdf}}
  + \raisebox{-0.35\height}{\includegraphics[width=0.20\textwidth]{Graph5.pdf}}
 &
   \raisebox{-0.35\height}{\includegraphics[width=0.20\textwidth]{Graph4.pdf}}
  \\
 \raisebox{-0.35\height}{\includegraphics[width=0.20\textwidth]{Graph3.pdf}} 
 &       
 \raisebox{-0.35\height}{\includegraphics[width=0.20\textwidth]{Graph2.pdf}}
 %%
 \end{array}
   \right] \,,
 \end{equation}
\end{comment}
% 
\begin{comment}
\begin{figure}
    \centering
    \includegraphics[width=0.3\textwidth]{Graph1.pdf}\quad
    % \includegraphics[width=0.18\textwidth]{Graph2.pdf}
    \includegraphics[width=0.3\textwidth]{Graph3.pdf}
    % \includegraphics[width=0.18\textwidth]{Graph4.pdf}
    % \includegraphics[width=0.18\textwidth]{Graph5.pdf}
    \caption{Caption}
    \label{fig:my_label}
\end{figure}

\begin{equation}
     G  = \left[
\begin{array}{cc}
% \mathcal{D}_a+\mathcal{X}_c & \mathcal{X}_d\\
% \mathcal{X}_e & \mathcal{D}_b
\text{Fig}.~8a + \text{Fig}.~8b & \text{Fig}.~8c\\
\text{Fig}.~8d & \text{Fig}.~8e
%%
\end{array}
   \right] \,,
\end{equation}
\end{comment}   
where suppressed
$\decay{\Dstarz}{\Dp\pim}$~transition
is neglected. 
%%
%% {\color{red}
The~$\Dstarp\Dz \leftrightarrow \Dstarz\Dp$ rescattering 
occurs due to 
%% is enforced by the
non-diagonal element of the~$K$-matrix (contact interaction) and
non-diagonal elements of the~$G$~matrix (long-range interaction).
The~matrix $G$ 
and the~self\nobreakdash-energy function 
$\Sigma(s)$ from 
Eqs.~\eqref{eq:imSigma}
and~\eqref{eq:reSigma},
are related as 
%% is related to the~self\nobreakdash-energy function 
%% $\Sigma(s)$ from Eq.~\eqref{eq:Sigma_real} as
\begin{equation}
\left| g\right|^2 \Sigma(s) = 
   %%\begin{pmatrix} \phantom{-}g\\-g  \end{pmatrix}^{\dagger} 
   \begin{pmatrix} g^* & -g^*  \end{pmatrix} 
   G 
   \begin{pmatrix} \phantom{-}g\\-g  \end{pmatrix}\,.
\end{equation}

Similar to the~Flatt\'e~function~\cite{Flatte:1976xu}
for large values of the~$\left|g\right|$~parameter,  
the~$\mathfrak{F}^{\mathrm{U}}$~signal profile
exhibits a~scaling 
property~\cite{Baru:2004xg,LHCb-PAPER-2020-008}.  
For large values of the~$\left|g\right|$~parameter 
the width approaches asymptotic behaviour,
see 
%% Extended Data 
Supplementary~Fig.~\ref{fig:SCALING}.
%%Information
%%Fig.~7.
%% 
The~unitarised three\nobreakdash-body
Breit\protect\nobreakdash--Wigner 
function 
$\mathfrak{F}^{\mathrm{U}}$ for 
\mbox{${\decay{\Tcc}{\Dz\Dz\pip}}$}~decays
with parameters $m_{\mathrm{U}}$ and $\left|g\right|$ 
obtained
from the~fit to data   
is shown in 
%% Extended Data 
Supplementary~Fig.~\ref{fig:BWU_shape}.  
%%Information
%%Fig.~8.
The~inset illustrates
 the similarity of the~profile with 
 the~single\nobreakdash-pole 
 profile in the~vicinity of the~pole 
\begin{equation}
     \sqrt{ \hat{s} } 
     = \mathfrak{m} - \frac{i}{2}\mathfrak{w}\,,
\end{equation}
where $\mathfrak{m}$ and $\mathfrak{w}$
are the~mode and full width at half maximum, 
respectively.

\subsection*{Analytic continuation}

Equations~\eqref{eq:reSigma} define 
$\Sigma(s)$ and the amplitude $\mathcal{A}_{\mathrm{U}}(s)$ 
for real values of~$s$.
Analytic continuation  
to the~whole first Riemann sheet is calculated as
\ifthenelse{\boolean{wordcount}}{}{
\begin{align} \label{eq:irho.analytic}
\Sigma\left(s\right)= 
    \dfrac{s}{2\pi}\int\limits_{s_\text{th}^*}^{+\infty} 
    \dfrac{\varrho_\text{tot}(s')}{s'\left(s'-s\right)}\,ds' 
    - \xi\left( m_{\mathrm{U}}^2\right)
    \,,
\end{align}
}
where the integral is understood as in Eq.~\eqref{eq:sth.inf.limits}.
The search for the resonance pole 
requires knowledge of the~amplitude 
on the~second Riemann sheet denoted 
by $\mathcal{A}^{I\! I}_{\mathrm{U}}$.
According to the~optical theorem~\cite{Gribov:2009zz},
the~discontinuity of the~inverse amplitude 
across the~unitarity cut is given 
by $i \left|g\right|^2 \varrho_\text{tot}(s)$:
\ifthenelse{\boolean{wordcount}}{}{
\begin{align}
    \dfrac{1}{\mathcal{A}^{I\! I}_{\mathrm{U}}\left(s \right)} &=  
    m_{\mathrm{U}}^2 - s  - 
    \left|g\right|^2\Sigma(s) + 
    i\left|g\right|^2\varrho_\text{tot}(s)\,.  \label{eq:ABWII}
\end{align}
}
For the~complex-values $s$, 
the~analytic continuation of $\varrho_\text{tot}(s)$ is needed:
the~phase space integral in Eq.~\eqref{eq:rho_i} is performed over 
a~two\nobreakdash-dimensional complex manifold 
$\mathcal{D}$\,(see discussion on 
the~continuation in~Ref.\cite{Mikhasenko:2018bzm}):
\ifthenelse{\boolean{wordcount}}{}{
\begin{equation} \label{eq:three.body.phase.space}
    \int\limits_{\mathcal{D}} \left|\mathfrak{M}\right|^2\, d \Phi_3 = 
    \dfrac{1}{2\pi (8\pi)^2 s}
    \int\limits_{{(m_2+m_3)^2}}^{{(\sqrt{s}-m_1)^2}} d s_{23}
    \int\limits_{s_{12}^-(s,s_{23})}^{s_{12}^+(s,s_{23})} 
    \left|\frak{M}\right|^2\,d s_{12}\,,
\end{equation}
}
where the limits of the second integral represent  
the~Dalitz plot borders~\cite{Byckling},
\ifthenelse{\boolean{wordcount}}{}{
\begin{equation} \label{eq:s12limits}
\begin{split}
s_{12}^{\pm}\left(s,s_{23}\right)  = m_1^2 + m^2_2 
& - 
\dfrac{ \left(s_{23} - s + m_1^2\right)
\left(s_{23} + m_2^2 - m_3^2\right)}{ 2s_{23}} \\ 
  & \pm  
 \dfrac{
 \uplambda^{\nicefrac{1}{2}}\left(s_{23},s,m^2_1\right)
  \uplambda^{\nicefrac{1}{2}}\left(s_{23},m^2_2,m^2_3\right)}{2s_{23}}
  \,.
\end{split} 
\end{equation}
}
The integration is performed along 
straight lines connecting the~end points in 
the~complex plane.

\subsection*{$\D\D$~spectra from
\mbox{$\decay{\Tcc}{\D\D\Ppi/\g}$}~decays}\label{seq:DD_spectra}

The~shapes of the~$\Dz\Dz$ and $\Dp\Dm$~mass spectra 
from \mbox{$\decay{\Tcc}{\D\D\Ppi/\g}$}~decays 
are obtained 
via integration
of the~$\left|\mathfrak{M}_f\right|^2$ expressions 
from 
%% Eqs.~\eqref{eq:msq_pions} 
%% and~\eqref{eq:msq_gamma}
Eqs.~\eqref{eq:ampsqpip}--\eqref{eq:msq_gamma_two}
over  the~$s$ and $s_{12}$~variables
with the \Tcc\nobreakdash amplitude squared, 
$\left| \mathcal{A}_{\mathrm{U}} \left(s\right)\right|^2$, from Eq.~\eqref{eq:bwU}: 
\ifthenelse{\boolean{wordcount}}{}{
\begin{equation}
R_f\left(s_{23}\right) \equiv 
\int\limits_{
% m_1^2 - 2m_2m_3+s_{23}
\left(m_1 + \sqrt{s_{23}}\right)^2
}^{+\infty}ds\, 
% \mathfrak{F}_f^{\mathrm{U}}(s)
\left| \mathcal{A}_{\mathrm{U}} \left(s\right)\right|^2
f_C(s) 
% \dfrac{1}{s^{3/2}}
\,\dfrac{1}{s}
\int\limits_{s^-_{12}(s,s_{23})}^{s^+_{12}(s,s_{23})}ds_{12}\,
% \sum 
\left|\mathfrak{M}_f(s,s_{12},s_{23})\right|^2\,, 
\label{eq:r_case_one}
\end{equation}
}
where the~lower and upper integration limits 
for $s_{12}$ at fixed $s$ and $s_{23}$
are~given in Eq.~\eqref{eq:s12limits}. 
The~function $f_C(s)$ is introduced 
to perform a~smooth cutoff of 
the~long tail of the~$\Tcc$~profile.
Cutoffs are chosen to suppress the~profile
for regions   $\left| \sqrt{s} - \mathfrak{m} \right| \gg  \mathfrak{w}$,
where $\mathfrak{m}$ and $\mathfrak{w}$ 
are the~mode and FWHM for 
the~$\mathfrak{F}^{\mathrm{U}}(s)$~distribution.
%% from Table~\ref{tab:DATA_visible}. 
Two cut-off functions $f_C(s)$ are studied:  
\begin{enumerate}
    \item A Gaussian cut-off $f^G_C(s)$ defined as 
\ifthenelse{\boolean{wordcount}}{}{
\begin{equation} \label{eq:cutoff1gauss}
    f^G_C(s\,|\,x_c, \upsigma_{c}) = \begin{dcases} 
    1 
    & \text{for } \sqrt{s} \le x_c \,; \\ 
    \mathrm{e}^{-\frac{\left( \sqrt{s}-x_c\right)^2}{2\upsigma_c^2}} 
    & \text{for }  \sqrt{s} > x_c \,. \end{dcases} 
\end{equation}
}
\item A power-law cut off $f^P_C(s)$ defined as
\ifthenelse{\boolean{wordcount}}{}{
\begin{equation}\label{sq:cutoffpow}
   f^P_C(s\,|\,x_c,\upsigma_{c}, \upnu_c) = \begin{dcases} 
    1 
    & \text{for } \sqrt{s} \le x_c \,; \\
    \left(1 + \frac{1}{\upnu_c} \frac{(\sqrt{s}-x_c)^2}{\upsigma_c^2} \right)^{- \frac{\upnu_c+1}{2}}
    & \text{for } \sqrt{s} > x_c \,.\end{dcases} 
\end{equation}
}
\end{enumerate}
Fits to the~background\nobreakdash-subtracted 
$\Dz\Dz\pip$~mass spectrum 
using a~signal profile of 
the~form \mbox{$\mathfrak{F}^{\mathrm{U}}(s)\times f_C(s)$}
show that the~parameter $\updelta m_{\mathrm{U}}$ is insensitive 
to the~choice of cut\nobreakdash-off function 
when \mbox{$x_c \ge m_{\Dstarz}+m_{\Dp}$} and \mbox{$\upsigma_c\ge1\mevcc$}.
The~power\nobreakdash-law cut\nobreakdash-off 
function $f^P_C(s)$ with parameters
$x_c = m_{\Dstarz}+m_{\Dp}$ 
and $\upsigma_c = 1\mevcc$ is chosen. 
The~shapes for the~$\Dz\Dz$ and  
$\Dp\Dz$~mass distributions are defined as 
%% \begin{subequations} \label{eq:dd_spectra}
\begingroup
\allowdisplaybreaks
    \begin{eqnarray}
       F_{\Dz\Dz} (m) & = & m  R_{\Dz\Dz\pip} (m^2) \,, 
       \label{eq:fDzDz} \\ 
       F_{\Dp\Dz} (m) & = & m  R_{\Dp\Dz\pip} (m^2) +  m  R_{\Dp\Dz\g} (m^2) \,.
       \label{eq:fDpDz}
    \end{eqnarray}
\endgroup
%% \end{subequations}

\subsection*{Low-energy scattering amplitude}\label{sec:lowenergy}

The~unitarized  Breit\nobreakdash--Wigner amplitude
is formally similar to the~low\nobreakdash-energy expansion 
given by Eq.~\eqref{eq:A_NR}
once the factor $\frac{1}{2}\left|g\right|^2$ is divided out
%% \begin{subequations}
\begingroup
\allowdisplaybreaks
\begin{eqnarray} \label{eq:inverse.amplitude.nonrel}
    \mathcal{A}_\text{NR}^{-1} 
    & = & 
    \dfrac{1}{a}+ r\dfrac{k^2}{2} 
    - i k 
    + \mathcal{O}(k^4) \,,
    \\ \label{eq:inverse.amplitude.advancedbw}
     \dfrac{2}{\left|g\right|^2} 
     \mathcal{A}_{\mathrm{U}}^{-1} 
    & = & 
        - \left[\xi(s)-\xi(m_{\mathrm{U}}^2)\right]
        + 2\dfrac{m_{\mathrm{U}}^2-s}{\left|g\right|^2}
        - i \varrho_\text{tot}(s)\,.
\end{eqnarray}
\endgroup
%% \end{subequations}
The~function $i\varrho_\text{tot}(s)$ matches $ik$ up 
to a~slowly varying energy factor that 
can be approximated by
a~constant in the~threshold region.
The~proportionality factor $w$ has the dimension of an inverse mass 
and is found by matching the~decay probability 
to the~two\nobreakdash-body phase\nobreakdash-space expression:
\begin{align}\label{eq:w_scattering}
    w = 
    % \dfrac{3}{2} 
    % \dfrac{16\pi}{\sqrt{s_\mathrm{th}}} 
    \dfrac{24\pi}{m_\Dstarp+m_\Dz} 
    \dfrac{1}{c_1} \,,
\end{align}
where $c_1$ is a~coefficient computed in Eq.~\eqref{eq:c_one}.
The~comparison of 
$\mathcal{A}_\text{NR}^{-1}$ and
% $w\mathcal{A}_{\mathrm{U}}^{-1}$
$\mathcal{A}_{\mathrm{U}}^{-1}
\times 2w/\left|g\right|^2$
that validates 
the~matching is 
shown in 
%% Extended Data 
Supplementary~Fig.~\ref{fig:MATCHING}.
%%Information 
%%Fig.~9.

The inverse scattering length is defined 
as the~value of the~amplitude in 
Eq.~\eqref{eq:inverse.amplitude.nonrel} 
at the~$\Dstarp\Dz$ threshold:
\ifthenelse{\boolean{wordcount}}{}{
\begin{align} \label{eq:scatt.length}
    \frac{1}{a} &= -\dfrac{1}{w}\Bigl\{
        \left[\xi(s_\text{th}) - \xi(m_{\mathrm{U}}^2)\right]
        + i \varrho_\text{tot}(s_\text{th})
        \Bigr\}\,.
\end{align}
}
The~imaginary part is fully determined by 
the~available decay channels,
while the~real part depends on the~constant $\xi(m_{\mathrm{U}}^2)$ adjusted in 
the~fit.
The~quadratic term, $k^2$ in Eq.~\eqref{eq:inverse.amplitude.nonrel},  
corresponds to the~linear correction 
in $s$ since $k^2 = (s-s_\text{th})/4$ for 
the~non\nobreakdash-relativistic case.
Hence, the~slope of the~linear term in the
$\mathcal{A}_{\mathrm{U}}^{-1}$~amplitude 
is related to the~effective range as follows:
\ifthenelse{\boolean{wordcount}}{}{
\begin{align}\label{eq:eff_range}
    r = -\dfrac{1}{w}\frac{16}{\left|g\right|^2}\,.
\end{align}
}

\subsection*{Mass splitting
for the~$\hat{\PT}_{\cquark\cquark}$~isotriplet}\label{sec:isotriplet}

While the degrees of freedom of the~light diquark for the~isoscalar 
\Tcc~state are similar to those for the~\Lcbar~state, 
for the~$\hat{\PT}_{\cquark\cquark}$~isotriplet ($\hat{\PT}^{0}_{\cquark\cquark}$,
$\hat{\PT}^{+}_{\cquark\cquark}$,
$\hat{\PT}^{++}_{\cquark\cquark}$)
the~light diquark degrees of freedom 
would be similar to those for the~$\Sigmacbar$~(anti)triplet. 
Assuming that the~difference in the light quark masses, 
the~Coulomb interaction of light quarks in 
the~diquark, and the~Coulomb interaction of 
the~light diquark with 
the~$\cquark$\nobreakdash-quark are 
responsible for the~observed mass splitting 
in the~\Sigmac~isotriplet,  
the~masses for the~\Sigmac~states can be written as 
%% 
%% \begin{subequations}
\begingroup
\allowdisplaybreaks
\begin{eqnarray}
m_{\Sigmacpp} 
& = & m_\Sigma + m_{\uquark} + m_{\uquark} 
- a\,q_{\uquark} q_{\uquark} 
- b\,q_{\cquark} \left(q_{\uquark}+q_{\uquark}\right)  \,,  
\\
m_{\Sigmacp} 
& = & m_\Sigma + m_{\uquark} +m_{\dquark} 
- a\,q_{\uquark}q_{\dquark} 
- b\,q_{\cquark} \left(q_{\uquark}+q_{\dquark}\right) \,,
\\
m_{\Sigmacz} 
& = & m_\Sigma + m_{\dquark} + m_{\dquark} - 
a\,q_{\dquark}q_{\dquark} -
b\,q_{\cquark} \left(q_{\dquark}+q_{\dquark}\right) \,,
\end{eqnarray}
\endgroup
%% \end{subequations}
%%
where $m_\Sigma$~is a common mass parameter;  
the~second and third terms describe the contribution 
from the light quark masses, $m_{\uquark}$ and $m_{\dquark}$, 
into the mass splitting;
terms proportional to $a$ describe Coulomb 
interactions of light quarks in the diquark; 
terms proportional to $b$ describe 
the~Coulomb interactions of the~diquark 
with the $\cquark$\nobreakdash-quark; 
and $q_{\quark}$~denotes 
the~charge of the~$\quark$\nobreakdash-quark. 
Similar expressions can be written for 
the~$\hat{\PT}_{\cquark\cquark}$~isotriplet:  
%%
%% \begin{subequations} \label{eq:sigc_exp}
\begingroup
\allowdisplaybreaks
\begin{eqnarray}
m_{\hat{\PT}^0_{\cquark\cquark}} 
& = & m_{\hat{\PT}_{\cquark\cquark}} + m_{\uquark} + m_{\uquark} 
- a^\prime\,q_{\uquarkbar}q_{\uquarkbar} 
- b^\prime\,q_{\cquark\cquark} \left(q_{\uquarkbar}+q_{\uquarkbar}\right)  \,,  
\\
m_{\hat{\PT}^+_{\cquark\cquark}} 
& = & m_{\hat{\PT}_{\cquark\cquark}} + m_{\uquark} +m_{\dquark} 
- a^\prime\,q_{\uquarkbar}q_{\dquarkbar} 
- b^\prime\,q_{\cquark\cquark} \left(q_{\uquarkbar}+q_{\dquarkbar}\right) \,,
\\
m_{\hat{\PT}^{++}_{\cquark\cquark}} 
& = & m_{\hat{\PT}_{\cquark\cquark}} + m_{\dquark} + m_{\dquark} 
- a^\prime\,q_{\dquarkbar}q_{\dquarkbar} 
- b^\prime\,q_{\cquark\cquark} \left(q_{\dquarkbar}+q_{\dquarkbar}\right) \,,
\end{eqnarray}
\endgroup
%% \end{subequations}
%%
where $m_{\hat{\PT}_{\cquark\cquark}}$ is the~common mass parameter,
$q_{\quarkbar}=-q_{\quark}$ and 
$q_{\cquark\cquark}=2q_{\cquark}$~is 
the~charge of a $\cquark\cquark$~diquark. 
Using the~known masses of the light quarks 
and \Sigmac~states~\cite{PDG2021}
and taking $a^{\prime}=a$ and $b^{\prime}=b$, the~mass splitting 
for the~$\hat{\PT}_{\cquark\cquark}$~isotriplet is estimated to be 
%%
%% \begin{subequations} \label{eq:tcc_init}
\begingroup
\allowdisplaybreaks
\begin{eqnarray}
m_{\hat{\PT}^0_{\cquark\cquark}}  
- m_{\hat{\PT}^+_{\cquark\cquark}}  
& = &  -5.9 \pm 1.3\mevcc 
\,,  \label{eq:mTccone_init} \\ 
m_{\hat{\PT}^{++}_{\cquark\cquark}}  
- m_{\hat{\PT}^+_{\cquark\cquark}}  
& = & \phantom{-}  7.9 \pm 1.0\mevcc
\,. \label{eq:mTcctwo_init}
\end{eqnarray}
\endgroup
%% \end{subequations}
%%
The validity of this approach is 
tested by comparing the~calculated 
mass splitting between 
$\Sigmabp$ and $\Sigmabm$~states
of \mbox{$-6.7 \pm 0.7\mevcc$} 
with the~measured value of 
\mbox{$-5.1 \pm 0.2\mevcc$}~\cite{PDG2021}. 
Based on the~small observed difference, 
an~addition uncertainty of 0.8\mevcc 
is  added in quadrature  to the~results from 
%% Eq.~\eqref{eq:tcc_init}, 
Eqs.~\eqref{eq:mTccone_init}
and ~\eqref{eq:mTcctwo_init}, 
and finally one gets 
%% 
%%\begin{subequations} \label{eq:tcc}
\begingroup
\allowdisplaybreaks
\begin{eqnarray}
m_{\hat{\PT}^0_{\cquark\cquark}}  - m_{\hat{\PT}^+_{\cquark\cquark}}  
& = &  -5.9 \pm 1.5\mevcc  \label{eq:mTccone}
\,,  \\ 
m_{\hat{\PT}^{++}_{\cquark\cquark}}  - m_{\hat{\PT}^+_{\cquark\cquark}}  
& = & \phantom{-}  7.9 \pm 1.3\mevcc
\,. \label{eq:mTcctwo} 
\end{eqnarray}
\endgroup
%%\end{subequations}
%%
These results agree
within  the~assigned uncertainty
with results based 
on a~more advanced model 
from Ref.~\cite{Karliner:2019lau}.

\section*{Acknowledgements}
%
% These Acknowledgements valid from 3-May-2019
%
\noindent 
This paper is dedicated 
to the~memory of our 
dear friend and colleague
Simon Eidelman, 
whose contributions to 
improving the~quality of our 
papers were greatly appreciated.
%% 
%% We~thank 
%% M.~Karliner for inspiring 
%% and stimulating 
%% discussions. 
%%
We express our gratitude to our colleagues in the CERN
accelerator departments for the excellent performance of the LHC. We
thank the technical and administrative staff at the LHCb
institutes.
We acknowledge support from CERN and from the national agencies:
CAPES, CNPq, FAPERJ and FINEP (Brazil); 
MOST and NSFC (China); 
CNRS/IN2P3 (France); 
BMBF, DFG and MPG (Germany); 
INFN (Italy); 
NWO (Netherlands); 
MNiSW and NCN (Poland); 
MEN/IFA (Romania); 
MSHE (Russia); 
MICINN (Spain); 
SNSF and SER (Switzerland); 
NASU (Ukraine); 
STFC (United Kingdom); 
DOE NP and NSF (USA).
We acknowledge the computing resources that are provided by CERN, IN2P3
(France), KIT and DESY (Germany), INFN (Italy), SURF (Netherlands),
PIC (Spain), GridPP (United Kingdom), RRCKI and Yandex
LLC (Russia), CSCS (Switzerland), IFIN-HH (Romania), CBPF (Brazil),
PL-GRID (Poland) and NERSC (USA).
We are indebted to the communities behind the multiple open-source
software packages on which we depend.
Individual groups or members have received support from
ARC and ARDC (Australia);
AvH Foundation (Germany);
EPLANET, Marie Sk\l{}odowska-Curie Actions and ERC (European Union);
A*MIDEX, ANR, IPhU and Labex P2IO, and R\'{e}gion Auvergne-Rh\^{o}ne-Alpes (France);
Key Research Program of Frontier Sciences of CAS, CAS PIFI, CAS CCEPP, 
Fundamental Research Funds for the Central Universities, 
and Sci. \& Tech. Program of Guangzhou (China);
%Key Research Program of Frontier Sciences of CAS, CAS PIFI,
%Thousand Talents Program, and Sci. \& Tech. Program of Guangzhou (China);
RFBR, RSF and Yandex LLC (Russia);
GVA, XuntaGal and GENCAT (Spain);
the Leverhulme Trust, the Royal Society
 and UKRI (United Kingdom).

\subsection*{Data Availability Statement} 

LHCb data used in this analysis will be released according 
to the~LHCb external data access
policy, that can be downloaded from 
\href{http://opendata.cern.ch/record/410/files/LHCb-Data-Policy.pdf}
    {\tt{http://opendata.cern.ch/record/410/files/LHCb-Data-Policy.pdf}}.
The~raw data in all of the~figures of this manuscript
can be downloaded from 
\href{https://cds.cern.ch/record/2780001}{\tt{https://cds.cern.ch/record/2780001}},
where no access codes are required. 
In~addition, 
the~unbinned background\nobreakdash-subtracted data, 
shown in Figs.~\ref{fig:DATA_BWU_fit},
 \ref{fig:DATA_BWU_Dpi} and 
  \ref{fig:DATA_DD} 
have been added to the~{\sc{HEPData}} record at
\href{https://www.hepdata.net/record/ins1915358} 
{\tt{https://www.hepdata.net/record/ins1915358}}.

\subsection*{Code Availability Statement} 
LHCb software used to process the~data analysed in this manuscript 
is available at {\sc{GitLab}} repository 
\href{https://gitlab.cern.ch/lhcb}{\tt{https://gitlab.cern.ch/lhcb}}.
The~specific software used in data analysis 
is available at {\sc{Zenodo}} repository
\href{https://zenodo.org/record/5595937}{\tt{DOI:10.5281/zenodo.5595937}}.

\subsection*{Author Contribution Statement} 
All contributing authors, as listed at the~end of
this manuscript, 
have contributed to the publication, 
being variously involved in the design 
and the~construction of the~detector, 
in writing software, calibrating sub-systems, 
operating the~detector and acquiring data 
and finally analysing the~processed data.

\subsection*{Competing Interests Statement}
The authors declare no competing interests.

\subsection*{Correspondence and requests for materials}
Correspondence and requests for materials
should be addressed to I.~Belyaev \href{mailto:Ivan.Belyaev@itep.ru}{Ivan.Belyaev@itep.ru}.

\clearpage
\addcontentsline{toc}{section}{References}
\bibliographystyle{LHCb}
\bibliography{main,standard,LHCb-PAPER,LHCb-CONF,LHCb-DP,LHCb-TDR}

\clearpage
%%\renewcommand{\figurename}{\bf Extended Data Fig.}
%% \section*{Extended Data}

\section*{Supplementary Information:
Study of  
the~doubly charmed tetraquark~\Tcc}
\renewcommand{\figurename}{\bf Supplementary Fig.}
\setcounter{figure}{0}
\setcounter{table}{0}
\setcounter{equation}{0}

\begin{figure}[htb]  
  \centering
  \includegraphics[width=\textwidth]{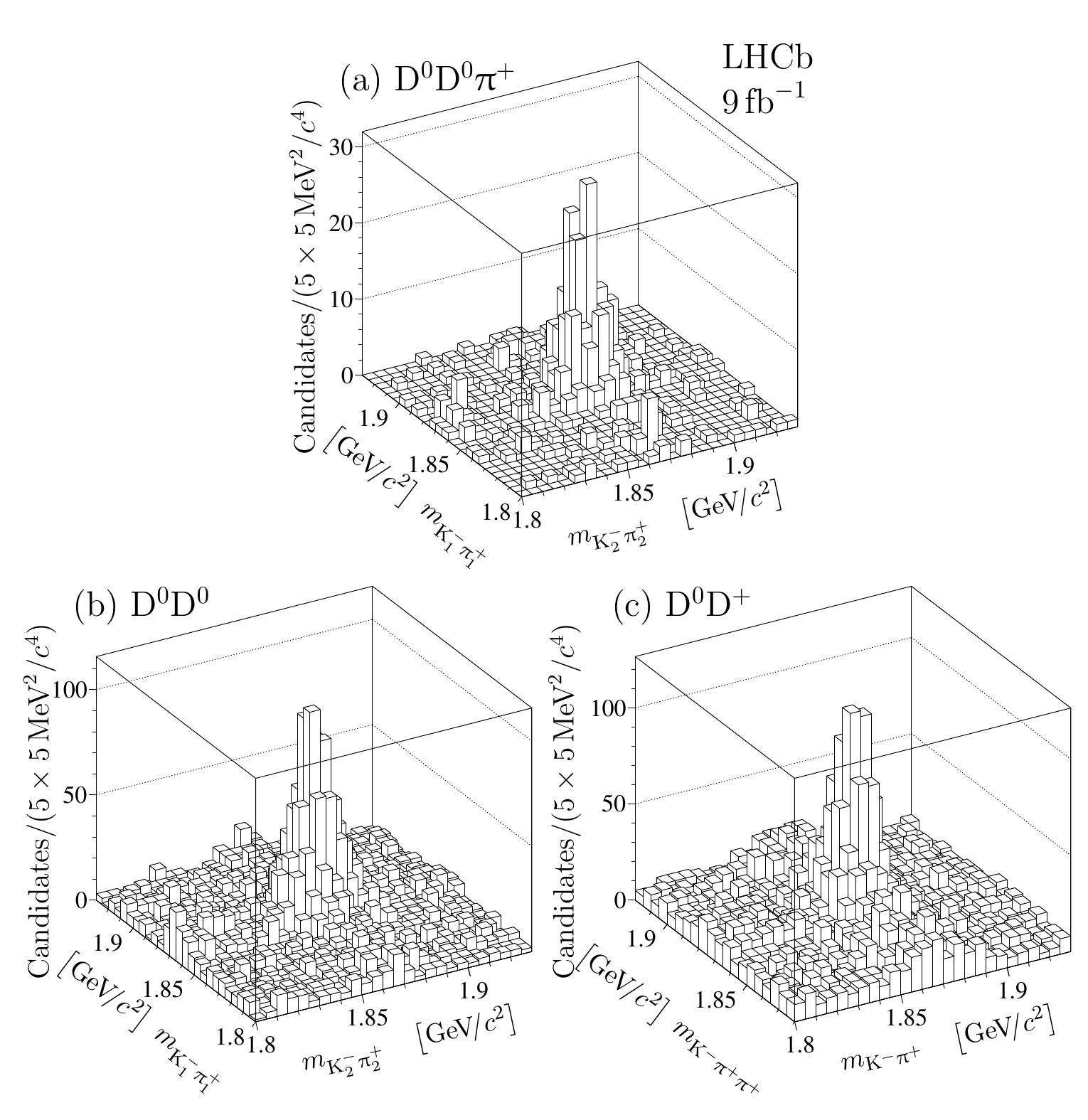}
  \caption {\small
  {\bf{Two-dimensional mass distributions for  selected $\Dz\Dz\pip$, 
  $\Dz\Dz$ and $\Dz\Dp$~combinations.}}
  Two\protect\nobreakdash-dimensional mass distributions 
  for $\Dz$ and $\Dp$~candidates
  from selected (a)~$\Dz\Dz\pip$, 
  (b)~$\Dz\Dz$ and (c)~$\Dz\Dp$~combinations.
  $\kaon_{1/2}$ and $\pion_{1/2}$ correspond 
  to daughters of the~first/second \Dz candidate 
  in $\Dz\Dz$ and $\Dz\Dz\pip$ final states.
  }
  \label{fig:DATA_2D}
\end{figure}

%%%%%%%%%%%%%%%%%%%%%%%%%%%
%%%%%%%%%%%%%%%%%%%%%%%%%%%

\begin{figure}[htb]
  \centering
  \includegraphics[width=\textwidth]{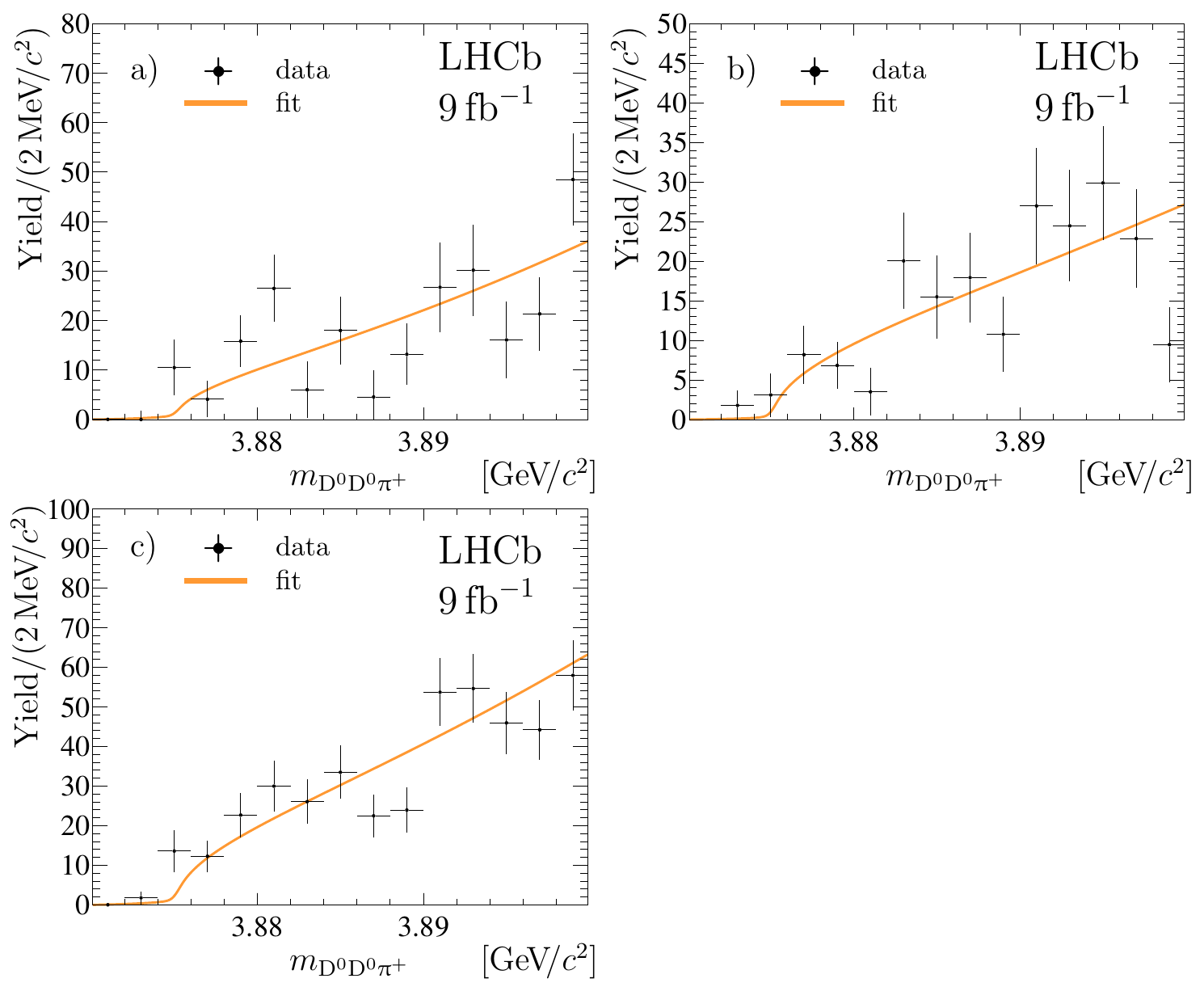}
  \caption { \small
    {\bf Mass distributions for $\Dz\Dz\pip$~combinations 
with fake \Dz~candidates.}
  Mass distributions for $\Dz\Dz\pip$~combinations 
with (a)~one true and one fake \Dz~candidate, 
(b)~two fake $\Dz$~candidates
and (c)~at least one fake $\Dz$~candidate.
Results of the~fits with 
background\protect\nobreakdash-only
functions are overlaid.
Uncertainties on the~data points are statistical
 only and represent one standard deviation, 
 calculated as a~sum in quadrature of the~assigned weights from 
 the~background\protect\nobreakdash-subtraction procedure.
 %% }}
 %%
  }
  \label{fig:SIDEBANDS}
\end{figure}

\begin{figure}[htb]
  \centering
  \includegraphics[width=\textwidth]{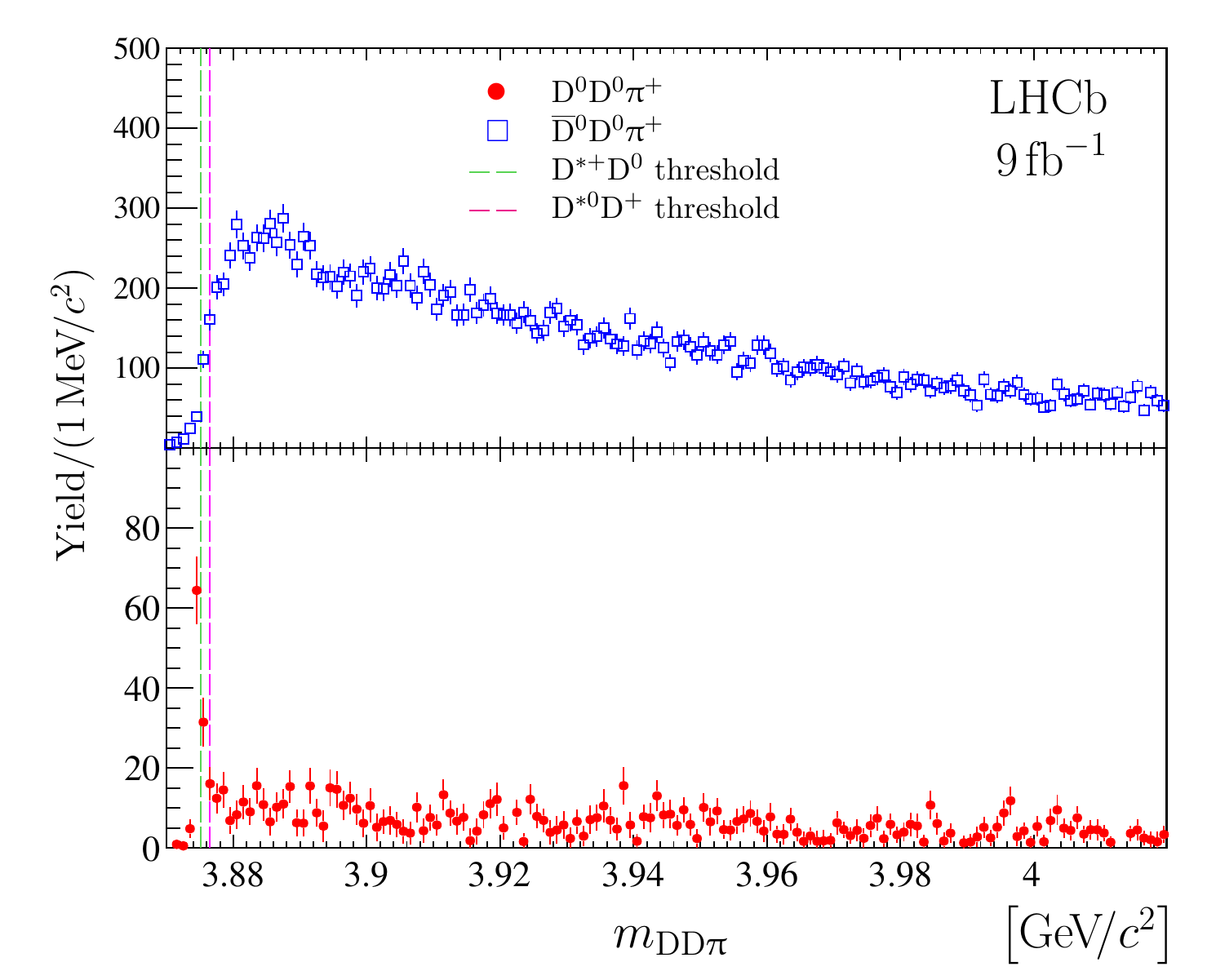}
  \caption {\small
  {\bf{Mass distributions for \Dz\Dz\pip and \Dz\Dzb\pim~candidates.}}
   Background\protect\nobreakdash-subtracted 
  \Dz\Dz\pip and \Dz\Dzb\pim~mass distributions.
  %%
   %% {\color{red}{
   Uncertainties on the~data points are statistical
 only and represent one standard deviation, 
 calculated as a~sum in quadrature of the~assigned weights from 
 the~background\protect\nobreakdash-subtraction procedure.
 %% }}
  %% 
  }
  \label{fig:D0D0piOS}
\end{figure}

\begin{figure}[htb]
  \centering
  \includegraphics[width=\textwidth]{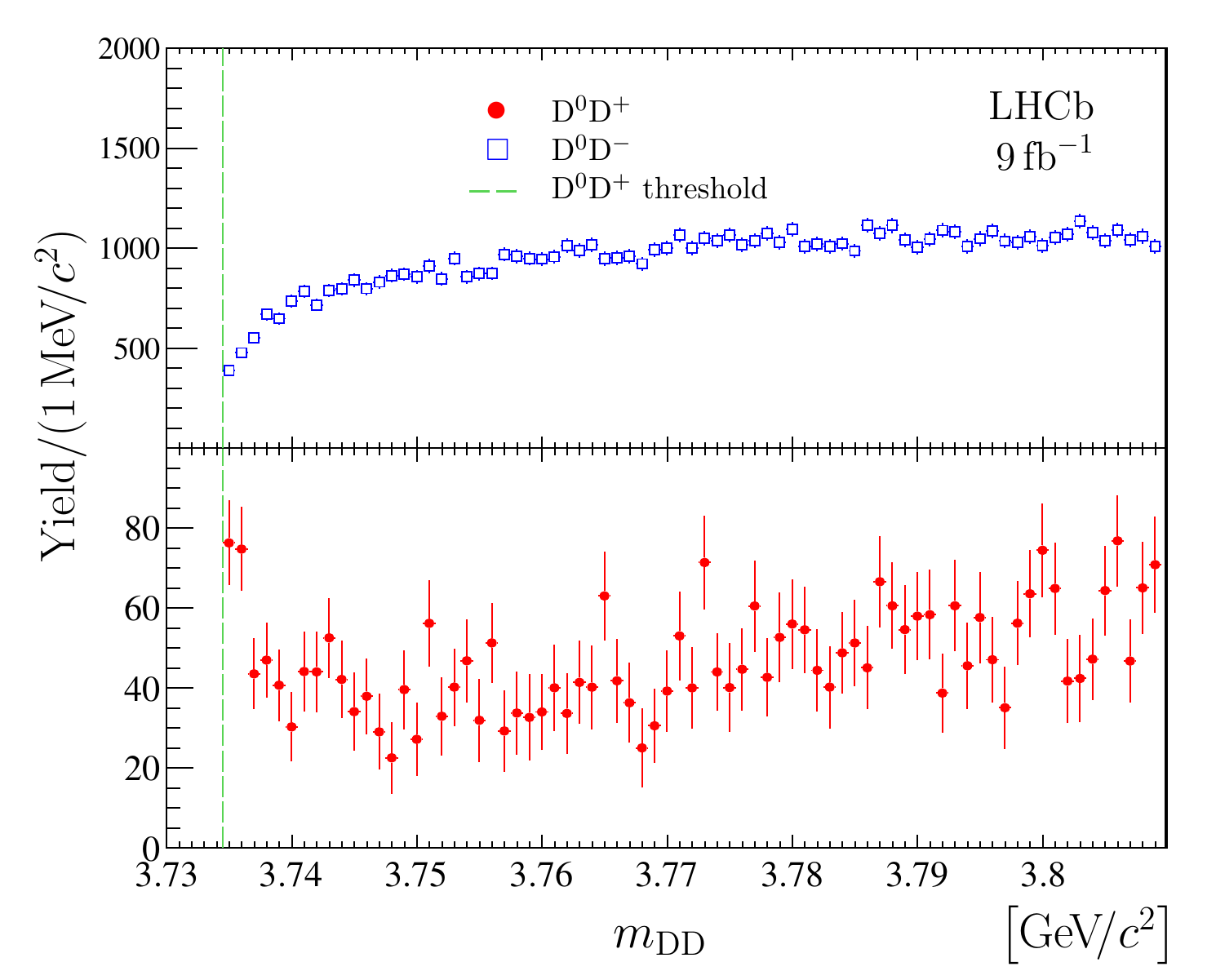}
  \caption {\small
   {\bf{Mass distributions for \Dz\Dp and \Dz\Dm~candidates.}}
   Background\protect\nobreakdash-subtracted 
  \Dz\Dp and \Dz\Dm~mass distributions.
  %%
   %% {\color{red}{
   Uncertainties on the~data points are statistical
 only and represent one standard deviation, 
 calculated as a~sum in quadrature of the~assigned weights from 
 the~background\protect\nobreakdash-subtraction procedure.
 %% }}
  %%
  }
  \label{fig:D0DpOS}
\end{figure}

%%%%%%%%%%%%%%%%%%%%%%%%%%%%%
%%%%%%%%%%%%%%%%%%%%%%%%%%%%%

\begin{figure}[htb]
  \centering
  \includegraphics[width=\textwidth]{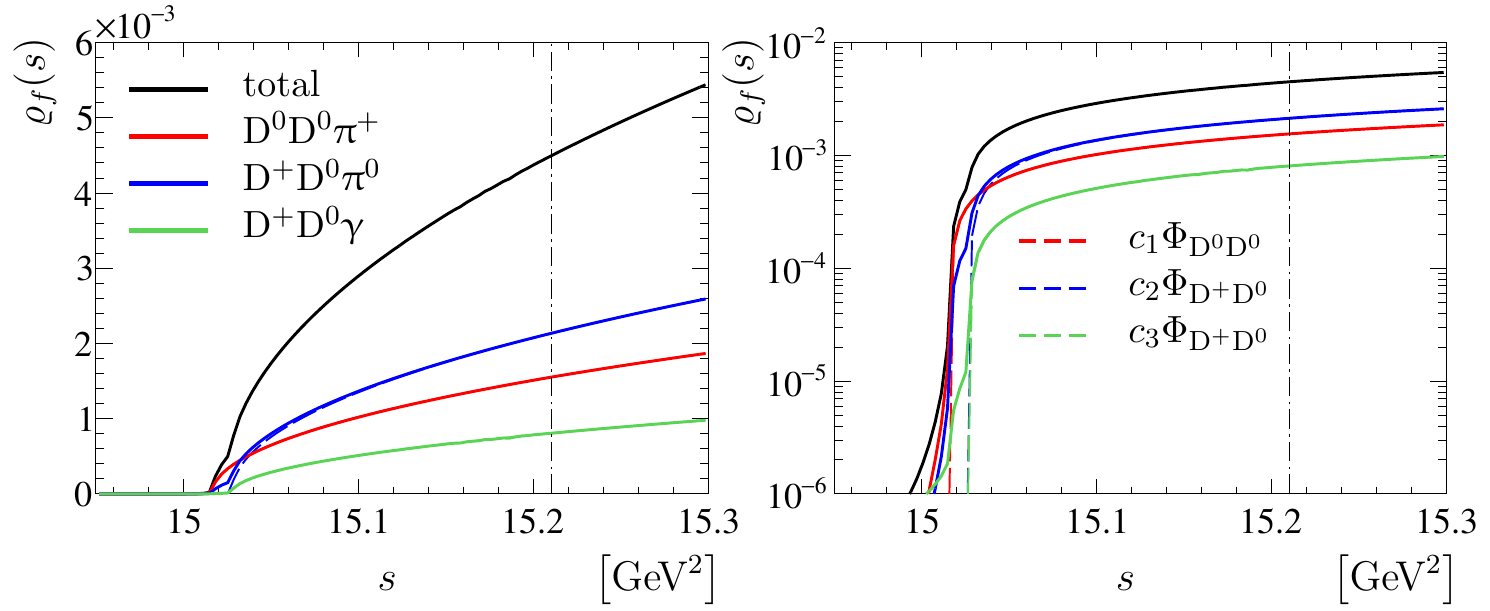}
  \caption { \small
   {\bf{Three\protect\nobreakdash-body phase\protect\nobreakdash-space functions  
   $\varrho_f(s)$.}}
   Three\protect\nobreakdash-body phase\protect\nobreakdash-space functions  
   $\varrho_f(s)$ with (left) linear and (right) logarithmic 
   vertical-axis scale:
   (red)~\mbox{$\decay{\Tcc}{\Dz\Dz\pip }$},
   (blue)~\mbox{$\decay{\Tcc}{\Dp\Dz\piz}$} and 
   (green)~\mbox{$\decay{\Tcc}{\Dp\Dz\g}$}.
   The~sum, $\varrho_{\mathrm{tot}}(s)$,  
   is shown with a~black line. 
   The~two\protect\nobreakdash-body $\Dstar\D$~phase-space shapes 
   are shown by the~dashed lines and are different 
   from the~$\varrho_f(s)$~functions only 
   in the~vicinity 
   of the~thresholds or below them. 
   Vertical dash-dotted line indicates 
   $\sqrt{s^*}=3.9\gev$.
  }
  \label{fig:RHOS}
\end{figure}

\begin{figure}[htb]
  \centering
  \includegraphics[width=\textwidth]{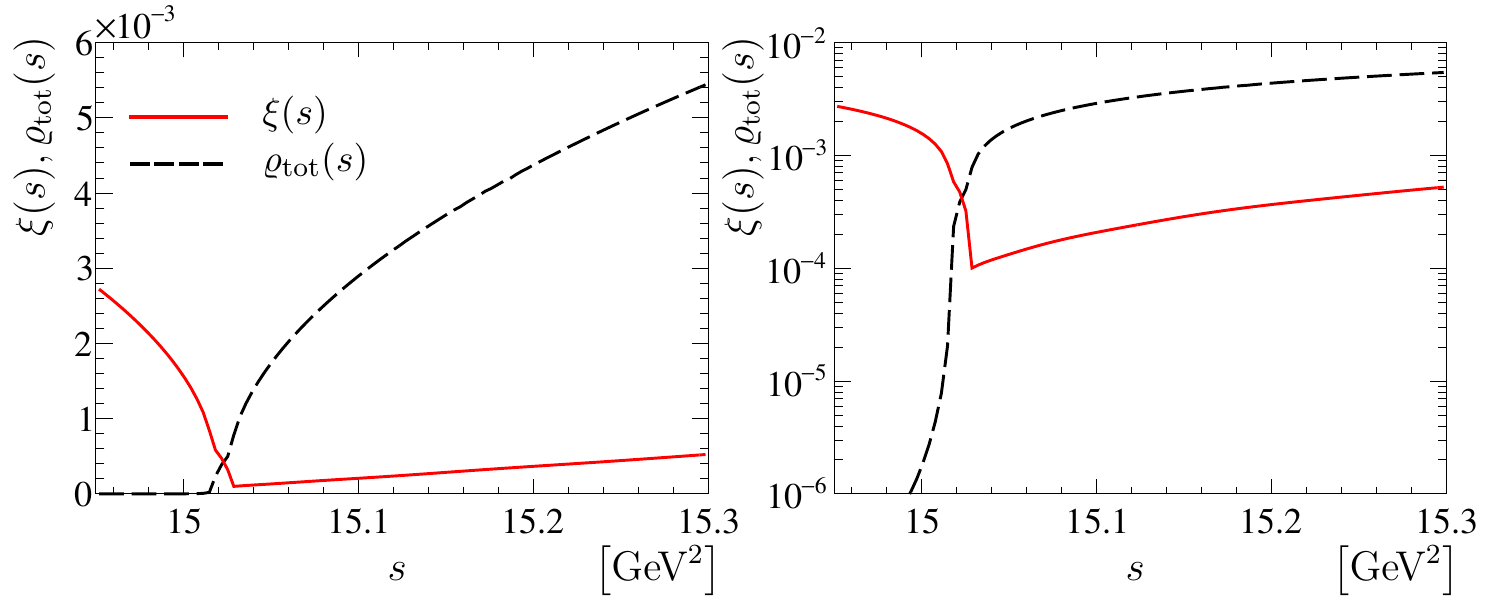}
  \caption { \small
  {\bf{Function $\xi(s)$.}} 
  Function $\xi(s)$ 
  with (left) linear and (right) logarithmic vertical-axis scale
  is shown with 
  a~red line.
  The~three\protect\nobreakdash-body phase-space function
   $\varrho_{\mathrm{tot}}(s)$ 
   is shown for comparison with a~dashed line. 
  }
  \label{fig:xi}
\end{figure}

\begin{figure}[htb]
  \centering
  \includegraphics[width=\textwidth]{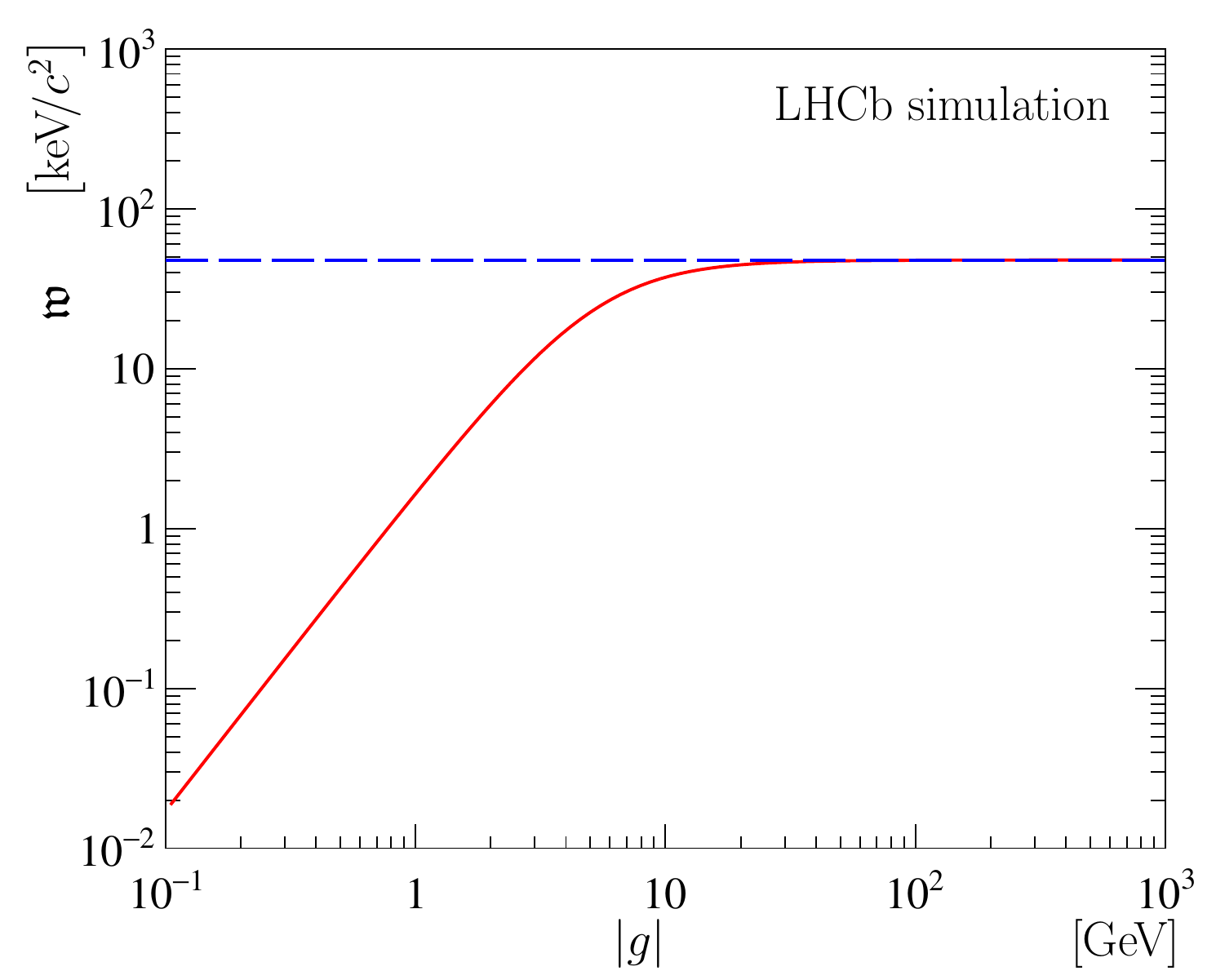}
  \caption { \small
  {\bf{Scaling behaviour of the~$\mathfrak{F}^{\mathrm{U}}$~profile.}}
  %%
  %% Scaling behaviour of the~$\mathfrak{F}^{\mathrm{U}}$~profile.
  The~full width at half maximum $\mathfrak{w}$ as a~function
  of the~$\left|g\right|$~parameter for
  a~fixed value of the~$\updelta m_{\mathrm{U}}$~parameter
  \mbox{$\updelta m_{\mathrm{U}}=-359\kevcc$}.
  The~horizontal dashed blue line indicates 
  the~value of $\mathfrak{w}$ corresponding 
  to the~best fit parameters.
  }
  \label{fig:SCALING}
\end{figure}

\begin{figure}[htb]
  \centering
  \includegraphics[width=\textwidth]{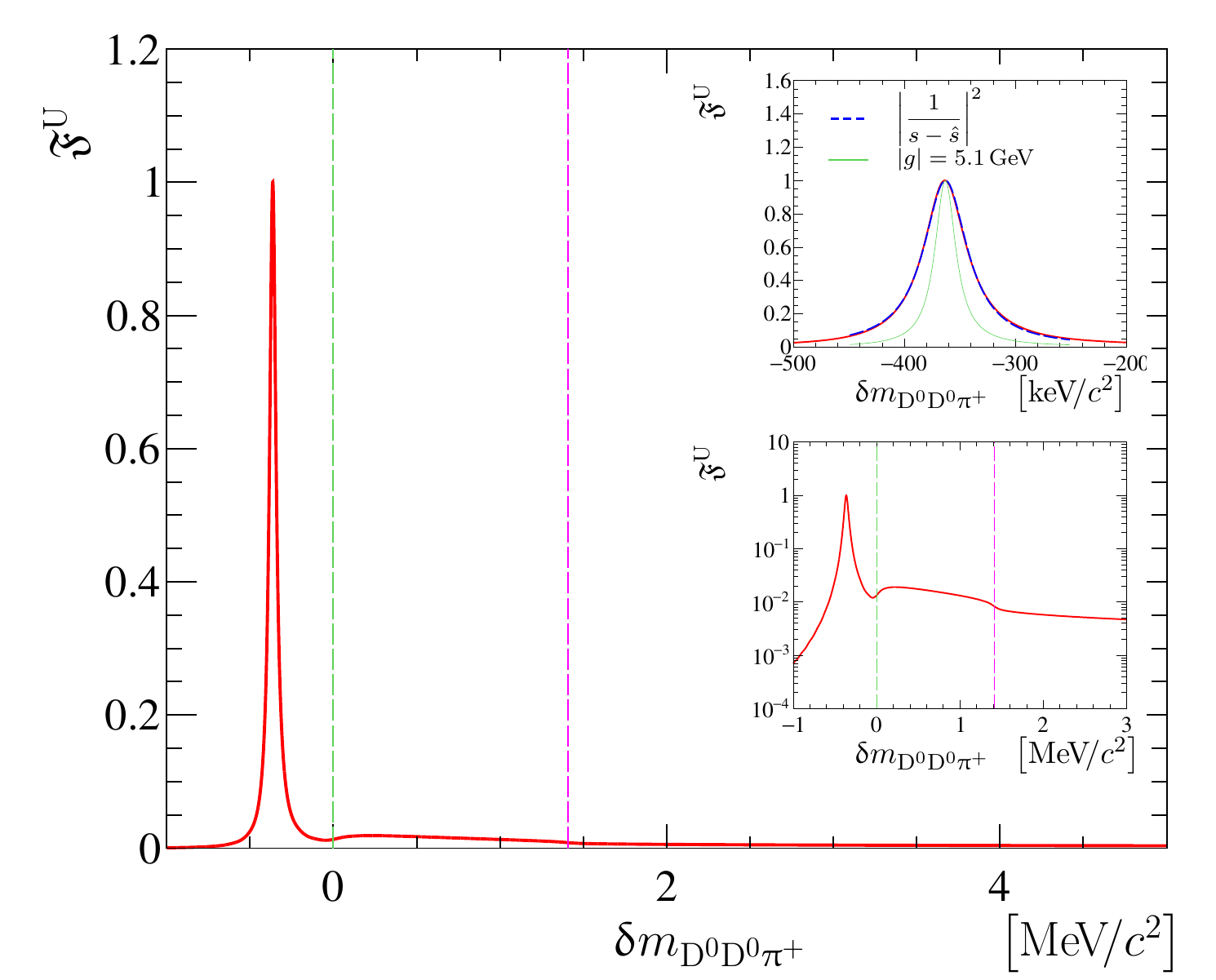}
  \caption { \small
  {\bf{Unitarised three-body Breit\protect\nobreakdash--Wigner 
  function %% $\mathfrak{F}^{\mathrm{U}}_{\decay{\Tcc}{\Dz\Dz\pip}}$.}}
  $\mathfrak{F}^{\mathrm{U}}$.}}
  Unitarised three\protect\nobreakdash-body Breit\protect\nobreakdash--Wigner 
  function $\mathfrak{F}^{\mathrm{U}}$ for 
  ${\decay{\Tcc}{\Dz\Dz\pip}}$~decays\,(red line)
  for a~large value of the~$\left|g\right|$~parameter
  and $\updelta m_{\mathrm{U}}=-359\kevcc$,
  normalized to unity for $\updelta m_{\Dz\Dz\pip}=\updelta m_{\mathrm{U}}$.
  Top inset shows a~zoomed region with 
  overlaid (blue dashed line)
  {\em{single\protect\nobreakdash-pole}} profile 
  with $\sqrt{\hat{s}}=\mathfrak{m}-\tfrac{i}{2}\mathfrak{w}$,
  and (thin green line)~three\protect\nobreakdash-body Breit\protect\nobreakdash--Wigner
  profile with $\left|g\right|=5.1\gev$.
  Bottom inset shows the~$\mathfrak{F}^{\mathrm{U}}$~profile 
  in log-scale. Vertical dashed lines indicate 
  (left\protect\nobreakdash-to\protect\nobreakdash-right) 
  $\Dstarp\Dz$ and $\Dstarz\Dp$~mass thresholds. 
  }
  \label{fig:BWU_shape}
\end{figure}

\begin{figure}[htb]
  \centering
  \includegraphics[width=\textwidth]{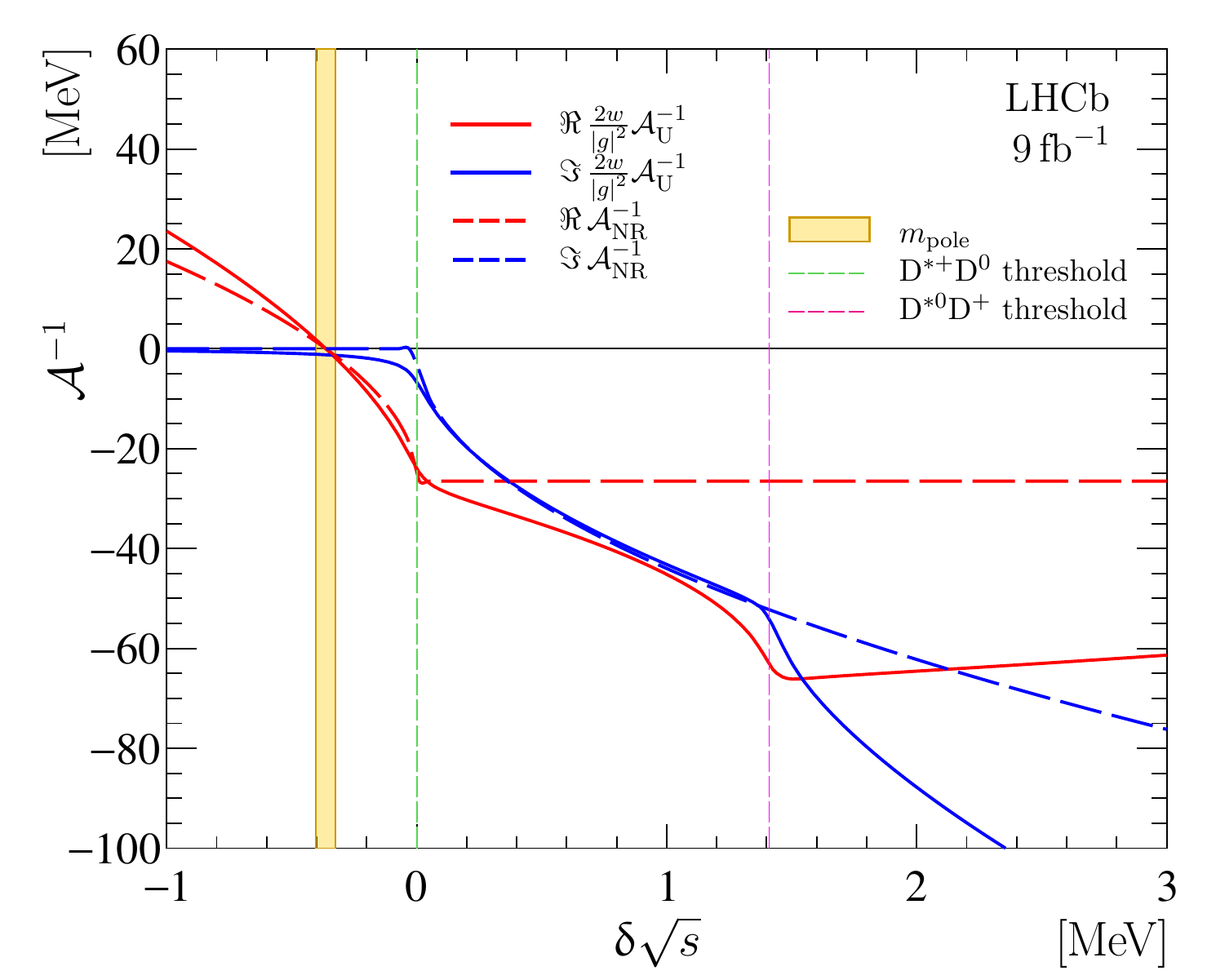}
  \caption { \small
  {\bf{Comparison of the~$\mathcal{A}_{\mathrm{U}}$
   and $\mathcal{A}_{\mathrm{NR}}$~amplitudes.}} 
  The~real and imaginary
  parts of the inverse $\mathcal{A}_{\mathrm{U}}$
   and $\mathcal{A}_{\mathrm{NR}}$~amplitudes.
   The yellow band correspond to the~pole position
   and vertical dashed lines show the
   $\Dstarp\Dz$ and $\Dstarz\Dp$~mass thresholds.
  }
  \label{fig:MATCHING}
\end{figure}

\begin{figure}[htb]
  \centering
  \includegraphics[width=\textwidth]{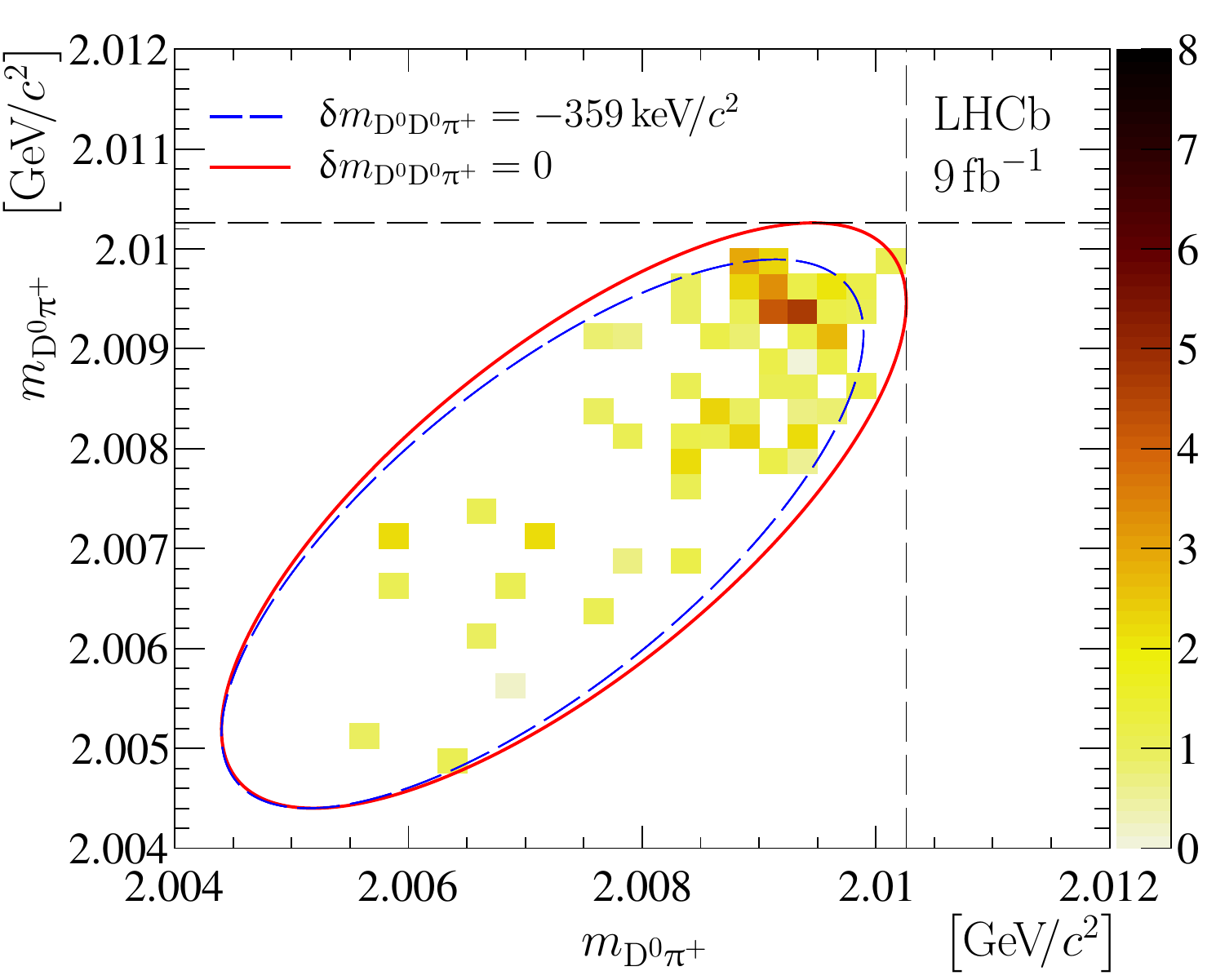}
  \caption { \small
  {\bf{Two-dimensional $\Dz\pip$ mass distribution.}}
  Background\protect\nobreakdash-sub\-tracted 
  two\protect\nobreakdash-dimensional $\Dz\pip$ mass distribution
  %{\em{Dalitz-like}} plot 
  for the~$\Dz\Dz\pip$~events with $\updelta m_{\Dz\Dz\pip}\le0$. 
  Dashed vertical and horizontal lines indicate the~known $\Dstarp$~mass.
  Red and dashed blue lines show  the~boundary 
  %of the Dalitz plot 
  corresponding to 
  \mbox{$\updelta m_{\Dz\Dz\pip}=0$}
  and \mbox{$\updelta m_{\Dz\Dz\pip}=-359\kevcc$}, respectively.
  }
  \label{fig:DALITZ}
\end{figure}

\clearpage 
\subsection*{Propagation matrix $G$}

The propagation matrix $G$
describes the~$\Dstar\D \to \Dstar\D$ 
rescattering via 
the~virtual loops 
including the~one\nobreakdash-particle exchange process
and expressed in a~symbolical way  as 
\begin{equation}
      G  = \left[
 \begin{array}{cc}
   \raisebox{-0.35\height}{\includegraphics[width=0.20\textwidth]{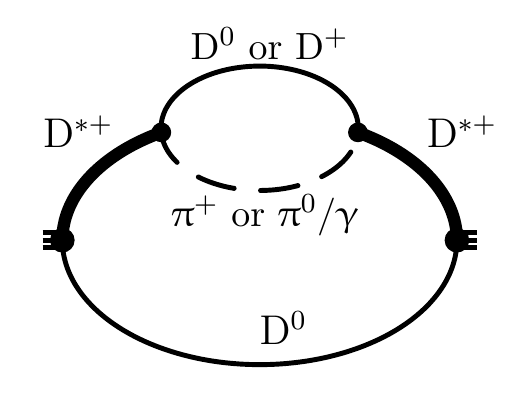}}
  + \raisebox{-0.35\height}{\includegraphics[width=0.20\textwidth]{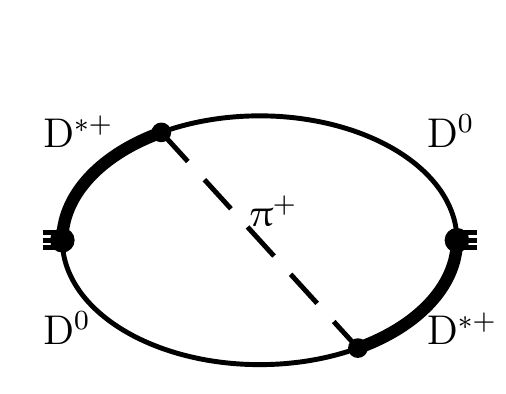}}
 &
   \raisebox{-0.35\height}{\includegraphics[width=0.20\textwidth]{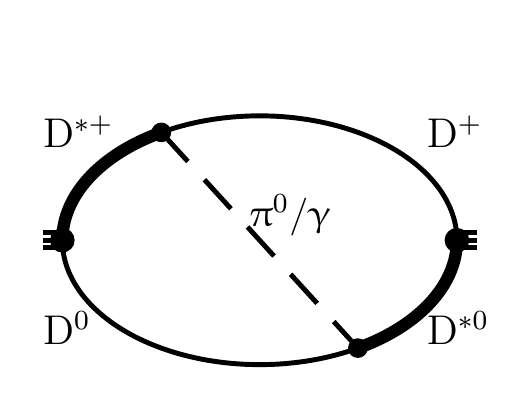}}
  \\
 \raisebox{-0.35\height}{\includegraphics[width=0.20\textwidth]{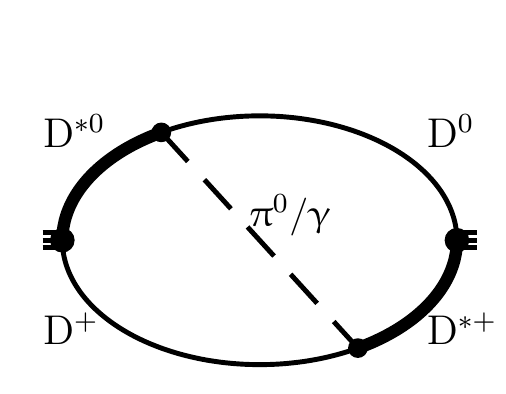}} 
 &       
 \raisebox{-0.35\height}{\includegraphics[width=0.20\textwidth]{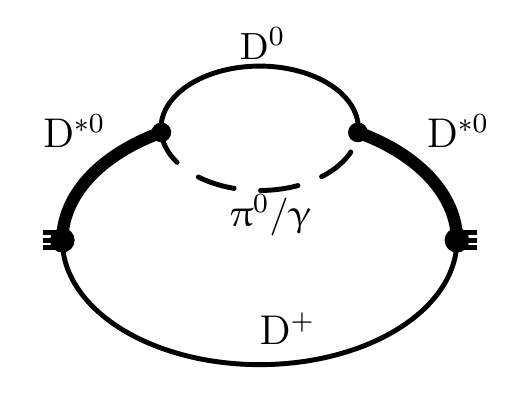}}
 \end{array}
   \right] \,, \label{eq:G}
 \end{equation}
where suppressed
$\decay{\Dstarz}{\Dp\pim}$~transition
is neglected.

%% \bibliographystyle{LHCb}
%% \input{refs_main}

%% \cite{FAKE} 

%% \setcounter{mybibstartvalue}{153}
%% \renewcommand\refname{Methods References}
%% \input{refs_methods}

%% \clearpage 
%% \input{extended} 

\clearpage
%%\input{Authorship_LHCb-PAPER-2021-032}
% LHCb collaboration author list
% Data extracted on July 8th, 2021 at 8:41pm for paper reference LHCb-PAPER-2021-032
\centerline
{\large\bf LHCb collaboration}
\begin
{flushleft}
\small
R.~Aaij$^{1}$,
A.S.W.~Abdelmotteleb$^{2}$,
C.~Abell{\'a}n~Beteta$^{3}$,
F.J.~Abudinen~Gallego$^{2}$,
T.~Ackernley$^{4}$,
B.~Adeva$^{5}$,
M.~Adinolfi$^{6}$,
H.~Afsharnia$^{7}$,
C.~Agapopoulou$^{8}$,
C.A.~Aidala$^{9}$
% ...

{\footnotesize \it

$^{1}$Nikhef National Institute for Subatomic Physics, Amsterdam, Netherlands\\
$^{2}$Department of Physics, University of Warwick, Coventry, United Kingdom\\
$^{3}$Physik-Institut, Universit{\"a}t Z{\"u}rich, Z{\"u}rich, Switzerland\\
$^{4}$Oliver Lodge Laboratory, University of Liverpool, Liverpool, United Kingdom\\
$^{5}$Instituto Galego de F{\'\i}sica de Altas Enerx{\'\i}as (IGFAE), Universidade de Santiago de Compostela, Santiago de Compostela, Spain\\
$^{6}$H.H. Wills Physics Laboratory, University of Bristol, Bristol, United Kingdom\\
$^{7}$Universit{\'e} Clermont Auvergne, CNRS/IN2P3, LPC, Clermont-Ferrand, France\\
$^{8}$LPNHE, Sorbonne Universit{\'e}, Paris Diderot Sorbonne Paris Cit{\'e}, CNRS/IN2P3, Paris, France\\
$^{9}$University of Michigan, Ann Arbor, United States, associated to $^{25}$\\
% ...
$^{25}$Syracuse University, Syracuse, NY, United States\\
% ...
}
\end{flushleft}

\end{document}